\documentclass[acmsmall]{acmart}
\usepackage{CJKutf8}

\usepackage{multirow}
\usepackage{lscape}
\usepackage[figuresright]{rotating}
\usepackage{float}
\usepackage{tabularx}
\usepackage{colortbl}

\usepackage[normalem]{ulem}
\useunder{\uline}{\ul}{}

\AtBeginDocument{%
  \providecommand\BibTeX{{%
    Bib\TeX}}}

\AtBeginDocument{%
  \providecommand\BibTeX{{%
    \normalfont B\kern-0.5em{\scshape i\kern-0.25em b}\kern-0.8em\TeX}}}

\setcopyright{acmcopyright}
\copyrightyear{}
\acmYear{2024}
\acmDOI{}

\usepackage[normalem]{ulem}
\useunder{\uline}{\ul}{}

\acmJournal{CSUR}
\acmMonth{2}

\begin{document}
\settopmatter{printacmref=false}
\title{A Systematic Survey of Deep Learning-based Single-Image Super-Resolution}






\author{Juncheng Li}
\affiliation{%
  \institution{Shanghai University}
  \city{Shanghai}
  \country{China}}
\email{junchengli@shu.edu.cn}

\author{Zehua Pei}
\affiliation{%
  \institution{The Chinese University of Hong Kong}
  \city{Hong Kong}
  \country{China}}
\email{pzehua2000@gmail.com}

\author{Wenjie Li}
\affiliation{%
  \institution{Beijing University of Posts and Telecommunications}
  \city{Beijing}
  \country{China}}
\email{lewj2408@gmail.com}

\author{Guangwei Gao}
\affiliation{%
  \institution{Nanjing University of Posts and Telecommunications}
  \city{Nanjing}
  \country{China}}
\email{csggao@gmail.com}

\author{Longguang Wang}
\affiliation{%
  \institution{Aviation University of Air Force}
  \city{Changchun}
  \country{China}}
\email{wanglongguang15@nudt.edu.cn}

\author{Yingqian Wang}
\affiliation{%
  \institution{National University of Defense Technology}
  \city{Changsha}
  \country{China}}
\email{wangyingqian16@nudt.edu.cn}

\author{Tieyong Zeng}
\affiliation{%
  \institution{The Chinese University of Hong Kong}
  \city{Hong Kong}
  \country{China}}
\email{zeng@math.cuhk.edu.hk}

\renewcommand{\shortauthors}{Juncheng Li et al.}

\begin{abstract}
  Single-image super-resolution (SISR) is an important task in image processing, which aims to enhance the resolution of imaging systems. Recently, SISR has made a huge leap and has achieved promising results with the help of deep learning (DL). In this survey, we give an overview of DL-based SISR methods and group them according to their design targets. Specifically, we first introduce the problem definition, research background, and the significance of SISR. Secondly, we introduce some related works, including benchmark datasets, upsampling methods, optimization objectives, and image quality assessment methods. Thirdly, we provide a detailed investigation of SISR and give some domain-specific applications of it. Fourthly, we present the reconstruction results of some classic SISR methods to intuitively know their performance. Finally, we discuss some issues that still exist in SISR and summarize some new trends and future directions. This is an exhaustive survey of SISR, which can help researchers better understand SISR and inspire more exciting research in this field. An investigation project for SISR is provided at \url{https://github.com/CV-JunchengLi/SISR-Survey}.
\end{abstract}



\maketitle


\begin{figure}[h]
   \begin{center}
   \includegraphics[width=0.55\linewidth]{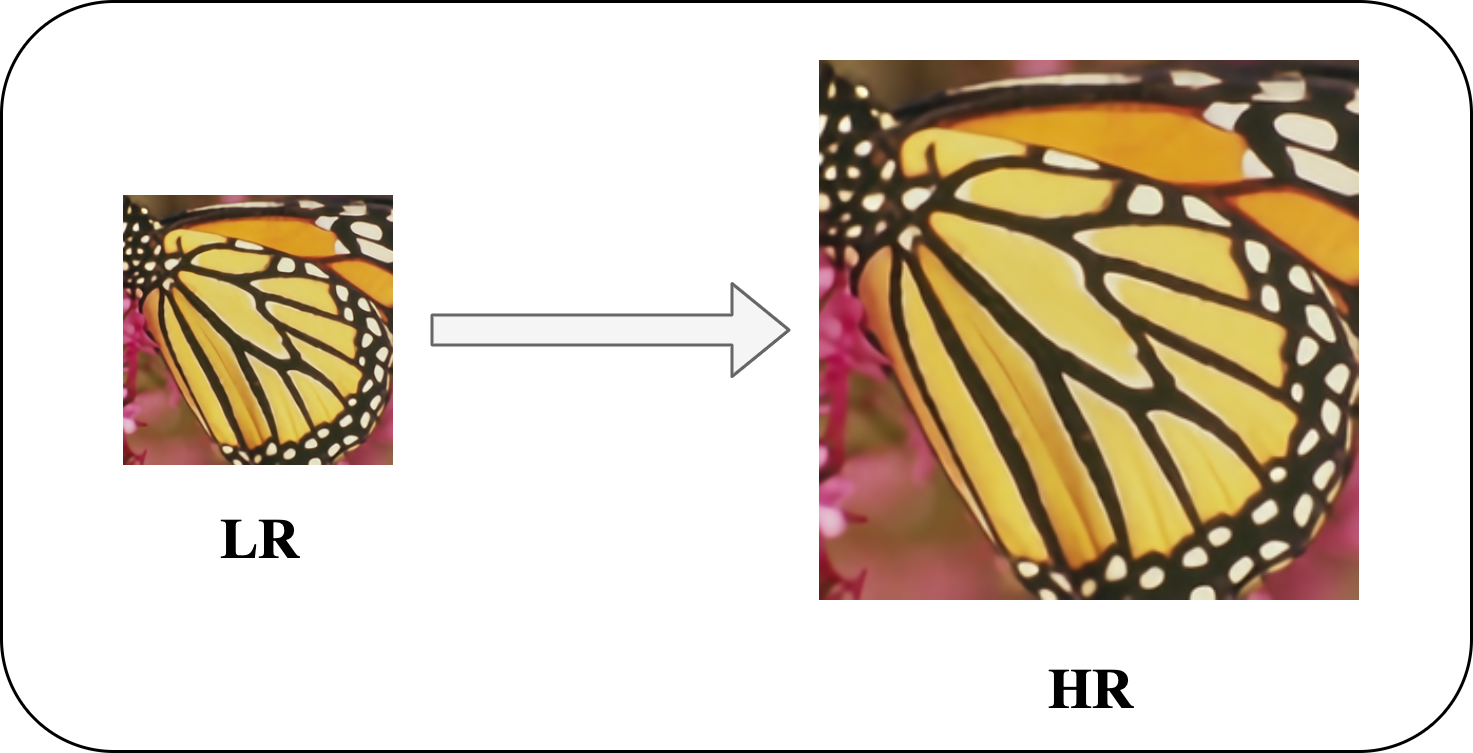}
   \end{center}
   \caption{SISR aims to reconstruct a high-resolution (HR) image from its degraded low-resolution (LR) one.}
   \label{SISR}
\end{figure}

\section{Introduction}
Image super-resolution (SR), especially single-image super-resolution (SISR), is one kind of image transformation task and has received increasing attention in academia and industry. As shown in Fig.~\ref{SISR}, SISR aims to reconstruct a high-resolution (HR) image from its degraded low-resolution (LR) one. It is widely used in various computer vision applications, including security and surveillance images, medical image reconstruction, video enhancement, and image segmentation.

Many SISR methods have been studied long before, such as bicubic interpolation and Lanczos resampling~\cite{duchon1979lan}, which are based on interpolation. However, SISR is an inherently ill-posed problem, and multiple HR images corresponding to the same LR image always exist. To solve this issue, some numerical methods (e.g.,  edge-based methods~\cite{sun2008grad} and image statistics-based methods~\cite{kim2010sparse}) utilize prior information to restrict the solution space. Meanwhile, there are some widely used learning-based methods, such as neighbor embedding methods~\cite{chang2004super} and sparse coding methods~\cite{yang2010image}, which learn a transformation between LR and HR patches.

Recently, deep learning (DL)~\cite{lecun2015deep} has demonstrated better performance than traditional machine learning models in many artificial intelligence fields, including computer vision~\cite{krizhevsky2012imagenet} and natural language processing~\cite{collobert2008unified}. With the rapid development of DL techniques, numerous DL-based methods have been proposed for SISR, continuously prompting the State-Of-The-Art (SOTA) forward. Like other image transformation tasks, the SISR task can generally be divided into three steps: feature extraction and representation, non-linear mapping, and image reconstruction~\cite{dong2015image}. In traditional numerical models, it is time-consuming and inefficient to design an algorithm satisfying all these processes. On the contrary, DL can transfer the SISR task to an almost end-to-end framework incorporating all these three processes, which can greatly decrease manual and computing expenses~\cite{dong2011image}. Additionally, given the ill-posed nature of SISR which can lead to unstable and hard convergence on the results, DL can alleviate this issue through efficient network architecture and loss functions design. Moreover, modern GPU enables deeper and more complex DL models to train fast, which shows greater representation power than traditional numerical models.

\begin{figure*}[t]
   \begin{center}
   \includegraphics[width=1\linewidth]{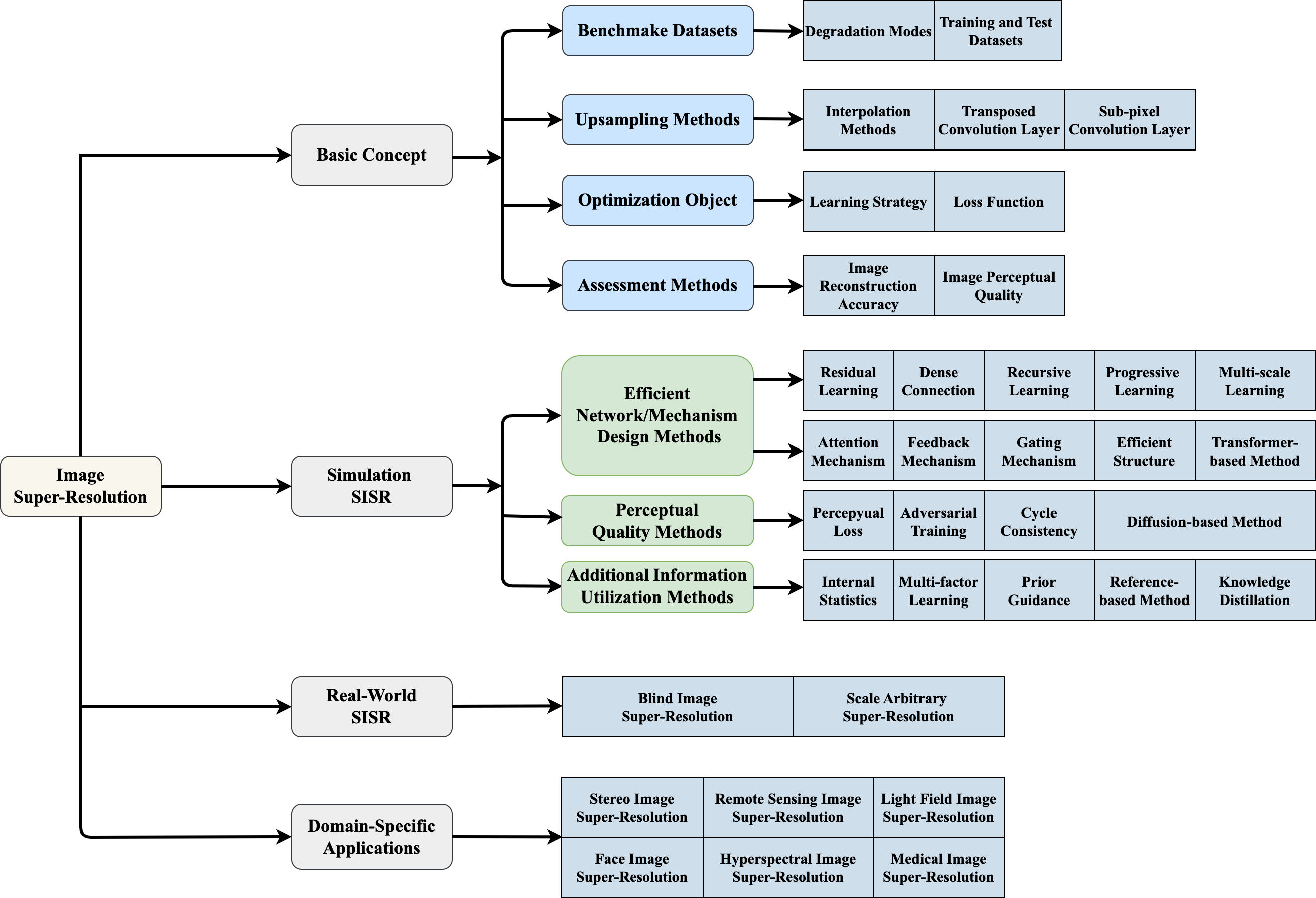}
   \end{center}
   \caption{The content and taxonomy of this survey. In this survey, we divide image super-resolution methods into three categories: Simulation SISR, Real-World SISR, and Domain-Specific Applications.}
   \label{Survey}
\end{figure*}

It is well known that DL-based methods can be divided into supervised and unsupervised methods. This is the simplest classification criterion, but the range of this classification criterion is too large and not clear. As a result, many technically unrelated methods may be classified into the same type while methods with similar strategies may be classified into completely different types. Different from previous SISR surveys~\cite{wang2020deep,anwar2019deep} that use supervision as the classification criterion or introduce the methods in a pure literature way, in this survey, we attempt to give a comprehensive overview of DL-based image super-resolution methods and categorize them according to their specific targets. In Fig.~\ref{Survey}, we show the content and taxonomy of this survey. We divide these methods into three categories: Simulation SISR, Real-World SISR, and Domain-Specific Applications. Additionally, we divide Simulation SISR methods into three categories: Efficient Network / Mechanism Design Methods, Perceptual Quality Methods, and Additional Information Utilization Methods, according to their specific targets. This target-based survey has a clear context hence it is convenient for readers to consult. 
Specifically, in this survey, we first introduce the problem definition, research background, and significance of SISR. Then, we introduce some related works, including benchmark datasets, upsample methods, optimization objectives, and assessment methods. After that, we provide a detailed investigation of SISR methods and provide the reconstruction results of them. Finally, we discuss some issues that still exist in SISR and provide some new trends and future directions. 
Overall, the main contributions of this survey are as follows:
\par (1). We give a thorough overview of DL-based SISR methods according to their targets. This is a new perspective that makes the survey clear in context and convenient.
\par (2). This survey covers more than 100 SR methods and introduces a series of new tasks and domain-specific applications extended by SISR in recent years.
\par (3). We provide a detailed comparison of reconstruction results, including classic, latest, and SOTA SISR methods, to help readers intuitively know their performance.
\par (4). We discuss some issues that still exist in SISR and look forward to the future trend and direction of SR.

\section{Problem Setting and Related Works}
\subsection{Problem Definition}

Image super-resolution is a classic technique to improve the resolution of an imaging system, which can be classified into single-image super-resolution (SISR) and multi-image super-resolution (MISR) according to the number of input LR images. Compared with MISR, SISR is much more challenging since MISR has extra information for reference while SISR only has information of a single input image for the missing image features reconstruction.

Define the low-resolution image as $I_x\in \mathbb{R}^{h \times w}$ and the ground-truth high-resolution image as $I_y\in \mathbb{R}^{H \times W}$, where $H>h$ and $W>w$. Typically, in an SISR framework, the LR image $I_x$ is modeled as $I_x=\mathcal{D}(I_y;\theta_\mathcal{D})$, where $\mathcal{D}$ is a degradation map $\mathbb{R}^{H \times W}\to \mathbb{R}^{h \times w}$ and $\theta_{\mathcal{D}}$ denotes the degradation factor. In most cases, the degradation process is unknown. Therefore, researchers are trying to model it. The most popular degradation mode is:
\begin{equation} \label{degration}
    \mathcal{D}(I_y;\theta_\mathcal{D})=(I_y\otimes \kappa)\downarrow_s + n,
\end{equation}
where $I_y\otimes \kappa$ represents the convolution between the blur kernel $\kappa$ and the HR image $I_y$, $\downarrow_s$ is a subsequent downsampling operation with scale factor $s$, and $n$ is usually the additive white Gaussian noise (AWGN) with standard deviation $\sigma$. In the SISR task, we need to recover an SR image $I_{SR}$ from the LR image $I_x$. Therefore, the task can be formulated as $I_{SR}=\mathcal{F}(I_x;\theta_\mathcal{F})$, where $\mathcal{F}$ is the SR algorithm and $\theta_\mathcal{F}$ is the parameter set of the SR process.

Recently, researchers have converted the SISR into an end-to-end learning task, relying on massive training data and effective loss functions. Meanwhile, more and more DL-based models have been proposed due to the powerful representation power of CNN and its convenience in both forward and backward computing. Therefore, SISR task can be transformed into the following optimization goal:
\begin{equation}
    \hat{\theta}_\mathcal{F}=\mathop{\arg\min}_{\theta_\mathcal{F}} \mathcal{L}(I_{SR},I_y)+\lambda\Phi(\theta),
\end{equation}
where $\mathcal{L}$ denotes the loss function between the generated SR image $I_{SR}$ and the HR image $I_y$, $\Phi(\theta)$ denotes the regularization term, and $\lambda$ is the trade-off parameter that is used to control the weight of the regularization term.

\begin{figure}[h]
   \begin{center}
   \includegraphics[width=0.75\linewidth]{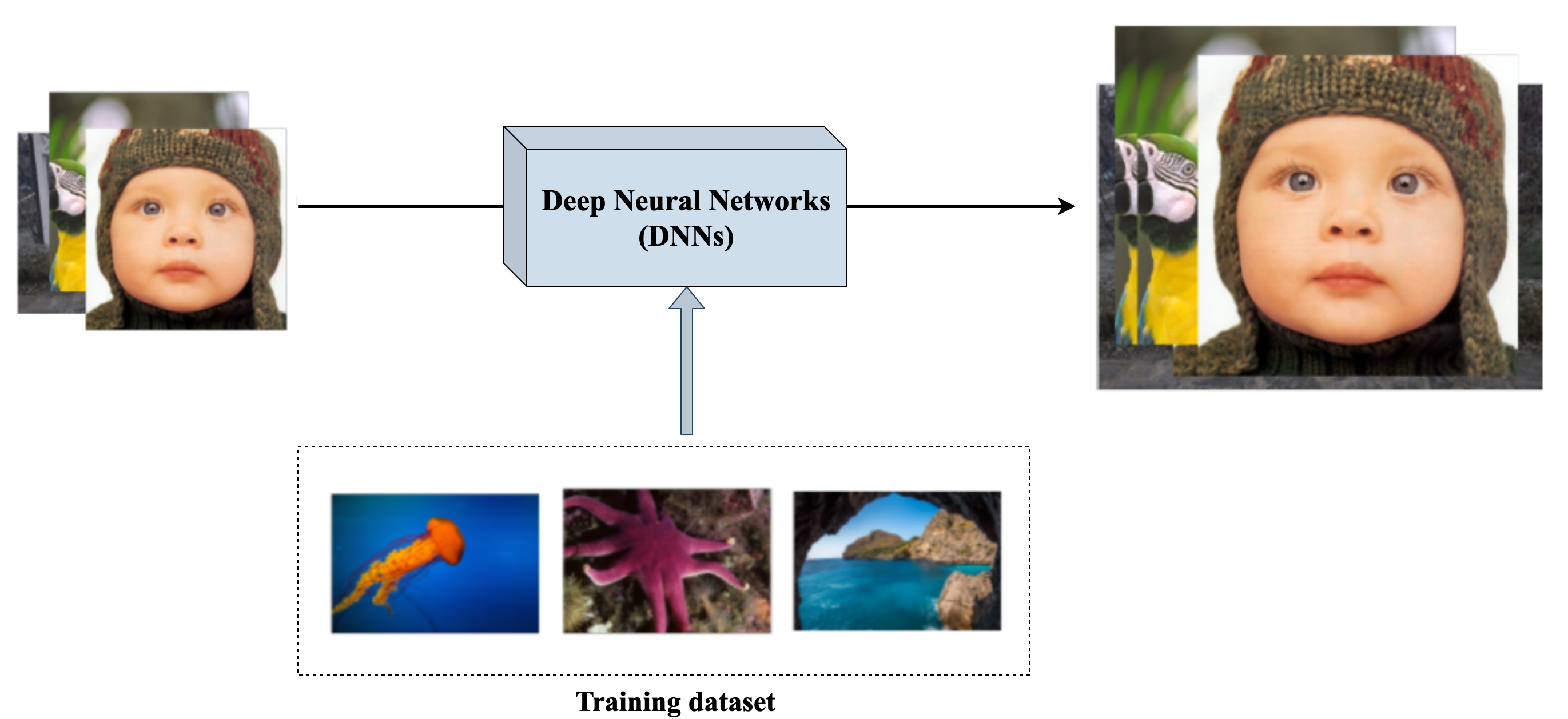}
   \end{center}
   \caption{The training process of data-driven based deep neural networks.} 
   \label{Train}
\end{figure}

\subsection{Benchmark Datasets}

Data is always essential for data-driven models, especially in the DL-based SISR models, to achieve promising reconstruction performance (Fig.~\ref{Train}). Nowadays, industry and academia have launched several available datasets for SISR.

\begin{table*}[t]
   \centering
   \setlength{\tabcolsep}{4.5mm}
   \renewcommand\arraystretch{1}
   \caption{Benchmarks datasets for SISR.}
   \scalebox{0.68}{
   \begin{tabular}{|c|c|c|c|c|}
   \hline
   \rowcolor{gray!70}\textbf{Name} & \textbf{Usage} & \textbf{Amount} & \textbf{Format} & \textbf{Description} \\
   \hline 
   General-100~\cite{dong2016accelerating} & Train & 100 & BMP & Common images with clear edges but fewer smooth regions \\
   \hline
   \rowcolor{gray!30}T91~\cite{yang2010image} & Train & 91 & PNG & Common Images \\
   \hline
   WED~\cite{ma2016waterloo} & Train & 4744 & MAT & Common images \\
   \hline
   \rowcolor{gray!30}Flickr2K~\cite{timofte2017ntire} & Train & 2650  & PNG & 2K images from Flickr \\
   \hline
   FFHQ~\cite{karras2019style} & Train & 70000  & PNG & A high-quality image dataset of human faces \\
   \hline
   \rowcolor{gray!30}CelebA-HQ~\cite{CelebAMask-HQ} & Train/Val & 30000  & PNG & A GAN Synthetic data of human faces \\
   \hline
   CelebA~\cite{liu2015deep} & Train/Val & 202600   & JPG & a large-scale face attributes dataset \\
   \hline
   \rowcolor{gray!30}DRealSR~\cite{wei2020component} & Train/Val & 31970   & PNG & a benchmark with diverse real-world degradation processes \\
   \hline
   DIV2K~\cite{agustsson2017ntire} & Train/Val & 1000  & PNG & High-quality dataset for CVPR NTIRE competition \\
   \hline
   \rowcolor{gray!30}BSDS300~\cite{martin2001database} & Train/Val & 300  & JPG & Common images \\
   \hline
   BSDS500~\cite{arbelaez2011} & Train/Val & 500  & JPG & Common images \\
   \hline 
   \rowcolor{gray!30}RealSR~\cite{cai2019toward} & Train/Val & 100 & PNG  & 100 real-world low and high resolution image pairs \\
   \hline 
   OutdoorScene~\cite{wang2018recovering} & Train/Val & 10624  & PNG & Images of outdoor scenes \\
   \hline 
   \rowcolor{gray!30}City100~\cite{chen2019camera} & Train/Test & 100  & RAW & Common images \\
   \hline
   Flickr1024~\cite{wang2019flickr1024} & Train/Test & 100  & RAW & Stereo images used for Stereo SR \\
   \hline
   \rowcolor{gray!30}SR-RAW~\cite{zhang2019image} & Train/Test & 7*500  & JPG/ARW & Raw images produced by real-world computational zoom \\
   \hline
   PIPAL~\cite{jinjin2020pipal} & Test & 200  & PNG & Perceptual image quality assessment dataset \\
   \hline
   \rowcolor{gray!30}Set5~\cite{bevilacqua2012low} & Test & 5 & PNG & Common images, only 5 images \\
   \hline
   Set14~\cite{zeyde2010single} & Test & 14  & PNG & Common images, only 14 images \\
   \hline 
   \rowcolor{gray!30}BSD100~\cite{martin2001database} & Test & 100 & JPG & A subset of BSDS500 for testing \\
   \hline
   Urban100~\cite{huang2015single} & Test & 100  & PNG & Images of real-world structures \\
   \hline
   \rowcolor{gray!30}Manga109~\cite{fujimoto2016manga109} & Test & 109 & PNG & Japanese manga\\
   \hline 
   L20~\cite{timofte2016seven} & Test & 20  & PNG & Common images, very high-resolution \\
   \hline
   \rowcolor{gray!30}PIRM~\cite{blau20182018} & Test & 200  & PNG & Common images, datasets for ECCV PIRM competition \\
   \hline
   \end{tabular}
   }
   \label{dataset}
\end{table*} 

\subsubsection{\textbf{Degradation Mode}}\label{DD}
Due to the particularity of the SISR task, it is difficult to construct a large-scale paired real SR dataset. Therefore, researchers often apply degradation patterns on the aforementioned datasets to obtain corresponding degraded images to construct paired datasets. However, images in the real world are easily disturbed by various factors (e.g., sensor noise, motion blur, and compression artifacts), resulting in the captured images being more complex than the simulated ones. To alleviate these problems and train a more effective and general SISR model, some works model the degradation mode as a combination of several operations (Eq.~\ref{degration}). Based on this degradation formula, the three most widely used degradation modes have been proposed: BI, BD, and DN. Among them, \textbf{BI} is the most widely used degraded mode to simulate LR images, which is essentially a bicubic downsampling operation. For \textbf{BD}, the HR images are blurred by a Gaussian kernel of size $7 \times 7$ with standard deviation \textit{1.6} and then downsampled with a scaling factor of 3. To obtain LR images under DN mode, the Bicubic downsampling is performed on the HR image with a scaling factor of 3, and then the Gaussian noise with a noise level of 30 is added to the image.

\subsubsection{\textbf{Training and Test Datasets}}

Recently, many datasets for the SISR task have been proposed, including BSDS300~\cite{martin2001database}, DIV2K~\cite{agustsson2017ntire}, and Flickr2K~\cite{timofte2017ntire}. Meanwhile, there are also many test datasets that can be used to effectively test the performance of the models, such as Set5~\cite{bevilacqua2012low}, Set14~\cite{zeyde2010single}, Urban100~\cite{huang2015single}, and Manga109~\cite{fujimoto2016manga109}. In Table~\ref{dataset}, we list a series of commonly used datasets and indicate their detailed attribute.

Among these datasets, DIV2K~\cite{agustsson2017ntire} is the most widely used dataset for model training, which is a high-quality dataset that contains 800 training images, 100 validation images, and 100 test images. Flickr2k is a large extended dataset, which contains 2650 2K images from Flickr. RealSR~\cite{cai2019toward} is the first truly collected real-world SISR dataset with paired LR and HR images. In addition to the listed datasets, some datasets widely used in other computer vision tasks are also used as supplementary training datasets for SISR, such as ImageNet~\cite{deng2009imagenet} and CelebA~\cite{liu2015deep}. In addition, combining multiple datasets (e.g., DF2K) for training to further improve the model performance has been also widely used.

\subsection{Upsampling Methods}

The purpose of SISR is to enlarge a smaller size image into a larger one and to keep it as accurate as possible. Therefore, enlargement operation, also called upsampling, is an important step in SISR. The current upsampling mechanisms can be divided into four types: pre-upsampling SR, post-upsampling SR, progressive upsampling SR, and iterative up-and-down sampling SR. In this section, we introduce several upsampling methods that support these upsampling mechanisms. 

\subsubsection{\textbf{Interpolation Methods}}

Interpolation is the most widely used upsampling method. The current mainstream of interpolation methods includes Nearest-neighbor Interpolation, Bilinear Interpolation, and Bicubic Interpolation. Being highly interpretable and easy to implement, these methods are still widely used today. Among them, \textbf{Nearest-neighbor Interpolation} is a simple and intuitive algorithm that selects the nearest pixel value for each position to be interpolated, which has fast execution time but has difficulty in producing high-quality results. \textbf{Bilinear Interpolation} sequentially performs linear interpolation operations on the two axes of the image. This method can obtain better results than nearest-neighbor interpolation while maintaining a relatively fast speed. \textbf{Bicubic Interpolation} performs cubic interpolation on each of the two axes. Compared with Bilinear, the results of Bicubic are smoother with fewer artifacts but slower than other interpolation methods. Interpolation is also the mainstream method for constructing SISR-paired datasets and is widely used in the data pre-processing of DL-based SISR models.

\subsubsection{\textbf{Transposed Convolutional Layers}}
As shown in Fig.~\ref{SBPixel} (a), researchers usually consider two kinds of transposed convolution operations: one adds padding around the input matrix and then applies the convolution operation, and the other adds padding between the values of the input matrix followed by the direct convolution operation. The latter is also called fractionally strided convolution since it works like performing convolution with a sub-pixel level stride. In the transposed convolutional layer, the upsampling level is controlled by the size of the padding, which is essentially opposite to the operation of the normal convolutional layer. The transposed convolutional layer is first proposed in FSRCNN~\cite{dong2016accelerating} and is widely used in DL-based SISR models.

\subsubsection{\textbf{Sub-pixel Convolutional Layer}}

\begin{figure}[t]
    \begin{center}
    \includegraphics[width=1\linewidth]{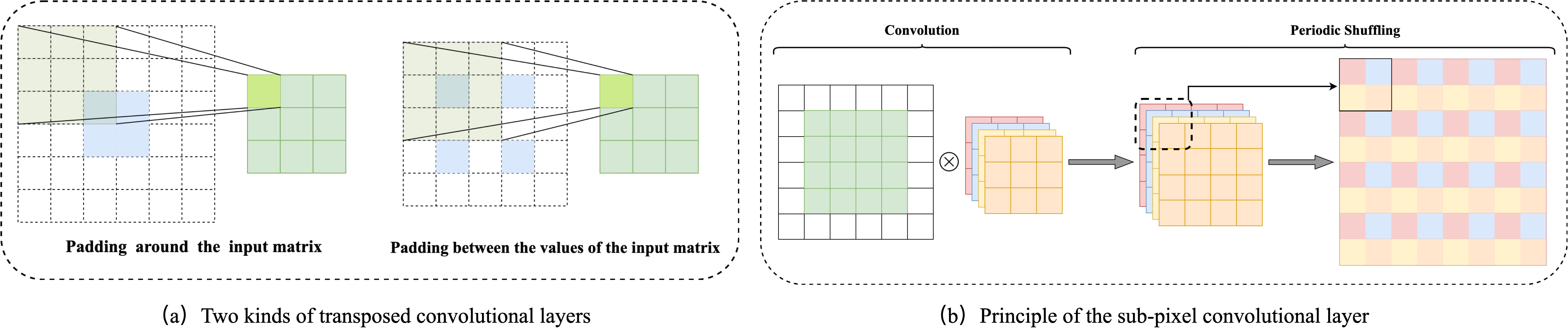}
    \end{center}
    \caption{Upsampling methods: (a) transposed convolutional layers (b) sub-pixel convolutional layer.} 
    \label{SBPixel}
\end{figure}

In ESPCN~\cite{shi2016real}, Shi \emph{et al.} proposed an efficient sub-pixel convolutional layer. Instead of increasing the resolution by directly increasing the number of LR feature maps, sub-pixel first increases the dimension of LR feature maps, i.e., the number of the LR feature maps, and then a periodic shuffling operator is used to rearrange these points in the expanded feature maps to obtain the HR output (Fig.~\ref{SBPixel} (b)). In detail, the formulation of the sub-pixel convolutional layer can be defined as follows:
\begin{equation}
    I_{SR}=f^L(I_x)=\mathcal{PS}(W_L*f^{L-1}(I_x)+b_L),
\end{equation}
where $\mathcal{PS}$ denotes the periodic shuffling operator, which transfers an $h\times w\times C\cdot r^2$ tensor to a tensor of shape $rh\times rw\times C$, and $rh \times rw$ is explicitly the size of the HR image, $C$ is the number of channels. In addition, the convolutional filter $W_L$ has the shape $n_{L-1}\times r^2C\times K_L\times K_L$, where $n_L$ is the number of feature maps in the ($L-1$) layer. Compared with the transposed convolutional layer, the sub-pixel convolutional layer exhibits better efficiency and thus is widely used in DL-based SISR models.

\subsection{Optimization Objective}

Evaluation and parameter up-gradation are the important steps in all DL-based models. In this section, we will introduce the necessary procedures during the model training.

\subsubsection{\textbf{Learning Strategy}}\label{Content} 
In this work, we use a common division method in the SR field, that is, whether paired LR-HR images are used for model training. It is worth noting that the HR image here refers to the additional introduced high-resolution image, not the image itself. In addition, learning strategy has no clear definitions in SISR. According to this criteria, the DL-based SISR models can be mainly divided into supervised learning methods and unsupervised learning methods.

\textbf{Supervised Learning}: In SISR, we often call the method of using pairs of LR-HR images for training a supervised learning paradigm. In simulated SISR, LR images are often obtained by downsampling HR images. In real SISR, LR images and HR images are obtained by adjusting the zoom of the camera. In general, the LR and HR images of this type of method have a one-to-one correspondence, and researchers compute the reconstruction error between the ground-truth image $I_y$ and the reconstructed image $I_{SR}$:
\begin{equation}
\hat{\theta}_\mathcal{F}=\mathop{\arg\min}_{\theta_\mathcal{F}} \mathcal{L}(I_{SR},I_y).
\end{equation}

Alternatively, researchers may sometimes search for a mapping $\Phi$, such as a pre-trained neural network, to transform the images or image feature maps to other space and then compute the error:
\begin{equation}
\hat{\theta}_\mathcal{F}=\mathop{\arg\min}_{\theta_\mathcal{F}} \mathcal{L}(\Phi(I_{SR}),\Phi(I_y)).
\end{equation}

Among them, $\mathcal{L}$ is the loss function that is used to minimize the distance between the reconstructed image and the ground-truth image. By using different loss functions, the model can achieve different performance. Therefore, an effective loss function is also crucial for SISR.

\textbf{Unsupervised Learning}: The simulated paired images have poor versatility, while the real paired images are difficult to collect. To address this issue, some methods began to try to no longer use paired LR-HR images for training. We often call this type of method an unsupervised learning method. This type of unsupervised method no longer uses paired LR-HR images for training but uses unpaired LR-HR images (GAN-based method) or itself (self-supervised learning method) for training. For example, ZSSR~\cite{shocher2018zero} uses the test image and its downscaling versions with the data augmentation approaches to build the "training dataset" and then applies the loss function to optimize the model. In addition, weakly-supervised learning also belongs to the unsupervised learning strategy. Among them, some researchers first learn the HR-to-LR degradation and use it to construct datasets for training the model, while other researchers design cycle-in-cycle models to learn the LR-to-HR and HR-to-LR mappings simultaneously. For instance, CinCGAN~\cite{yuan2018unsupervised} consists of two CycleGAN~\cite{zhu2017unpaired}, where one cycle is adopted for translating between the real LR and synthetic LR images while the other is used between the real LR and HR images.

\subsubsection{\textbf{Loss Function}} 

In the SISR task, the loss function is used to guide the iterative optimization process of the model by computing some kind of error. Meanwhile, compared with a single loss function, researchers find that combining multiple loss functions can better reflect the situation of image restoration. In this section, we briefly introduce several commonly used loss functions.

\textbf{Pixel Loss}: Pixel loss is the simplest and most popular loss function in SISR, which aims to measure the difference between two images on a pixel basis so that these two images can converge as close as possible. It mainly includes the L1 loss, Mean Square Error (MSE) Loss, and Charbonnier loss (a differentiable variant of the L1 loss): 
\begin{equation}
   \mathcal{L}_{L1}(I_{SR}, I_{y}) = \frac{1}{hwc} \sum_{i,j,k} \left | I_{SR}^{i,j,k} - I_{y}^{i,j,k} \right |,
\end{equation}
\begin{equation}
   \mathcal{L}_{MSE}(I_{SR}, I_{y}) = \frac{1}{hwc} \sum_{i,j,k} (I_{SR}^{i,j,k} - I_{y}^{i,j,k})^{2},
\end{equation}
\begin{equation}
   \mathcal{L}_{Char}(I_{SR}, I_{y}) = \frac{1}{hwc} \sum_{i,j,k} \sqrt{(I_{SR}^{i,j,k} - I_{y}^{i,j,k})^{2} + \epsilon^{2}},
\end{equation}
where, $h$, $w$, and $c$ are the height, width, and the number of channels of the image. $\epsilon$ is a numerical stability constant, usually being set to $10^{-3}$. Since most mainstream image evaluation indicators are highly correlated with pixel-by-pixel differences, pixel loss is still widely used. However, the images reconstructed by this type of loss function usually lack high-frequency details and thus perform inferior in visual effects.

\textbf{Content Loss}: Content loss is also termed perceptual loss, which uses a pre-trained classification network to measure the semantic difference between images, and can be further expressed as the Euclidean distance between the high-level representations of these two images:
\begin{equation}
   \mathcal{L}_{Cont}(I_{SR}, I_{y}, \phi) = \frac{1}{h_{l}w_{l}c_{l}} \sum_{i,j,k} (\phi^{i,j,k}_{(l)} (I_{SR}) - \phi^{i,j,k}_{(l)} (I_{y})),
\end{equation}
where $\phi$ represents the pre-trained classification network and $\phi_{(l)}(I_{HQ})$ represents the high-level representation extracted from the $l$ layer of the network. $h_{l}$, $w_{l}$, and $c_{l}$ are the height, width, and the number of channels of the feature map in the $l$-th layer, respectively. By using this loss, the visual effects of these two images can be as consistent as possible. Among them, VGG~\cite{simonyan2014very} and ResNet~\cite{ledig2017photo} are the most commonly used pre-training classification networks.

\textbf{Adversarial Loss}: To make the reconstructed SR image more realistic, Generative Adversarial Networks (GANs~\cite{goodfellow2014generative}) have been introduced into the SISR task. Specifically, GAN is composed of a generator and a discriminator. The generator is responsible for generating fake samples, and the discriminator is used to determine the authenticity of the generated samples. For example, the discriminative loss function based on cross-entropy is proposed by SRGAN~\cite{ledig2017photo}:
\begin{equation}
   \mathcal{L}_{Adversarial}(I_{x},G, D) = \sum_{n=1}^{N} -logD(G(I_{x})),
\end{equation}
where $G(I_{LQ})$ is the reconstructed SR image, $G$ and $D$ represent the Generator and the Discriminator, respectively. 

\textbf{Prior Loss:} Apart from the above loss functions, some prior knowledge can also be introduced into SISR models to participate in high-quality image reconstruction, such as sparse prior, gradient prior, and edge prior. Among them, gradient prior loss and edge prior loss are the most widely used prior loss functions, which are defined as follows:
\begin{equation}
\small
   \mathcal{L}_{TV}(I_{SR}) = \frac{1}{hwc} \sum_{i,j,k}\sqrt{(I_{SR}^{i,j+1,k}-I_{y}^{i,j,k})^{2} + (I_{SR}^{i+1,j,k}-I_{y}^{i,j,k})^{2}},
\end{equation}
\begin{equation}
   \mathcal{L}_{Edge}(I_{SR}, I_{y}, E) = \frac{1}{hwc} \sum_{i,j,k} \left | E(I_{SR}^{i,j,k}) - E(I_{y}^{i,j,k}) \right |,
\end{equation}
where $E$ is the image edge detector, and $E(I_{SR}^{i,j,k})$ and $E(I_{y}^{i,j,k})$ are the image edges extracted by the detector. The purpose of the prior loss is to optimize some specific information of the image toward the expected target so that the model can converge faster and the reconstructed image will contain more texture details.

\textbf{Fourier Space Loss:} The design of perceptual losses predominantly focuses on the spatial domain. However, SR is tightly coupled to the frequency domain, as only high frequencies are removed during the downsampling process. To solve this problem, Fuoli \emph{et al.}~\cite{fuoli2021fourier} propose a novel Fourier Space Loss by calculating the frequency components with the Fast Fourier Transform (FFT) for direct emphasis on the frequency content. Firstly, the image is transformed into Fourier space by applying the Fast Fourier transform (FFT). Then, the method calculates the amplitude difference $F_{f}$, $\left | . \right |$ and phase difference, $\angle$ of all frequency components between output image and ground truth image. The averaged differences are computed as the total frequency loss as follows:

\begin{equation}
\small
   L_{f}, \left | . \right | = \frac{2}{UV} \sum_{u=0}^{U/2-1}\sum_{v=0}^{V-1} \Big\vert \mid \hat{Y} \mid_{u,v} - \mid Y \mid_{u,v} \Big\vert,
\end{equation}
\begin{equation}
\small
   L_{f}, \angle = \frac{2}{UV} \sum_{u=0}^{U/2-1}\sum_{v=0}^{V-1} \Big\vert \angle \hat{Y}_{u,v} - \angle Y_{u,v} \Big\vert,
\end{equation}
\begin{equation}
\small
   L_{f} = \frac{1}{2} L_{f}, \left | . \right | + \frac{1}{2} L_{f}, \angle ,
\end{equation}
where $\hat{Y}_{u,v}$ represents the spectrum of the recovered image, and $Y_{u,v}$ represents the spectrum of the ground truth image.

\textbf{Mixed Loss:} In SISR, there are also some classic combinations of loss functions that are widely used to guide the network towards generating high-quality HR images. These combinations aim to balance the quality, details, and visual perception of the generated image. Here are some commonly used classic combinations of loss functions. 

L1 + Perceptual Loss: combining L1 loss with perceptual loss, such as the feature loss based on VGG networks, can generate images that are clearer and have better details. This combination can effectively reduce noise and distortion in the image; L1 + TV Loss: combining L1 loss with total variation (TV) loss can generate images with good edge and texture details. TV loss helps to reduce blocky artifacts in the image; Content Loss + Adaptive Loss: combining Content loss with adaptive loss can generate images with better visual coherence. Adaptive loss can adjust the loss weights based on the content of the image.

The choice of loss function combinations depends on the specific requirements of the SISR task, such as the desired balance between perceptual quality and computational efficiency. In practical applications, researchers may adjust the weights of the loss functions based on experimental results to find the combination that best suits a specific task.

\subsection{Assessment Methods}
The image quality assessment (IQA) can be generally divided into objective methods and subjective methods. Objective methods commonly use a specific formulation to compute the results, which are simple and fair, thus becoming the mainstream assessment method in SISR. However, they can only reflect the recovery of image pixels from a numerical point of view and are difficult to accurately measure the true visual effect of the image. In contrast, subjective methods are always based on human subjective judgments and are more related to evaluating the perceptual quality of the image. Based on the pros and cons of the two types of methods mentioned above, several assessment methods are briefly introduced in the following with respect to the aspects of image reconstruction accuracy, image perceptual quality, and reconstruction efficiency.

\subsubsection{\textbf{Image Reconstruction Accuracy}}

The assessment methods used for image reconstruction accuracy evaluation are also called \emph{Distortion measures}, which are full-reference. Specifically, given a distorted image $\hat{x}$ and a ground-truth reference image $x$, full-reference distortion quantifies the quality of $\hat{x}$ by measuring its discrepancy to $x$~\cite{blau2018perception} using different algorithms.

\textbf{Peak Signal-to-Noise Ratio (PSNR)}: PSNR is the most widely used IQA method in the SISR field, which can be easily defined via the mean squared error (MSE) between the ground truth image $I_y\in \mathbb{R}^{H \times W}$ and the reconstructed image $I_{SR}\in \mathbb{R}^{H \times W}$:
\begin{equation}
    MSE=\frac{1}{HW}\sum_{i=0}^{H-1}\sum_{j=0}^{W-1}(I_y(i,j)-I_{SR}(i,j))^2,
\end{equation}
\begin{equation}
    PSNR=10 \cdot \log_{10}(\frac{MAX^2}{MSE}),
\end{equation}
where MAX is the maximum possible pixel of the image. Since PSNR is highly related to MSE, a model trained with the MSE loss will be expected to have high PSNR scores. Although higher PSNR generally indicates that the construction is of higher quality, it just considers the per-pixel MSE, which makes it fail to capture the perceptual differences~\cite{wang2009mean}.

\textbf{Structural Similarity Index Measure (SSIM)}: SSIM~\cite{wang2004image} is another popular assessment method that measures the similarity between two images on a perceptual basis, including structures, luminance, and contrast. Different from PSNR, which calculates absolute pixel-level errors, SSIM suggests that there exist strong inter-dependencies among the spatially adjacent pixels. These dependencies carry important information related to the structures perceptually. Therefore, the SSIM can be expressed as a weighted combination of three comparative measures: 
\begin{equation}
    \begin{split}
    SSIM(I_{SR},I_y) &=(l(I_{SR},i_y)^\alpha \cdot c(I_{SR},I_y)^\beta \cdot s(I_{SR},I_y)^\gamma) \\
    &=\frac{(2\mu_{I_{SR}}\mu_{I_y}+c_1)(2\sigma_{I_{SR}I_y}+c_2)}{(\mu_{I_{SR}}^2+\mu_{I_y}^2+c_1)(\sigma_{I_{SR}}^2+\sigma_{I_y}^2+c_2)},
    \end{split}
\end{equation}
where $l$, $c$, and $s$ represent luminance, contrast, and structure between $I_{SR}$ and $I_y$, respectively. $\mu_{I_{SR}}$, $\mu_{I_{y}}$, $\sigma_{I_{SR}}^2$, $\sigma_{I_y}^2$, and $\sigma_{I_{SR}I_y}$ are the average value, variance, and covariance of the corresponding items, respectively.

A higher SSIM indicates higher similarity between two images, which has been widely used due to its convenience and stable performance in evaluating perceptual quality. In addition, there are also some variants of SSIM, such as Multi-Scale SSIM, which is conducted over multiple scales by a process of multiple stages of subsampling.

\subsubsection{\textbf{Image Perceptual Quality}}

Since the visual system of humans is complex and concerns many aspects to judge the differences between two images, i.e., the textures and flow inside the images, methods that pursue absolutely similar differences (PSNR/SSIM) will not always perform well. Although distortion measures have been widely used, the improvement in reconstruction accuracy is not always accompanied by an improvement in visual quality. In fact, researchers have shown that the distortion and perceptual quality are at odds with each other in some cases~\cite{blau2018perception}. The image perceptual quality of an image $\hat{x}$ is defined as the degree to which it looks like a natural image, which has nothing to do with its similarity to any reference image.

\textbf{Mean Opinion Score (MOS)}: MOS is a subjective method that can straightforwardly evaluate perceptual quality. Specifically, several volunteers rate their opinions on the quality of a set of images by Double-stimulus~\cite{mittal2012no}, i.e., every volunteer has both the source and test images. After all the volunteers finish ratings, the results are mapped onto numerical values, and the average scores will be the final MOS. MOS is a time-consuming and expensive method since it requires manual participation. Meanwhile, MOS is also doubted to be unstable, since the MOS differences may be not noticeable to the users. Moreover, this method is too subjective to guarantee fairness.

\textbf{Learned Perceptual Image Patch Similarity (LPIPS)}: LPIPS~\cite{zhang2018unreasonable} is a popular metric used to measure the perceived differences between differnet images, which not only focuses on the structure and content of an image but also reflects the sensitivity of the human eye to image differences. Specifically, the feature layers of different images are first extracted using a pre-trained model(e.g., VGG~\cite{simonyan2014very}), and then the LPIPS value can be obtained by calculating the weighted summed distance between the different feature spaces:
\begin{equation}
   \begin{small}
    LPIPS({I_{SR}},{I_y}) = \sum\limits_{l = 1}^N {{{\left\| {{\omega _l}.\left( {{\phi _l}\left( {{I_{SR}}} \right),{\phi _l}\left( {{I_y}} \right)} \right)} \right\|}_2}} ,
    \end{small}
\end{equation}
where $l$ is the lth feature layer of the pre-trained model, $N$ is the total number of feature layers of the pre-trained model, ${{\omega _l}}$ is the weight used to weigh the lth feature layers, ${{\phi _l}}$ is the lth feature extraction layer in the pre-trained model, and ${\left\| {} \right\|_2}$ is the L2 paradigms. However, LPIPS is obtained by learning from DL models, so its performance is affected by the training data, leading to the fact that LPIPS may lack generalization ability in some cases.

\textbf{Deep Image Structure and Texture Similarity (DISTS)}: DISTS~\cite{ding2020image} is the first complete reference image quality model that explicitly tolerates texture resampling, and it utilizes injective differentiable functions constructed from CNN to convert images to a multi-scale hyper-complete representation, a representation in which the spatial average of the feature maps captures the texture appearance and matches human ratings of image quality.
\begin{equation}
   \begin{small}
    l(I_{SR}^{(i)},I_y^{(i)}) = \frac{{2\mu _{{I_{SR}}}^{(i)}\mu _{{I_y}}^{(i)} + {c_1}}}{{{{(\mu _{{I_{SR}}}^{(i)})}^2} + {{(\mu _{{I_y}}^{(i)})}^2} + {c_1}}},
    \end{small}
\end{equation}
\begin{equation}
   \begin{small}
    s(I_{SR}^{(i)},I_y^{(i)}) = \frac{{2\sigma _{{I_{SR}}{I_y}}^{(i)} + {c_2}}}{{{{(\sigma _{{I_{SR}}}^{(i)})}^2} + {{(\sigma _{{I_y}}^{(i)})}^2} + {c_2}}},
    \end{small}
\end{equation}
\begin{equation}
   \begin{small}
    DISTS({I_{SR}},{I_y},\alpha ,\beta ) = 1 - \sum\limits_{i = 0}^m {\sum\limits_{j = 1}^{{n_i}} {\left( {{\alpha _{ij}}l(I_{SR}^{(i)},I_y^{(i)}) + {\beta _{ij}}s(I_{SR}^{(i)},I_y^{(i)})} \right)} } ,
    \end{small}
\end{equation}
where ${\mu _{{I_{SR}}}^{(i)}}$, ${\mu _{{I_y}}^{(i)}}$, ${\sigma _{{I_{SR}}}^{(i)}}$, ${\sigma _{{I_y}}^{(i)}}$ and ${\sigma _{{I_{SR}}{I_y}}^{(i)}}$ denote average value and variances of $I_{SR}^{(i)}$ and $I_y^{(i)}$, and covariance between  $I_{SR}^{(i)}$ and $I_y^{(i)}$, respectively. ${{c_1}}$ and ${{c_2}}$ are two small
positive constants. And $\left\{ {{\alpha _{ij}},{\beta _{ij}}} \right\}$ are learnable weights, satisfying $\sum {_{i = 0}^m} \sum {_{j = 1}^{{n_j}}({\alpha _{ij}} + {\beta _{ij}})}  = 1$. And despite its beneficial mathematical properties, the DISTS metric is still highly non-convex and therefore requires more iterations to recover from random noise using stochastic gradient descent methods than metrics such as SSIM.

\textbf{Natural Image Quality Evaluator (NIQE)}: NIQE~\cite{mittal2012making} is a completely blind image quality assessment method. Without the requirement of knowledge about anticipated distortions in the form of training examples and corresponding human opinion scores, NIQE only makes use of measurable deviations from statistical regularities observed in natural images. It extracts a set of local features from images based on a natural scene statistic (NSS) model, then fits the feature vectors to a multivariate Gaussian (MVG) model. The quality of a test image is then predicted by the distance between its MVG model and the MVG model learned from a natural image:
\begin{equation}
   \begin{small}
    D(\nu_1,\nu_2,\Sigma_1,\Sigma_2)=\sqrt{((\nu_1-\nu_2)^T(\frac{\Sigma_1+\Sigma_2}{2})^{-1}(\nu_1-\nu_2))},
    \end{small}
\end{equation}
where $\nu_1$, $\nu_2$, and $\Sigma_1$, $\Sigma_2$ are the mean vectors and covariance matrices of the HR and SR image's MVG model, respectively. Notice that, a higher NQIE index indicates lower image perceptual quality. Compared with MOS, NIQE is a more convenient perceptual evaluation method.

\textbf{Ma}: Ma \emph{et al.}~\cite{ma2017learning} proposed a learning-based no-reference image quality assessment. It is designed to focus on SR images, while other learning-based methods are applied to images degraded by noise, compression, or fast fading rather than SR images. It learns from perceptual scores based on human subject studies involving a large number of SR images. Then, it quantifies the SR artifacts through three types of statistical properties, i.e., local/global frequency variations and spatial discontinuity. Afterward, these features are modeled by three independent learnable regression forests respectively to fit the perceptual scores of SR images, $\hat{y}_n (n=1,2,3)$. The final predicted quality score is $\hat{y}=\sum_n \lambda_n\cdot\hat{y}_n$, and the weight $\lambda$ is learned by minimizing $\lambda^*=\mathop{\arg\min}_{\lambda} (\sum_n \lambda_n\cdot\hat{y}_n-y)^2$.

Ma performs well on matching the perceptual scores of SR images but is still limited as compared with other learning-based no-reference methods since it can only assess the quality degradation arising from the distortion types on which they have been trained.

\textbf{Perception Index (PI)}: In the 2018 PIRM Challenge on Perceptual Image Super-Resolution~\cite{blau20182018}, perception index (PI) is first proposed to evaluate perceptual quality. It is a combination of the no-reference image quality measures Ma and NIQE:
\begin{equation}
    PI=\frac{1}{2}((10-Ma)+NIQE).
\end{equation}
A lower PI indicates better perceptual quality. This is a new image quality evaluation standard, which has been greatly promoted and used in recent years.

Apart from the aforementioned evaluation methods, some new methods have also been proposed over these years. For example, Zhang \emph{et al.}~\cite{zhang2019ranksrgan} proposed $Ranker$ to learn the ranking orders of NR-IQA methods (i.e., NIQE) on the results of some perceptual SR models. Zhang \emph{et al.}~\cite{zhang2018unreasonable} introduced a new dataset of human perceptual similarity judgments. Meanwhile, a perceptual evaluation metric, Learned Perceptual Image Patch Similarity (LPIPS), is constructed by learning the perceptual judgment in this dataset. Ramsauer \emph{et al.}~\cite{heusel2017gans} proposed Fréchet Inception Distance (FID), which quantifies the quality of SISR images by comparing the difference between the data distribution of SISR results and the true data distribution. In summary, how to measure the perceptual quality of SR images more accurately and efficiently is an important issue that needs to be explored.

\section{Image Super-Resolution}
In 2014, Dong \emph{et al.}~\cite{dong2015image} proposed the Super-Resolution Convolutional Neural Network (SRCNN). SRCNN is the first CNN-based SISR model. It shows that a deep CNN model is equivalent to the sparse-coding-based method, which is an example-based method for SISR. Recently, more and more SR models treat it as an end-to-end learning task. Therefore, building a deep neural network to directly learn the mapping between LR and HR images has become the mainstream method in SR. After that, CNN-based SR methods are blooming and constantly refreshing the best results. 

In this part, we divide DL-based imgae super-resolution methods into three categories: Simulation SISR, Real-World SISR, and Domain-Specific Applications.

\subsection{Simulation SISR}
In recent years, the field of SISR has developed rapidly, and a large number of excellent models have emerged. However, it is worth noting that most of these models use simulated datasets for testing and training, we call this method simulated SISR. In other words, the low-resolution images used in this type of method are usually obtained by applying some fixed degradation modes to the high-resolution images. This will affect the performance of the model in practical applications. However, it is undeniable that the emergence of these methods has enriched and promoted the development of SISR. According to different design targets, we divide these methods into three categories: efficient network/mechanism design methods, perceptual quality methods, and additional information utilization methods.

\subsubsection{\textbf{Efficient Network / Mechanism Design Methods}}
Most of the methods that have emerged in recent years focus on efficient and accurate network structure and mechanism design, which enable the model to achieve better performance with fewer parameters. In this section, we will discuss some methods that contribute to efficient and accurate network design.

\begin{figure}[h]
\centering
\includegraphics[width=8cm]{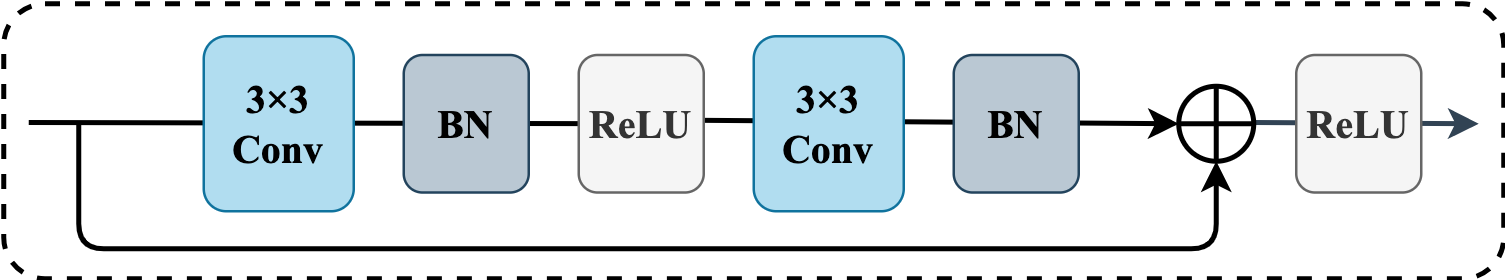}
\caption{Sketch of residual learning architecture / residual block.}
\label{RL}
\end{figure}

\textbf{Residual Learning:} In SRCNN, researchers find that better results can be obtained by adding more convolutional layers to increase the receptive field. However, directly stacking the layers will cause vanishing/exploding gradients and degradation problems~\cite{he2015convolutional}. Meanwhile, adding more layers will lead to a higher training error and more expensive computational costs.

In ResNet~\cite{he2016deep}, He \emph{et al.} proposed a residual learning framework, where a residual mapping is desired instead of fitting the whole underlying mapping (Fig.~\ref{RL}). In SISR, as the LR image and HR image share most of the same information, it is easy to explicitly model the residual image between LR and HR images. Residual learning enables deeper networks and remits the problem of gradient vanishing and degradation. With the help of residual learning, Kim \emph{et al.}~\cite{kim2016accurate} proposed a very deep super-resolution network, also known as VDSR. For the convenience of network design, the residual block~\cite{he2016deep} has gradually become the basic unit in the network structure. The convolutional branch, usually has two $3 \times 3$ convolutional layers, two batch normalization layers, and one ReLU activation function in between. It is worth noting that the batch normalization layer is often removed in the SISR task since Lim \emph{et al.}~\cite{lim2017enhanced} point out that the batch normalization layer consumes more memory but will not improve the model performance.

\textit{Global and Local Residual Learning}: Global residual learning is a skip-connection from input to the final reconstruction layer, which helps improve the transmission of information from input to output and reduces the loss of information to a certain extent. However, as the network becomes deeper, a significant amount of image details are inevitably lost after going through so many layers. Therefore, local residual learning is proposed, which is performed in every few stacked layers instead of from input to output. In this approach, a multi-path mode is formed and rich image details are carried and also help gradient flow. Furthermore, many new feature extraction modules have introduced local residual learning to reinforce strong learning capabilities~\cite{Li_2018_ECCV,Zhang_2018_ECCV}. Of course, combining local residual learning and global residual learning is also highly popular now~\cite{ledig2017photo,lim2017enhanced,Zhang_2018_ECCV}.

\textit{Residual Scaling}: In EDSR, Lim \emph{et al.}~\cite{lim2017enhanced} found that increasing the feature maps, i.e., channel dimension, above a certain level would make the training procedure numerical unstable. To solve such issues, they adopted the residual scaling technique~\cite{szegedy2017inception}, where the residuals are scaled down by multiplying a constant between 0 and 1 before adding them to the main path. With the help of this residual scaling method, the model performance can be further improved.

\begin{figure}[h]
\centering
\includegraphics[width=8cm]{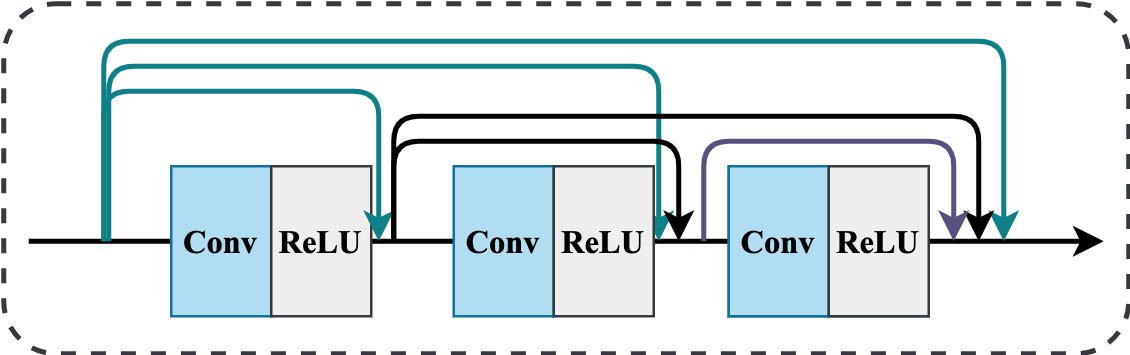}
\caption{The structure of the dense connection module.}
\label{DC}
\end{figure}

\textbf{Dense Connection:} A dense connection mechanism was proposed in DenseNet~\cite{huang2017densely}, which is widely used in computer vision tasks in recent years. Different from the structure that only sends the hierarchical features to the final reconstruction layer, each layer in the dense block receives the features of all preceding layers (Fig.~\ref{DC}). Short paths created between most of the layers can help alleviate the problem of vanishing/exploding gradients and strengthen the deep information flow through layers, thereby further improving the reconstruction accuracy.

Motivated by the dense connection mechanism, Tong \emph{et al.}~\cite{tong2017image} proposed an SRDenseNet. SRDenseNet uses not only the layer-level dense connections but also the block-level ones, where the output of each dense block is connected by dense connections. In this way, the low-level features and high-level features are combined and fully used to conduct the reconstruction. In RDN~\cite{zhang2018residual}, dense connections are combined with the residual learning to form the residual dense block (RDB), which allows low-frequency features to be bypassed through multiple skip connections, making the main branch focusing on learning high-frequency information. Apart from the aforementioned models, the dense connection is also applied in MemNet~\cite{tai2017memnet}, RPMNet~\cite{mei2019deep}, MFNet~\cite{shen2019multipath}, etc. With the help of a dense connection mechanism, the information flow among different depths of the network can be fully used, thus yielding better reconstruction results.

\begin{figure}[h]
\centering
\includegraphics[width=8cm]{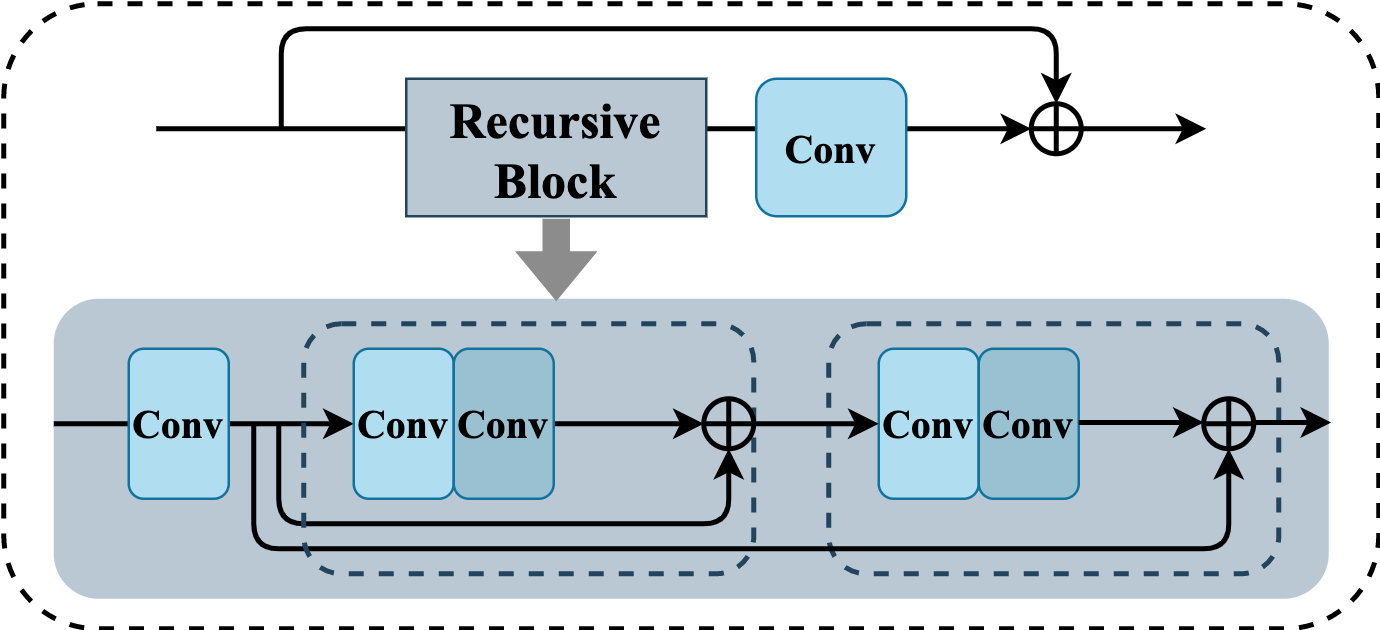}
\caption{The structure of DRRN, where the shaded part denotes the recursive block and the parameters in the dashed box are sharing.}
\label{RNN}
\end{figure}

\textbf{Recursive Learning:} To obtain a large receptive field without increasing model parameters, recursive learning is proposed for SISR, where the same sub-modules are repetitively applied in the network, and share the same parameters. In other words, a recursive block is a collection of recursive units, where the corresponding structures among these recursive units share the same parameters. For instance, the same convolutional layer is applied 16 times in DRCN~\cite{kim2016deeply}, resulting in a 41 $\times$ 41 size receptive field. However, too many stacked layers in the recursive learning-based model will still cause the problem of vanishing/exploding gradient. Therefore, in DRRN~\cite{tai2017image}, the recursive block is conducted based on residual learning (Fig.~\ref{RNN}). Recently, more and more models have introduced the residual learning strategy in their recursive units, such as MemNet~\cite{tai2017memnet}, CARN~\cite{ahn2018fast}, and SRRFN~\cite{Li_2019_ICCV}.

\textbf{Progressive Learning:} Progressive learning refers to gradually increasing the difficulty of the learning task. For some sequence prediction tasks or sequential decision-making problems, progressive learning is used to reduce the training time and improve the generalization performance. Since SISR is an ill-posed problem that is always confronted with great learning difficulty due to some adverse conditions such as large scaling factors, unknown degradation kernels, and noise, it is suitable to utilize progressive learning to simplify the learning process and improve the reconstruction efficiency. 

In LapSRN~\cite{lai2017deep}, the method is applied to progressively reconstruct the sub-band residuals of high-resolution images. In ProSR~\cite{wang2018fully}, each level of the pyramid is gradually blended in to reduce the impact on the previously trained layers, and the training pairs of each scale are incrementally added. In SRFBN~\cite{li2019feedback}, the strategy is applied to solve the complex degradation tasks, where targets of different difficulties are ordered for progressive learning. With the help of progressive learning, complex problems can be decomposed into multiple simple tasks, hence accelerating model convergence and obtaining better reconstruction results.

\begin{figure}[h]
\centering
\includegraphics[width=8cm]{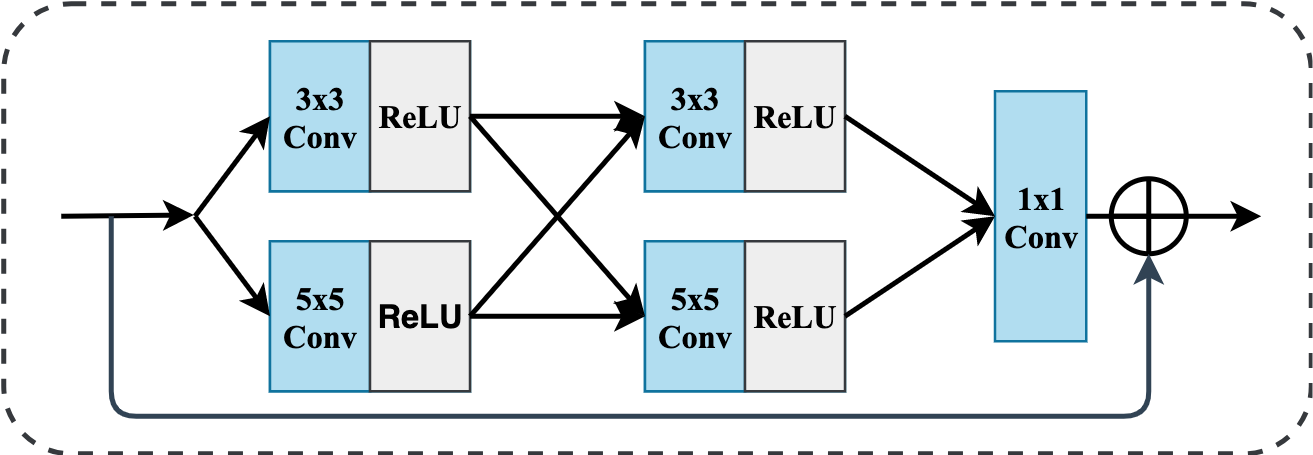}
\caption{The structure of multi-scale residual block (MSRB~\cite{Li_2018_ECCV}).}
\label{MSRB}
\end{figure}

\textbf{Multi-scale Learning:} Rich and accurate image features are essential for SR image reconstruction. Meanwhile, plenty of research works~\cite{szegedy2016rethinking, chollet2017xception, lai2017deep} have pointed out that images may exhibit different characteristics at different scales and thus making full use of these features can further improve model performance. Inspired by the inception module~\cite{chollet2017xception}, Li \emph{et al.}~\cite{Li_2018_ECCV} proposed a multi-scale residual block (MSRB, Fig.~\ref{MSRB}) for feature extraction. MSRB integrates different convolution kernels in a block to adaptively extract image features at different scales. After that, Li \emph{et al.}~\cite{li2020mdcn} further optimized the structure and proposed a more accurate multi-scale dense cross block (MDCB) for feature extraction. MDCB is essentially a dual-path dense network that can effectively detect local and multi-scale features.

Recently, more and more multi-scale SISR models have been proposed. For instance, Qin \emph{et al.}~\cite{qin2020multi} proposed a multi-scale feature fusion residual network (MSFFRN) to fully exploit image features for SISR. Chang \emph{et al.}~\cite{chang2019multi} proposed a multi-scale dense network (MSDN) by combining multi-scale learning with the dense connection. Cao \emph{et al.}~\cite{cao2019single} developed a new SR approach called multi-scale residual channel attention network (MSRCAN), which introduced the channel attention mechanism into the MSRB. All the above examples indicate that the extraction and utilization of multi-scale image features are of increasing importance to further improve the quality of the reconstructed images. 

\textbf{Attention Mechanism:} Attention mechanism can be considered as a tool that can allocate available resources to the most informative part of the input. To improve the efficiency during the learning procedure, some works are proposed to guide the network to pay more attention to the regions of interest. For instance, Hu \emph{et al.}~\cite{hu2018squeeze} proposed a squeeze-and-excitation (SE) block to model channel-wise relationships in the image classification task. Wang \emph{et al.}~\cite{wang2018non} proposed a non-local attention neural network for video classification by incorporating non-local operations. Motivated by these methods, an attention mechanism has also been introduced into SISR.

\begin{figure}[h]
\centering
\includegraphics[width=8cm]{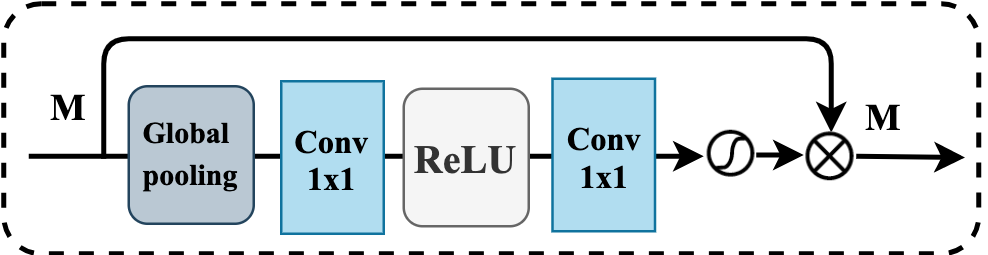}
\caption{The principle of channel attention mechanism (CAM).}
\label{CAM}
\end{figure}

\textit{Channel Attention}: In SISR, we mainly want to recover as much valuable high-frequency information as possible. However, common CNN-based methods treat channel-wise features equally, which lacks flexibility in dealing with different types of information. To solve this problem, many methods~\cite{Zhang_2018_ECCV,mei2018effective} introduce the SE mechanism in the SISR model. For example, Zhang \emph{et al.}~\cite{Zhang_2018_ECCV} proposed a new module based on the SE mechanism, named residual channel attention block (RCAB). As shown in Fig.~\ref{CAM}, a global average pooling layer followed by a Sigmoid function is used to rescale each feature channel, allowing the network to concentrate on the more useful channels and enhancing discriminative learning ability. In SAN~\cite{dai2019second}, second-order statistics of features are explored to conduct the attention mechanism based on covariance normalization. A great number of experiments have shown that second-order channel attention can help the network obtain more discriminative representations, leading to higher reconstruction accuracy.

\textit{Non-Local Attention}: When CNN-based methods conduct convolution in a local receptive field, the contextual information outside this field is ignored, while the features in distant regions may have a high correlation and can provide effective information. Given this issue, non-local attention has been proposed as a filtering algorithm to compute a weighted mean of all pixels of an image. In this way, distant pixels can also contribute to the response of a position in concern. For example, the non-local operation is conducted in a limited neighborhood to improve the robustness in NLRN~\cite{liu2018non}. A non-local attention block is proposed in RNAN~\cite{zhang2019residual}, where the attention mechanisms in both channel- and spatial-wise are used simultaneously in its mask branch to better guide feature extraction in the trunk branch. Meanwhile, a holistic attention network is proposed in HAN~\cite{niu2020single}, which consists of a layer attention module and a channel-spatial attention module, to model the holistic interdependence among layers, channels, and positions. In CSNLN~\cite{mei2020image}, a cross-scale non-local attention module is proposed to mine long-range dependencies between LR features and large-scale HR patches within the same feature map. To mitigate the noise pollution caused by non-local attention, ENLCA~\cite{xia2022efficient} utilizes efficient non-local attenuation and sparse aggregation to focus on useful information with contrast learning to separate irrelevant features. All these methods show the effectiveness of non-local attention, which can further improve the model performance.

\textbf{Feedback Mechanism:} The feedback mechanism refers to carrying a notion of output to the previous states, allowing the model to have a self-correcting procedure. It is worth noting that the feedback mechanism is different from recursive learning since in the feedback mechanism the model parameters keep self-correcting and do not share. Recently, the feedback mechanism has been widely used in many 
computer vision tasks~\cite{carreira2016human,cao2015look}, which is also beneficial for SR image reconstruction. Specifically, the feedback mechanism allows the network to carry high-level information back to previous layers and refine low-level information, thus fully guiding the LR image to recover high-quality SR images. 

In DBPN~\cite{haris2019deep}, iterative up- and down-sampling layers are provided to achieve an error feedback mechanism for projection errors at each stage. In DSRN~\cite{han2018image}, a dual-state recurrent network is proposed, where recurrent signals are exchanged between these states in both directions via delayed feedback. In SFRBN~\cite{li2019feedback}, a feedback block is proposed, in which the input of each iteration is the output of the previous one as the feedback information. Followed by several projection groups sequentially with dense skip connections, low-level representations are refined and become more powerful high-level representations. 

\begin{figure}[h]
\centering
\includegraphics[width=8cm]{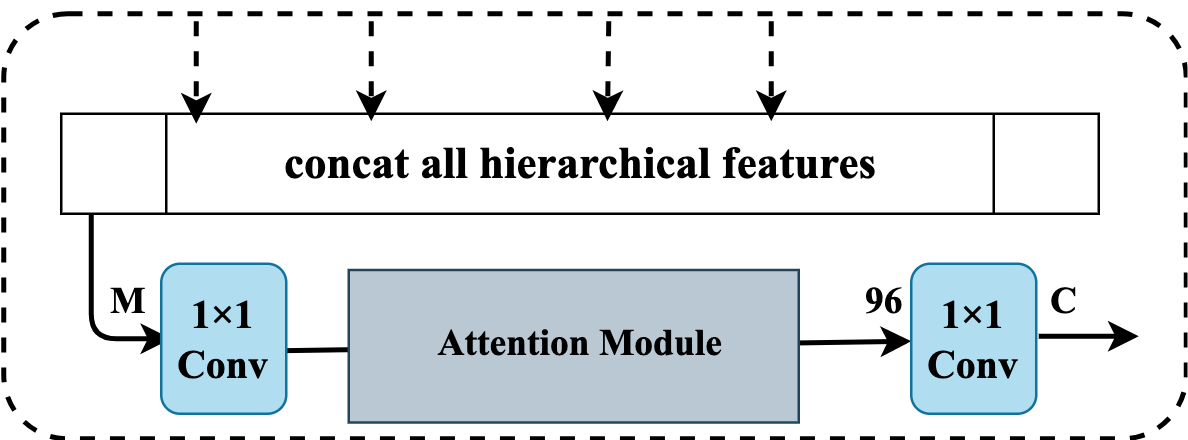}
\caption{The structure of the hierarchical feature distillation block (HFDB).}
\label{HFDB}
\end{figure}

\textbf{Gating Mechanism:} Skip connection in the above residual learning tends to make the channel dimension of the output features extremely high. If such a high-dimension channel remains the same in the following layers, the computational cost will be terribly large and therefore will affect the reconstruction efficiency and performance. Intuitively, the output features after the skip connection should be efficiently re-fused instead of simply concatenated. 

To solve this issue, researchers recommend using the gating mechanism to adaptively extract and learn more efficient information. Most of the time, a $1\times1$ convolutional layer is adopted to accomplish the gating mechanism, which can reduce the channel dimension and leave more effective information. In SRDenseNet~\cite{tong2017image} and MSRN~\cite{Li_2018_ECCV}, such $1\times1$ convolutional layer acts as a bottleneck layer before the reconstruction module. In MemNet~\cite{tai2017memnet}, it is a gate unit at the end of each memory block to control the weights of the long-term memory and short-term memory. Note that, the gate is not only able to serve as bottlenecks placed at the end of the network, but also continuously conducted in the network. For example, in MemNet~\cite{tai2017memnet} and CARN~\cite{ahn2018image}, the gating mechanism is used in both global and local regions. Sometimes, it can be combined with other operations, such as the attention mechanism, to construct a more effective gate module to achieve feature distillation. For instance, Li \emph{et al.}~\cite{li2020mdcn} proposed a hierarchical feature distillation block (Fig.~\ref{HFDB}) by combining $1 \times 1$ convolutional layer and attention mechanism.

\textbf{Efficient Structure:} There is no doubt that increasing the depth of the model is the easiest way to improve the model performance. However, due to the huge computational overhead of deep and large models, it is difficult to be applied to mobile devices with limited computing capabilities. To address this issue, more and more lightweight and efficient SISR methods have been proposed in recent years. For instance, Ahn \emph{et al.}~\cite{ahn2018fast} designed an architecture (CARN) that implements a cascading mechanism upon the residual network, which achieved fast, accurate, and lightweight SR. Hui \emph{et al.}~\cite{hui2018fast} proposed a novel Information Distillation Network (IDN) with lightweight parameters and computational complexity by using the information distillation strategy. After that, the author further proposed a Lightweight Information Multi-Distillation Network (IMDN) by constructing the cascaded information multi-distillation blocks. Liu \emph{et al.}~\cite{liu2020residual} proposed a RFDN, enhances the efficiency of single image super-resolution (SISR) by incorporating a lighter feature distillation connection operation. Zhou \emph{et al.}~\cite{zhou2022efficient} have developed VapSR, which refines attention mechanisms to create a more efficient super-resolution network. Li \emph{et al.}~\cite{sun2022shufflemixer} introduced ShuffleMixer, a technique that investigates the use of large convolutions and channel splitting shuffle operations to make the network more mobile-compatible. Li \emph{et al.}~\cite{li2022blueprint} proposed a Blueprint Separable Residual Network (BSRN) containing two efficient designs, blueprint separable convolution and more effective attention modules. Li \emph{et al.}~\cite{li2023cross} proposed a novel Cross-receptive Field Guided Transformer (CFGT) to enable the selection of contextual information required for reconstruction by using a modulated convolutional kernel. In addition, some hardware-friendly SISR methods have emerged. For example, Luo \emph{et al.}~\cite{luo2022adjustable} proposed an Individual Kernel Sparsity (IKS) method for memory-efficient and sparsity-adjustable image SR, which enables deep networks can be deploymented in memory-limited devices. Ye \emph{et al.}~\cite{ye2023hardware} proposed a Hardware-friendly Scalable SR (HSSR) with progressively structured sparsity. This model can cover multiple SR models with different sizes by a single scalable model, without extra retraining or post-processing. Lin \emph{et al.}~\cite{lin2023memory} proposed a Memory-friendly Scalable dynamic SR (MSSR) lightweight model via rewinding, which can be easily generalized to different SR models. Choi \emph{et al.}~\cite{choi2023n} introduced a NGswin, which boosts performance in SISR by broadening the receptive field of window-based self-attentive methods. Wang \emph{et al.}~\cite{wang2023omni} proposed a Omni-SR, enhancing the capabilities of lightweight models by replicating pixel interactions across both spatial and channel dimensions. Li \emph{et al.}~\cite{li2023dlgsanet} introduced a DLGSANet, which streamlines SISR efficiency by employing sparse global self-attention modules to pinpoint the most pertinent similarity values.

\begin{figure}[h]
\centering
\includegraphics[width=8cm]{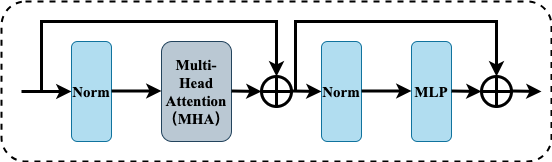}
\caption{The structure of classic Transformer. The key component is the multi-head attention (MHA) module.}
\label{Transformer}
\end{figure}

\textbf{Transformer-based Method:} The key idea of the Transformer is the “self-attention” mechanism, which can capture long-term information between sequence elements. Recently, Transformer~\cite{vaswani2017attention} (Fig.~\ref{Transformer}) has achieved brilliant results in NLP tasks. For example, the pre-trained deep learning models (e.g., BERT~\cite{devlin2018bert}, GPT~\cite{radford2019language}) have shown effectiveness over conventional methods. Inspired by this, more and more researchers have begun to explore the application of Transformers in computer vision tasks and have achieved breakthrough results in many tasks. In image restoration, Transformer is often used to capture the global information of the image to further improve the quality of the reconstructed image.

In recent years, more and more Transformer-based models have been proposed. For example, Chen \emph{et al.} proposed the Image Processing Transformer (IPT~\cite{chen2021pre}) which was pre-trained on large-scale datasets. In addition, contrastive learning is introduced for different image-processing tasks. Therefore, the pre-trained model can efficiently be employed on the desired task after finetuning. However, IPT~\cite{chen2021pre} relies on large-scale datasets and has a large number of parameters (over 115.5M parameters), which greatly limits its application scenarios. To solve this issue, Liang \emph{et al.} proposed the SwinIR~\cite{liang2021swinir} for image restoration based on the Swin Transformer~\cite{liu2021swin}. Specifically, the Swin Transformer blocks (RSTB) are proposed for feature extraction and DIV2K+Flickr2K is used for training. To improve the lack of direct interaction between different windows in SwinIR. Zamir~\cite{zamir2022restormer} \emph{et al.} proposed Restormer to reconstruct high-quality images by embedding CNNs within Transformer and performing local-global learning at multiple scales. Chen \emph{et al.} proposed CAT~\cite{chen2022cross} to extend the attention region and aggregate features across different windows. Then, to activate more of the pixels that Transformer focuses on, Chen \emph{et al.} proposed HAT~\cite{chen2205activating}, which uses overlapping cross-attention modules in conjunction with a pre-training strategy to enhance Transformer model potential. Li~\cite{li2023efficient} \emph{et al.} proposed GRL to explicitly model the image hierarchy at global, regional, and local scales by integrating various attentions within the Transformer. As for the application on the lightweight SISR model, Lu \emph{et al.}~\cite{lu2021efficient} proposed an Efficient Super-Resolution Transformer (ESRT) for fast and accurate SISR which achieves competitive results with fewer parameters and low computing costs. Zhang \emph{et al.}~\cite{zhang2022efficient} proposed ELAN with a shared self-attention mechanism to reduce model complexity and accelerate the Transformer-based model. Wang \emph{et al.}~\cite{wang2022uformer} proposed the Uformer, a general and superior U-shaped Transformer, which can reduce the computational complexity on high-resolution feature map while capturing
local context and multi-scale features. Zamir \emph{et al.}~\cite{zamir2022restormer} proposed an efficient Restormer that can capture
long-range pixel interactions while remaining applicable to large images. Li \emph{et al.}~\cite{li2023cross} proposed a Cross-receptive Focused Inference Network (CFIN) that can incorporate contextual modeling to achieve good performance with limited computational resources. Zhu \emph{et al.}~\cite{zhu2023attention} designed an Attention Retractable Frequency Fusion Transformer (ARFFT) to strengthen the representation ability and extend the receptive field to the whole image. Li \emph{et al.}~\cite{li2023lightweight} proposed a concise and powerful Pyramid Clustering Transformer Network (PCTN) for lightweight SISR. Chen \emph{et al.}~\cite{chen2023dual} proposed a novel Dual Aggregation Transformer (DAT) for SISR, which aggregates features across spatial and channel dimensions, in the interblock and intra-block dual manner. Zhou \emph{et al.}~\cite{zhou2023srformer} proposed a SRFormer, elevates the performance of window-based Transformer approaches by effectively integrating self-attentive channel and spatial information. Li \emph{et al.}~\cite{li2023efficient} achieves optimal performance across multiple scenarios by developing GRL, a hierarchical Transformer-based model for image upscaling that operates on global, regional, and local scales. ATDSR, brought forth by Zhang \emph{et al.}~\cite{zhang2024transcending}, enriches the SR Transformer with an auxiliary set of adaptive token dictionaries, thereby enhancing the precision of SISR. Adaptive token sparsifcation transformer (AdaFormer) proposed by Luo \emph{et al.}~\cite{luo2024adaformer} speeds up model inference for images by incorporating sparsity strategies. Although the performance of the Transformer-based method has greatly improved, the attention mechanism used in Transform will occupy a large amount of GPU memory. Therefore, how to further reduce the GPU memory of Transformer-based methods is worth further exploration.

\subsection{\textbf{Perceptual Quality Methods}}

Most methods simply seek to reconstruct SR images with high PSNR and SSIM. However, the improvement in reconstruction accuracy is not always accompanied by an improvement in visual quality. Blau \emph{et al.}~\cite{blau2018perception} pointed out that there was a perception-distortion trade-off. It is only possible to improve either perceptual quality or distortion while improving one must be at the expense of the other. Hence, in this section, we provide methods to ease this trade-off problem, hoping to provide less distortion while maintaining the good perceptual quality of the image. 

\textbf{Perceptual Loss:} Although pixel-wise losses, i.e., L1 and MSE loss, have been widely used to achieve high image quality, they do not capture the perceptual differences between the SR and HR images. In order to address this problem and allow the loss functions to better measure the perceptual and semantic differences between images, content loss, texture loss, and targeted perceptual loss are proposed. Among them, the content loss is widely used to keep the image consistent with the target~\cite{ledig2017photo,wang2018recovering}, which has been introduced in Sec.~\ref{Content}. Apart from obtaining more similar content, the same style, such as colors, textures, common patterns, and semantic information are also needed. Therefore, other perceptual losses need to be considered.

\textit{Texture Loss}: Texture loss, also called style reconstruction loss, is proposed by Gatys \emph{et al.}~\cite{gatys2015texture,gatys2015neural}, which can make the model reconstruct high-quality textures. The texture loss is defined as the squared Frobenius norm of the difference between the Gram matrices $G_j^{\phi}(x)$ of the output and the ground truth images:
\begin{equation}
    \mathcal{L}^{\phi,j}_{texture}(I_{SR},I_y)=||G_j^{\phi}(I_{SR})-G_j^{\phi}(I_y)||^2_F.
\end{equation}

With the help of the texture loss, the model tends to produce images that have the same local textures as the HR images during training~\cite{johnson2016perceptual}.

\textit{Targeted Perceptual Loss:} The conventional perceptual loss estimates the reconstruction error for an entire image without considering semantic information, resulting in limited capability. Rad \emph{et al.}~\cite{rad2019srobb} proposed a targeted perceptual loss that penalized images at different semantic levels based on the labels of object, background, and boundary. Therefore, more realistic textures and sharper edges can be obtained to reconstruct realistic SR images.

\textbf{Adversarial Training:} In 2014, the Generative Adversarial Networks (GANs) were proposed by Goodfellow \emph{et al.}~\cite{goodfellow2014generative}, which has been widely used in compute vision tasks, such as style transfer and image inpainting. The GANs consist of a generator and a discriminator. When the discriminator is trained to judge whether an image is true or false, the generator aims at fooling the discriminator rather than minimizing the distance to a specific image, hence it tends to generate outputs that have the same statistics as the training set.

Inspired by GAN, Ledig \emph{et al.}~\cite{ledig2017photo} proposed the Super-Resolution Generative Adversarial Network (SRGAN). In SRGAN, the generator $G$ is essentially an SR model that is trained to fool the discriminator $D$, and $D$ is trained to distinguish SR images from HR images. Therefore, the generator can learn to produce outputs that are highly similar to HR images, and then reconstruct more real and natural SR images. The generative loss $\mathcal{L}_{Gen}(I_x)$ can be defined as:
\begin{equation}
    \mathcal{L}_{Gen}=-\log D_{\theta_D}(G_{\theta_G}(I_x)),
\end{equation}
and the loss in terms of the discriminator is:
\begin{equation}
    \mathcal{L}_{Dis}=-\log (D_{\theta_D}(I_y))-\log (1-D_{\theta_D}(G_{\theta_G}(I_x))).
\end{equation}
Therefore, we need to solve the following problem:
\begin{equation}
\begin{split}
    \min_{\theta_G}\ \max_{\theta_D}\ &\mathbb{E}_{I_{y \sim p_{data}(I_y)}}(\log D_{\theta_D}(I_y))\ + \\
    &\mathbb{E}_{I_{x \sim p_{G}(I_x)}}(\log(1-D_{\theta_D}(G_{\theta_G}(I_x)))).
\end{split}
\end{equation}

In SRGAN~\cite{ledig2017photo}, the generator is the SRResNet and the discriminator uses the architecture proposed by Radford \emph{et al.}~\cite{radford2015unsupervised}. In ESRGAN~\cite{wang2018esrgan}, Wang \emph{et al.} made two modifications to the SRResNet: (1) replace the original residual block with the residual-in-residual dense block; (2) remove the BN layers to improve the generalization ability of the model. In SRFeat~\cite{park2018srfeat}, Park \emph{et al.} indicated that the GAN-based SISR methods tend to produce less meaningful high-frequency noise in reconstructed images. Therefore, they adopted two discriminators: an image discriminator and a feature discriminator, where the latter is trained to distinguish SR images from HR images based on the intermediate feature map extracted from a VGG network. In ESRGAN~\cite{wang2018esrgan}, Wang \emph{et al.} adopted the Relativistic GAN~\cite{jolicoeur2018relativistic}, where the standard discriminator was replaced with the relativistic average discriminator to learn the relatively realistic between two images. This modification helps the generator to learn sharper edges and more detailed textures. Wang \emph{et al.}~\cite{wang2023high} proposed a novel
GAN inversion framework that utilizes the powerful generative ability of StyleGAN-XL, which shows preferable quantitative and qualitative results in SISR.

\textbf{Cycle Consistency:} Cycle consistency assumes that there exist some underlying relationships between the source and target domains, and tries to make supervision at the domain level. To be precise, we want to capture some special characteristics of one image collection and figure out how to translate these characteristics into the other image collection. To achieve this, Zhu \emph{et al.}~\cite{zhu2017unpaired} proposed the cycle consistency mechanism, where not only the mapping from the source domain to the target domain is learned, but also the backward mapping is combined. Specifically, given a source domain $X$ and a target domain $Y$, we have a translator $G: X \rightarrow Y$ and another translator $F: Y \rightarrow X$ that is trained simultaneously to guarantee both an $adversarial\ loss$ that encourages $G(X)\approx Y$ and $F(Y)\approx X$ and a $cycle\ consistency\ loss$ that encourages $F(G(X))\approx X$ and $G(F(Y))\approx Y$. 

In SISR, the idea of cycle consistency has also been widely discussed. Given the LR images domain $X$ and the HR images domain $Y$, we not only learn the mapping from LR to HR but also the backward process. Researchers have shown that learning how to perform image degradation first without paired data can help generate more realistic images~\cite{bulat2018learn}. In CinCGAN~\cite{yuan2018unsupervised}, a cycle-in-cycle network is proposed, where the noisy and blurry input is mapped to a noise-free LR domain first and then upsampled with a pre-trained model. In DRN~\cite{guo2020closed}, the mapping from HR to LR images is learned to estimate the down-sampling kernel and reconstruct LR images, which forms a closed loop to provide additional supervision. DRN also gives us a novel approach in unsupervised learning SR, where the model is trained with both paired and unpaired data. 

\textbf{Diffusion-based Method:} Derived from the recent inspiration in the denoising diffusion probability model (DDPM)~\cite{ho2020denoising}, a new conditional image generation method is incorporated into the SISR task. Compared with the GAN-based SISR method, the diffusion model-based SISR methods~\cite{li2022srdiff, saharia2022image,shang2023resdiff,gao2023implicit} have better fidelity and reduce the generation of artifacts.

SRDiff~\cite{li2022srdiff} is the first diffusion-based SISR model, which provides diverse and realistic SISR predictions by gradually converting Gaussian noise into SISR images with LR as the input condition through Markov chains. SR3~\cite{saharia2022image} iteratively refines the pure Gaussian noise input using a model trained for denoising at various noise levels. Compared to GAN-based methods, it can output more realistic photos. IDM~\cite{gao2023implicit} integrates implicit neural representation and denoising diffusion model end-to-end and employs implicit neural representation to learn continuous image resolution representation during decoding. DR2~\cite{wang2023dr2} utilizes DDPM to coarsely reduce more complex low-quality face images and then uses the enhancement module to fully restore them to high-resolution (HR) face images. There is also a class of methods that aim to utilize the prior diffusion-based models to aid SISR. For example, StableSR~\cite{wang2023exploiting} and DiffBIR~\cite{lin2023diffbir} achieve real-world SISR by fine-tuning with prior knowledge from a pre-trained text-to-image diffusion model, such as Stable diffusion~\cite{rombach2022high}. DiffIR~\cite{xia2023diffir} utilizes a pre-trained model trained on ground-truth images to incorporate the prior into the SISR model, which can result in accurate estimates using fewer iterations than traditional DDPM. However, diffusion-based SISR models still need a large number of new samples and the slow convergence rate of the model limits their use scenarios. Therefore, how to overcome these drawbacks is still worthy of study.

\subsection{\textbf{Information Utilization Methods}}
In the aforementioned part, we have introduced the way to design an efficient SISR model, as well as obtaining high reconstruction accuracy and high perceptual quality for SR images. Although the current SISR model has made a significant breakthrough, how to use the information inside and outside of the image to further improve the performance of the model is still worth exploring.

\textbf{Internal Statistics:} In~\cite{zontak2011internal}, Zontak \emph{et al.} found that some patches exist only in a specific image and can not be found in any external database of examples. Therefore, SR methods trained on external images can not work well on such images due to the lack of patch information, while methods based on internal statistics may have a good performance. Meanwhile, Zontak \emph{et al.} pointed out that the internal entropy of patches inside a single image was much smaller than the external entropy of patches in a general collection of natural images. Therefore, using the internal image statistics to further improve model performance is a good choice. 

In ZSSR~\cite{shocher2018zero}, the property of internal image statistics is used to train an image-specific CNN, where the training examples are extracted from the test image itself. In the training phase, several LR-HR pairs are generated by using data augmentation, and a CNN is trained with these pairs. In test time, the LR image $I_{LR}$ is fed to the trained CNN as input to get the reconstructed image. In this process, the model makes full use of internal statistics of the image itself for self-learning. In SinGAN~\cite{shaham2019singan}, an unconditional generative model with a pyramid of fully convolutional GANs is proposed to learn the internal patch distribution at different scales of the image. To make use of the recurrence of internal information, they upsampled the LR image several times (depending on the final scale) to obtain the final SR output. 

\textbf{Multi-factor Learning:} Typically, in SISR, we often need to train specific models for different upsampling factors and it is difficult to arise at the expectation that a model can be applied to multiple upsampling factors. To solve this issue, some models have been proposed for multiple upsampling factors. Surprisingly, researchers found that this method can fully exploit the inter-scale correlation between different upsampling factors, which can further improve model performance.

In LapSRN~\cite{lai2017fast}, LR images are progressively reconstructed in the pyramid networks to obtain the large-scale results, where the intermediate results can be taken directly as the corresponding multiple factors results. In~\cite{lim2017enhanced}, Lim \emph{et al.} found the inter-related phenomenon among multiple scales tasks, i.e., initializing the high-scale model parameters with the pre-trained low-scale network can accelerate the training process and improve the performance. Therefore, they proposed the scale-specific processing modules at the head and tail of the model to handle different upsampling factors. To further exploit the inter-scale correlation between different upsampling factors, Li \emph{et al.} further optimized the strategy in MDCN~\cite{li2020mdcn}. Different from MDSR which introduces the scale-specific processing strategy both at the head and tail of the model, MDCN can maximize the reuse of model parameters and learn the inter-scale correlation.

\textbf{Prior Guidance:} Most methods tend to build end-to-end CNN models to achieve SISR since it is simple and easy to implement. However, it is rather difficult for them to reconstruct realistic high-frequency details due to plenty of useful features have been lost or damaged. To solve this issue, a priors-guided SISR framework has been proposed. Extensive experiments have shown that with the help of image priors, the model can converge faster and achieve better reconstruction accuracy. Recently, many image priors have been proposed, such as total variation prior, sparse prior, and edge prior.

Motivated by this, Yang \emph{et al.}~\cite{yang2017deep} integrated the edge prior with recursive networks and proposed a Deep Edge Guided Recurrent Residual Network (DEGREE) for SISR. After that, Fang \emph{et al.}~\cite{fang2020soft} proposed an efficient and accurate Soft-edge Assisted Network (SeaNet). Different from DEGREE, which directly applies the off-the-shelf edge detectors to detect image edges, SeaNet automatically learns more accurate image edges from the constructed EdgeNet. Meanwhile, they find that more accurate priors can lead to more significant performance. Additionally, image priors are also beneficial for GAN-based models. For example, the semantic categorical prior is used to generate richer and more realistic textures with the help of spatial feature transform (SFT) in SFTGAN\cite{wang2018recovering}. With this information from high-level tasks, similar LR patches can be easily distinguished and more natural textual details can be generated. In SPSR~\cite{ma2020structure}, the authors utilized the gradient maps to guide image recovery to solve the problem of structural distortions in the GAN-based methods. Among them, the gradient maps are obtained from a gradient branch and integrated into the SR branch to provide structure prior. With the help of gradient maps, we know which region should be paid more attention to, so as to guide image generation and reduce geometric distortions. In FeMaSR~\cite{chen2022fe}, the authors use discrete features obtained by VQ-GAN~\cite{yu2021vector} pre-training in HR images as prior information to performing image recovery by matching distorted LR image features with distortion-free HR features from the pre-trained HR prior.

\textbf{Reference-based Method:} In contrast to SISR where only a single LR image is used as input, reference-based SISR (RefSR) takes a reference image to assist the SR process. The reference images can be obtained from various sources like photo albums, video frames, and web image searches. Meanwhile, there are several approaches proposed to enhance image textures, such as image aligning and patch matching. Recently, some RefSR methods~\cite{yue2013landmark,zheng2018crossnet} chose to align the LR and reference images with the assumption that the reference image possesses similar content as the LR image. For instance, Yue \emph{et al.}~\cite{yue2013landmark} conducted global registration and local matching between the reference and LR images to solve an energy minimization problem. In CrossNet~\cite{zheng2018crossnet}, optical flow is proposed to align the reference and LR images at different scales, which are later concatenated into the corresponding layers of the decoder. However, these methods assume that the reference image has a good alignment with the LR image. Otherwise, their performance will be significantly influenced. Different from these methods, Zhang \emph{et al.}~\cite{zhang2019image} applied patch matching between VGG features of the LR and reference images to adaptively transfer textures from the reference images to the LR images. In TTSR~\cite{yang2020learning}, Yang \emph{et al.} proposed a texture transformer network to search and transfer relevant textures from the reference images to the LR image.

\textbf{Knowledge Distillation:} Knowledge distillation refers to a technique that transfers the representation ability of a large (Teacher) model to a small one (Student) for enhancing the performance of the student model. Hence, it has been widely used for network compression or to further improve the performance of the student model, which has shown effectiveness in many computer vision tasks. Meanwhile, there are mainly two kinds of knowledge distillation, soft label distillation, and feature distillation. In soft label distillation, the softmax outputs of a teacher model are regarded as soft labels to provide informative dark knowledge to the student model~\cite{hinton2015distilling}. In feature distillation, the intermediate features maps are transferred to the student model~\cite{ahn2019variational,romero2014fitnets}.

Inspired by this, some works introduce the knowledge distillation technique to SISR to further improve the performance of lightweight models. For instance, in SRKD~\cite{gao2018image}, a small but efficient student network is guided by a deep and powerful teacher network to achieve similar feature distributions to those of the teacher. In~\cite{lee2020learning}, the teacher network leverages the HR images as privileged information, and the intermediate features of the decoder of the teacher network are transferred to the student network via feature distillation so that the student can learn high-frequency details from the Teacher which is trained with the HR images. Subsequently, JDSR~\cite{luo2021boosting} explored a joint distillation learning that effectively improves the distillation performance of lightweight models by using distillation of HR's privileged information in conjunction with internal self-distillation. CSD~\cite{wang2021towards} combines the contrast learning and distillation tasks to further reduce the solution space of SISR. In addition, to solve the model compression problem for unsupervised issues, ~\cite{zhang2021data} used a generator to synthesize training samples close to the original data after using a progressive distillation scheme to improve student model performance.

\subsection{Real-World Image Super-Resolution}
The degradation modes are complex and unknown in real-world scenarios~\cite{chen2022real}, where downsampling is usually performed after anisotropic blurring and sometimes signal-dependent noise is added. It is also affected by the in-camera signal processing (ISP) pipeline. Therefore, simulation SISR models exhibit poor performance when handling real-world images. Meanwhile, most of the aforementioned models can only be applied to some specific integer upsampling factors. This greatly limits the practical application and promotion of these models. To solve these problems, some interesting methods have been proposed. Based on the problems they intend to solve, we divide them into two major categories: Blind Image Super-Resolution and Scale Arbitrary Super-Resolution.

\subsubsection{\textbf{Blind Image Super-Resolution}} 
Blind SISR has attracted increasing attention due to its significance in real-world applications, which aim to super-resolve LR images with unknown degradation. It is worth noting that blind SISR has no clear definitions. More details about blind SISR can be found in~\cite{liu2022blind}. In this work, we simply divided them into two categories: explicit degradation modeling methods and implicit degradation modeling methods, according to the ways of degradation modeling.
  
\textbf{Explicit Degradation Modeling:} Blind SISR methods with explicit modeling of the degradation process are mainly based on the classical degradation model, where the blur kernel and additive noise are two main degradation factors. According to whether the degradation process is estimated, this type of method can be further divided into two categories: image-specific adaptation without degradation estimation and image-specific adaptation with degradation estimation. Among them, the first type of method often uses an external method to perform degradation estimation before the SR process, thus adapting the framework to the blind setting. The second type of method often uses an internal module for degradation estimation and outputs the degradation representation. For example, Zhang \emph{et al.}~\cite{zhang2018learning} proposed a simple and scalable deep CNN framework (SRMD) for multiple degradations learning. In SRMD, the concatenated LR image and degradation maps are taken as input of the network to achieve image super-resolution under different degradations. Based on SRMD, Xu \emph{et al.}~\cite{xu2020unified} proposed the UDVD, which uses dynamic convolution to process different degradations in different areas in the image. This type of method often relies on reliable degradation estimation methods to quickly obtain satisfactory SR output. Hence, a method that incorporates degradation estimation into the SR framework will obtain more stable and reliable results. Towards filling this gap, growing attention has been paid to image-specific adaptation method with degradation estimation. This type of method combines degradation estimation and SR processes into a unified model, in which kernel estimation is the main research work. For example, in IKC~\cite{gu2019blind}, the iterative kernel correction procedure is proposed to help the blind SISR task find more accurate blur kernels. Inspired by it, Luo \emph{et al.}~\cite{luo2020unfolding} adopted an alternating optimization algorithm and proposed a Deep Alternating Network (DAN) to estimate blur kernel and restore SR image in a single network, which makes the restorer and estimator well compatible with each other, and thus achieves good results in kernel estimation. Although such methods are more robust than using off-the-shelf estimation algorithms, such iterative schemes often consume more inference time and may lead to SR failure due to large estimation errors. To address this issue, some works introduced more accurate degradation estimation methods. In~\cite{wang2021unsupervised}, the author suggested learning abstract representations to distinguish various degradations in the representation space and introduced a Degradation-Aware SR (DASR) network with flexible adaption to various degradations based on the learned representations. There are also some methods proposed to estimate more realistic kernels from real images. For instance, in~\cite{cai2019toward}, the RealSR dataset is proposed, where paired LR-HR images on the same scene are captured by adjusting the focal length of a digital camera.

\textbf{Implicit Degradation Modeling:} Blind SISR methods with implicit modeling of degradation process aim to model the degradation through learning with external dataset. This type of method usually learns data distribution by a GAN framework, and one or more discriminators are used to distinguish generated images from real ones. For example, Yuan \emph{et al.}~\cite{yuan2018unsupervised} proposed an unsupervised image SR using Cycle-in-Cycle Generative Adversarial Networks (CinCGAN). CinCGAN first mapped the noisy and blurry input to a noise-free low-resolution space, and then the intermediate image was up-sampled with a pre-trained model. Finally, these two modules are fine-tuned in an end-to-end manner to get SR output. Bulat \emph{et al.}~\cite{bulat2018learn} believed that low-resolution images in the real world constitute a specific distribution in high-dimensional space, and use a generative adversarial network to generate low-resolution images consistent with this distribution from high-resolution images. After that, Yuan \emph{et al.}~\cite{yuan2018unsupervised} and Maeda \emph{et al.}~\cite{maeda2020unpaired} further proposed a unified framework, which can simultaneously learn the generation of pseudo-low-resolution images and the reconstruction of high-resolution images, achieving better results in actual scenes. Wei \emph{et al.}~\cite{wei2021unsupervised} further considered the domain difference between pseudo-low-resolution images and real low-resolution images, and proposed a domain adaptation mechanism to improve model performance. Wolf \emph{et al.}~\cite{wolf2021deflow} proposed the DeFlow framework, which uses the stochastic modeling ability of the flow model to enhance the diversity of pseudo-low-resolution images, and further improves the image super-resolution performance in real scenes. 

\subsubsection{\textbf{Scale Arbitrary}}

In real application scenarios, in addition to processing real images, it is also important to handle arbitrary scale factors with a single model. To achieve this, Hu \emph{et al.} proposed two simple but powerful methods termed Meta-SR~\cite{hu2019meta} and Meta-USR~\cite{hu2020meta}. Among them, Meta-SR is the first SISR method that can be used for arbitrary scale factors and Meta-USR is an improved version that can be applied to arbitrary degradation mode (including arbitrary scale factors). Although Meta-SR and Meta-USR achieve promising performance on non-integer scale factors, they cannot handle SR with asymmetric scale factors. To alleviate this problem, Wang \emph{et al.}~\cite{wanglearning} suggested learning the scale-arbitrary SISR model from scale-specific networks and developed a plug-in module for existing models to achieve scale-arbitrary SR. Specifically, the proposed plug-in module uses conditional convolution to dynamically generate filters based on the input scale information, thus the networks equipped with the proposed module achieve promising results for arbitrary scales with only a single model.

\subsection{Domain-Specific Applications} \label{DSA}
\begin{figure*}[h]
\begin{center}
\includegraphics[width=1\linewidth]{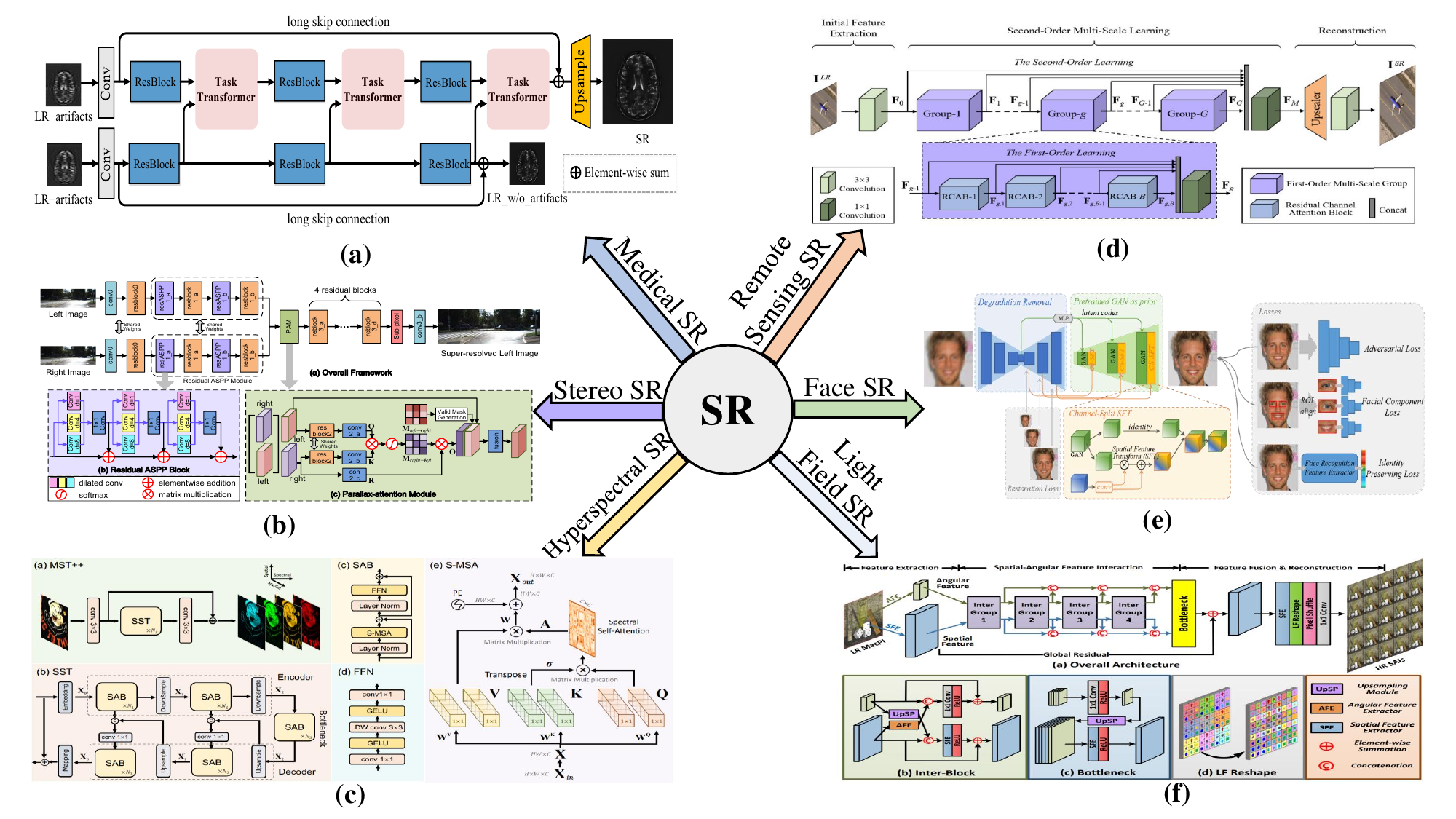}
\end{center}
   \caption{Examples of various popular SR tasks. (a) T2Net~\cite{feng2021task} for Medical SR, (b) PASSRNet~\cite{wang2020parallax} for Stereo SR, (c) MST++~\cite{cai2022mst++} for Hyperspectral SR, (d) SMSR~\cite{dong2020remote} for Remote Sensing SR, (e) GFPGAN~\cite{wang2021towards} for Face SR, (f) LF-InterNet~\cite{wang2020spatial} for Light Field SR.}
\label{SR_methods}
\end{figure*}

The technology of image super-resolution has been widely used in many application scenarios. As shown in Fig.~\ref{SR_methods}, we introduce various applications of SR in this section.

\subsubsection{\textbf{Stereo Image Super-Resolution}}
The dual camera has been widely used to estimate depth information. Meanwhile, stereo imaging can also be applied in image restoration. In this task, we have two images with a disparity much larger than one pixel. Therefore, full use of these two images can enhance spatial resolution. 

In StereoSR~\cite{jeon2018enhancing}, Jeon \emph{et al.} proposed a method that learned a subpixel parallax prior to enhancing the spatial resolution of the stereo images. However, the number of shifted right images is fixed in StereoSR, which makes it fail to handle different stereo images with large disparity variations. To handle this problem, Wang \emph{et al.}~\cite{wang2019learning,wang2020parallax} proposed a parallax-attention mechanism with a global receptive field along the epipolar line, which can generate reliable correspondence between the stereo image pair and improve the quality of the reconstructed SR images. In~\cite{wang2019flickr1024}, a dataset named Flickr1024 is proposed for stereo image super-resolution, which consists of 1024 high-quality stereo image pairs. In~\cite{ying2020stereo}, a stereo attention module is proposed to extend pre-trained SISR networks for stereo image SR, which interacts with stereo information bi-directionally in a symmetric and compact manner. In~\cite{wang2021symmetric}, a symmetric bi-directional parallax attention module and an inline occlusion handling scheme are proposed to effectively interact with cross-view information. In~\cite{dai2021feedback}, a Stereo Super-Resolution and Disparity Estimation Feedback Network (SSRDE-FNet) is proposed to simultaneously handle the stereo image super-resolution and disparity estimation in a unified framework. In~\cite{chu2022nafssr}, in addition to extracting single image features from the left and right views separately using NAFNet~\cite{chen2022simple}, a stereo cross-attention module is introduced to fuse the image features from the left and right views.

\subsubsection{\textbf{Remote Sensing Image Super-Resolution}}
With the development of satellite image processing, remote sensing has become more and more important. However, due to the limitations of current imaging sensors and complex atmospheric conditions, such as limited spatial resolution, spectral resolution, and radiation resolution, we are facing huge challenges in remote sensing applications. 

Recently, many methods have been proposed for remote sensing image super-resolution. For example, a new unsupervised hourglass neural network is proposed in~\cite{haut2018new} to super-resolved remote sensing images. The model uses a generative random noise to introduce a higher variety of spatial patterns, which can be promoted to a higher scale according to a global reconstruction constraint. In~\cite{gu2019deep}, a Deep Residual Squeeze and Excitation Network (DRSEN) are proposed to overcome the problem of the high complexity of remote sensing image distribution. In~\cite{zhang2020remote}, a mixed high-order attention network (MHAN) is proposed, which consists of a feature extraction network for feature extraction and a feature refinement network with the high-order attention mechanism for detail restoration. In~\cite{dong2020remote}, the authors developed a Dense-Sampling Super-Resolution Network (DSSR) to explore the large-scale SR reconstruction of the remote sensing imageries. In~\cite{lei2021hybrid}, the authors proposed a new Hybrid-scale Self-similarity Exploitation Network (HSENet), which can simultaneously exploit single and cross-scale similarities for high-quality image reconstruction; In~\cite{wang2023remote}, Wang \emph{et al.} proposed a Multi-scale Enhancement Network (MEN), which uses multi-scale features of remote sensing images to enhance the network’s reconstruction capability; In~\cite{liu2022dual}, Liu \emph{et al.} proposed a Dual Learning-based Graph Neural Network (DLGNN), in which the graph neural network (GNN) is utilized to consider the self-similarity patches in remote sensing imagery by aggregating cross-scale neighboring feature patches. All these methods achieve excellent results in remote sensing image super-resolution.

\subsubsection{\textbf{Light Field Image Super-Resolution}}
A light field (LF) camera is a camera that can capture information about the light field emanating from a scene and can provide multiple views of a scene. Recently, the LF image has become more and more important since it can be used for post-capture refocusing, depth sensing, and de-occlusion. However, LF cameras are faced with a trade-off between spatial and angular resolution~\cite{wang2020spatial}. To solve this issue, SR technology is introduced to achieve a good balance between spatial and angular resolution.

In~\cite{yoon2017light}, a cascade convolution neural network is introduced to simultaneously up-sample both the spatial and angular resolutions of a light field image. Meanwhile, a new light field image dataset is proposed for training and validation. To reduce the dependence of accurate depth or disparity information as priors for the light-field image super-resolution, Sun \emph{et al.}~\cite{wang2018lfnet} proposed a bidirectional recurrent convolutional neural network and an implicitly multi-scale fusion scheme for SR images reconstruction. In~\cite{wang2020spatial}, Wang \emph{et al.} proposed a spatial-angular interactive network (LF-InterNet) for LF image SR. Meanwhile, they designed an angular deformable alignment module for feature-level alignment and proposed a deformable convolution network (LF-DFnet~\cite{wang2020light}) to handle the disparity problem of LF image SR. In~\cite{wang2022disentangling}, Wang \emph{et al.} further proposed a generic light field disentangling mechanism to achieve state-of-the-art performance in spatial SR, angular SR and disparity estimation, respectively. In~\cite{van2023light}, Duong \emph{et al.} proposed a light field SR model via joint spatial-angular and epipolar information, which can simultaneously exploit information from three different types of 4D LF representation.

\subsubsection{\textbf{Face Image Super-Resolution}}
Face image super-resolution is the most famous field in which SR technology to domain-specific images. Due to the potential applications in facial recognition systems such as security and surveillance, face image SR has become an active area of research. 

Recently, DL-based methods have achieved remarkable progress in face image SR. In~\cite{zhou2015learning}, a dubbed CPGAN is proposed to address face hallucination and illumination compensation together, which is optimized by the conventional face hallucination loss and a new illumination compensation loss. In~\cite{zhu2016deep}, Zhu \emph{et al.} proposed to jointly learn face hallucination and facial spatial correspondence field estimation. In~\cite{yu2017hallucinating}, spatial transformer networks are used in the generator architecture to overcome problems related to the misalignment of input images. In~\cite{zhang2018super,dogan2019exemplar}, the identity loss is utilized to preserve the identity-related features by minimizing the distance between the embedding vectors of SR and HR face images. In~\cite{gao2023jdsr}, the mask occlusion is treated as image noise, and a joint and collaborative learning network (JDSR-GAN) is constructed for the masked face super-resolution task. These methods~\cite{zhu2022blind,gu2022vqfr,zhou2022towards} for reconstructing high-quality face images with photo-realistic textures from very low-resolution inputs are mainly based on the generative prior of GAN.

\subsubsection{\textbf{Hyperspectral Image Super-Resolution}}
In contrast to human eyes that can only be exposed to visible light, hyperspectral imaging is a technique for collecting and processing information across the entire range of electromagnetic spectrum\cite{rickard1993hydice}. The hyperspectral system is often compromised due to the limitations of the amount of incident energy, hence there is a trade-off between the spatial and spectral resolution. Therefore, hyperspectral image super-resolution is studied to solve this problem.

In~\cite{mei2017hyperspectral}, a 3D fully convolutional neural network is proposed to extract the feature of hyperspectral images. In~\cite{li2018single}, Li \emph{et al.} proposed a grouped deep recursive residual network by designing a group recursive module and embedding it into a global residual structure. In~\cite{fu2019hyperspectral}, an unsupervised CNN-based method is proposed to effectively exploit the underlying characteristics of the hyperspectral images. In~\cite{jiang2020learning}, Jiang \emph{et al.} proposed a group convolution and progressive upsampling framework to reduce the size of the model and make it feasible to obtain stable training results under small data conditions. In~\cite{liu2021spectral}, a Spectral Grouping and Attention-Driven Residual Dense Network is proposed to facilitate the modeling of all spectral bands and focus on the exploration of spatial-spectral features. In~\cite{cai2022mst++}, the quality of reconstructed images is improved from coarse to fine by using the spectral-wise multi-headed self-attention, which is based on the HSI spatially sparse while spectrally selfsimilar nature to compose the basic unit. In~\cite{zhang2023essaformer}, Zhang \emph{et al.} proposed an efficient Transformer for hyperspectral image super-resolution via a novel and efficient SCC-kernel-based self-attention method.

\subsubsection{\textbf{Medical Image Super-Resolution}}
Medical imaging methods such as Computational Tomography (CT) and Magnetic Resonance Imaging (MRI) are essential to clinical diagnoses and surgery planning. Hence, high-resolution medical images are desirable to provide necessary visual information about the human body. In recent years, many DL-based methods have also been proposed for medical image SR.

For instance, Chen \emph{et al.} proposed a Multi-level Densely Connected Super-Resolution Network (mDCSRN~\cite{chen2018efficient}) with GAN-guided training to generate high-resolution MR images, which can train and infer quickly. In~\cite{wang2019ct}, a 3D Super-Resolution Convolutional Neural Network (3DSRCNN) is proposed to improve the resolution of 3D-CT volumetric images. In~\cite{zhao2019channel}, Zhao \emph{et al.} proposed a deep Channel Splitting Network (CSN) to ease the representational burden of deep models and further improve the SR performance of MR images. In~\cite{peng2020saint}, Peng \emph{et al.} introduced a Spatially-Aware Interpolation Network (SAINT) for medical slice synthesis to alleviate the memory constraint that volumetric data posed. In ~\cite{feng2021task}, Feng \emph{et al.} proposed a Task Transformer Network (T2Net) to allow the network to share representation and feature transfer between the two tasks of reconstruction and super-resolution. In~\cite{georgescu2023multimodal}, Georgescu \emph{et al.} performed medical image super-resolution using a multimodal low-resolution input and propose a novel multimodal multi-head convolutional attention mechanism for multi-contrast medical image SR.

All of these methods are the cornerstone of building the smart medical system and have great research significance and value.

\begin{table*}
    \centering
    \doublerulesepcolor{gray}
    \setlength{\tabcolsep}{7.3mm}
    \renewcommand\arraystretch{1.2}
    \caption{PSNR/SSIM comparison on Set5 ($\times 4$), Set14 ($\times 4$), and Urban100 ($\times 4$). Meanwhile, the training datasets and the number of model parameters are provided. It is worth noting that the upper part of the table is lightweight models with parameters less than 1M (M=million) and they are sorted in ascending order by PSNR results on Set5. Meanwhile, the best results are \textbf{highlighted}.}
    \scalebox{0.62}{
    \begin{tabular}{|c|c|c|c|c|c|c|}
    \hline
    \rowcolor{gray!70} \textbf{Models} & \begin{tabular}[c]{@{}c@{}}\textbf{Set5}\\ \textbf{PSNR/SSIM}\end{tabular} & \begin{tabular}[c]{@{}c@{}}\textbf{Set14}\\ \textbf{PSNR/SSIM}\end{tabular} & \begin{tabular}[c]{@{}c@{}}\textbf{Urban100}\\ \textbf{PSNR/SSIM}\end{tabular} & \textbf{Training Datasets} & \textbf{Parameters} \\
    \hline
    SRCNN~\cite{yoon2015learning} & 30.48/0.8628 & 27.50/0.7513 & 24.52/0.7221 & T91+ImageNet & 57K   \\
    \hline
    \rowcolor{gray!30}ESPCN~\cite{shi2016real} & 30.66/0.8646 & 27.71/0.7562 & 24.60/0.7360 & T91+ImageNet & 20K   \\
    \hline 
    FSRCNN~\cite{dong2016accelerating} & 30.71/0.8660 & 27.59/0.7550 & 24.62/0.7280 & T91+General-100 & 13K   \\
    \hline 
    \rowcolor{gray!30}VDSR~\cite{kim2016accurate} & 31.35/0.8838 & 28.02/0.7680 & 25.18/0.7540 & BSD+T91 & 665K   \\
    \hline
    LapSRN~\cite{lai2017deep} & 31.54/0.8855 & 28.19/0.7720 & 25.21/0.7560 & BSD+T91 & 812K   \\
    \hline
    \rowcolor{gray!30}DRRN~\cite{tai2017image} & 31.68/0.8888 & 28.21/0.7721 & 25.44/0.7638 & BSD+T91 & 297K   \\
    \hline
    MemNet~\cite{tai2017memnet} & 31.74/0.8893 & 28.26/0.7723 & 25.50/0.7630 & BSD+T91 & 677K  \\
    \hline 
    \rowcolor{gray!30}AWSRN-S~\cite{wang1904lightweight} & 31.77/0.8893 & 28.35/0.7761 & 25.56/0.7678 & DIV2K & 588K  \\
    \hline 
    IDN~\cite{hui2018fast} & 31.82/0.8903 & 28.25/0.7730 & 25.41/0.7632 & BSD+T91 & 678K  \\
    \hline 
    \rowcolor{gray!30}NLRN~\cite{liu2018non} & 31.92/0.8916 & 28.36/0.7745 & 25.79/0.7729 & BSD+T91 & 330K   \\
    \hline
    ECBSR~\cite{zhang2021edge} & 31.92/0.8946 & 28.34/0.7817 & 25.81/0.7773 & DIV2K &  682K  \\
    \hline
    \rowcolor{gray!30}CARN-M~\cite{ahn2018fast} & 31.92/0.8903 & 28.42/0.7762 & 25.62/0.7694 & DIV2K & 412K   \\
    \hline
    SMSR~\cite{wang2021exploring} & 32.12/0.8932 & 28.55/0.7808 & 26.11/0.7868 & DIV2K &  1006K  \\
    \hline
    \rowcolor{gray!30}RFDN~\cite{liu2020residual} & 32.18/0.8948 & 28.58/0.7812 & 26.04/0.7848 & DIV2K & 441K   \\
    \hline
    ESRT~\cite{lu2021efficient} & 32.19/0.8947 & 28.69/0.7833 & 26.39/0.7962 & DIV2K & 751K   \\
    \hline
    \rowcolor{gray!30}IMDN~\cite{hui2019lightweight} & 32.21/0.8949 & 28.58/0.7811 & 26.04/0.7838 & DIV2K &  715K  \\
    \hline
    FDIWN~\cite{gao2022feature} & 32.23/0.8955 & 28.66/0.7829 & 26.28/0.7919 & DIV2K & 664K  \\
    \hline
    \rowcolor{gray!30}MAFFSRN~\cite{muqeet2020multi} & 32.24/0.8952 & 28.61/0.7819 & 26.11/0.7858 & DIV2K & 550K   \\
    \hline
    MSFIN~\cite{wang2021lightweight} & 32.28/0.8957 & 28.57/0.7813 & 26.13/0.7865 & DIV2K &  682K  \\
    \hline
    \rowcolor{gray!30}LBNet~\cite{gao2022lightweight} & 32.29/0.8960 & 28.68/0.7832 & 26.27/0.7906 & DIV2K &  742K  \\
    \hline
    LatticeNet-CL~\cite{luo2022lattice} & 32.30/0.8958 & 28.65/0.7822 & 26.19/0.78555 & DIV2K &  777K  \\
    \hline
    \rowcolor{gray!30}HPUN-L~\cite{sun2022hybrid} & 32.31/0.8962 &  28.73/0.7842 & 26.27/0.7918 & DIV2K & 734K  \\
    \hline
    ELAN~\cite{zhang2022efficient} & 32.43/0.8975 &  \textbf{28.78}/\textbf{0.7858} & \textbf{26.47}/\textbf{0.7980} & DIV2K & 601K  \\
    \hline
    \rowcolor{gray!30}SwinIR-light~\cite{liang2021swinir} & 32.44/0.8976 & 28.77/\textbf{0.7858} & \textbf{26.47/0.7980} & DIV2K & 886K  \\
    \hline
    CFIN~\cite{li2023cross} & \textbf{32.49}/\textbf{0.8985} & 28.74/0.7849 & 26.39/0.7946 & DIV2K &  699K  \\
    \hline
    \hline 
    \rowcolor{gray!30}DSRN~\cite{han2018image} & 31.40/0.8830 &  28.07/0.7700 & 25.08/0.7470 & T91 & 1.2M  \\
    \hline
    DRCN~\cite{kim2016deeply} & 31.53/0.8838 & 28.02/0.7670 & 25.14/0.7510  & T91 & 1.8M   \\
    \hline
    \rowcolor{gray!30}MADNet~\cite{lan2020madnet} & 31.95/0.8917 & 28.44/0.7780 & 25.76/0.7746  & DIV2K & 1M   \\
    \hline
    SRMD~\cite{zhang2018learning} & 31.96/0.8925 & 28.35/0.7787 & 25.68/0.7731 & BSD+DIV2K+WED & 1.6M   \\
    \hline 
    \rowcolor{gray!30}SRDenseNet~\cite{tong2017image} & 32.02/0.8934 & 28.50/0.7782 & 26.05/0.7819 & ImageNet & 2.0M   \\
    \hline
    SRResNet~\cite{ledig2017photo} & 32.05/0.8910 & 28.49/0.7800 & -------/------- & ImageNet & 1.5M   \\
    \hline
    \rowcolor{gray!30}MSRN~\cite{Li_2018_ECCV} & 32.07/0.8903 & 28.60/0.7751 & 26.04/0.7896 & DIV2K & 6.3M   \\
    \hline
    CARN~\cite{ahn2018fast} & 32.13/0.8937 & 28.60/0.7806 & 26.07/0.7837 & BSD+T91+DIV2K & 1.6M   \\
    \hline
    \rowcolor{gray!30}SeaNet~\cite{fang2020soft} & 32.33/0.8970 & 28.81/0.7855 & 26.32/0.7942  & DIV2K &  7.4M  \\
    \hline
    CRN~\cite{ahn2018fast} & 32.34/0.8971 & 28.74/0.7855 & 26.44/0.7967  & DIV2K &  9.5M  \\
    \hline
    \rowcolor{gray!30}EDSR~\cite{lim2017enhanced} & 32.46/0.8968 & 28.80/0.7876 & 26.64/0.8033 & DIV2K & 43M   \\
    \hline
    RDN~\cite{zhang2018residual} & 32.47/0.8990 & 28.81/0.7871 & 26.61/0.8028 & DIV2K & 22.6M   \\
    \hline
    \rowcolor{gray!30}DBPN~\cite{haris2019deep} & 32.47/0.8980 & 28.82/0.7860 & 26.38/0.7946 & DIV2K+Flickr2K & 10M   \\
    \hline
    SRFBN~\cite{li2019feedback} & 32.47/0.8983 & 28.81/0.7868 & 26.60/0.8015  & DIV2K+Flickr2K & 3.63M   \\
    \hline
    \rowcolor{gray!30}MDCN~\cite{li2020mdcn} & 32.48/0.8985 & 28.83/0.7879 & 26.69/0.8049  & DIV2K &  4.5M  \\
    \hline
    RNAN~\cite{zhang2019residual} & 32.49/0.8982 & 28.83/0.7878 & 26.61/0.8023 & DIV2K & 7.5M   \\
    \hline
    \rowcolor{gray!30}SRRFN~\cite{Li_2019_ICCV} & 32.56/0.8993 & 28.86/0.7882 & 26.78/0.8071  & DIV2K & 4.2M   \\
    \hline
    RCAN~\cite{zhang2018image} & 32.63/0.9002 & 28.87/0.7889 & 26.82/0.8087 & DIV2K & 16M   \\
    \hline
    \rowcolor{gray!30}SAN~\cite{dai2019second} & 32.64/0.9003 & 28.92/0.7888 & 26.79/0.8068 & DIV2K & 15.7M  \\
    \hline 
    HAN~\cite{niu2020single} & 32.64/0.9002 & 28.90/0.7890 & 26.85/0.8094 & DIV2K & 16.1M  \\
    \hline
    \rowcolor{gray!30}RFANet~\cite{liu2020residuala} & 32.66/0.9004 & 28.88/0.7894 & 26.92/0.8112 & DIV2K & 11M   \\
    \hline
    ENLCN~\cite{xia2022efficient} & 32.67/0.9004 & 28.94/0.7892 &  27.12/0.8141 & DIV2K & 43.6M   \\
    \hline
    \rowcolor{gray!30}DRN-S~\cite{guo2020closed} & 32.68/0.9010 & 28.93/0.7900 & 26.84/0.8070 & DIV2K+Flickr2K & 4.8M   \\
    \hline
    CRAN~\cite{zhang2021context} & 32.72/0.9012 & 29.01/0.7918 & 27.13/0.8167 & DIV2K &  14.9M  \\
    \hline
    \rowcolor{gray!30}SwinIR~\cite{liang2021swinir} & 32.92/0.9044 & 29.09/0.7950 & 27.45/0.8254 & DIV2K+Flickr2K &  11.8M  \\
    \hline
    CAT-A~\cite{chen2022cross} & 33.08/0.9052 & 29.18/0.7963 &  27.89/0.8339 & DIV2K+Flickr2K & 16.6M   \\
    \hline
    \rowcolor{gray!30}GRL-B~\cite{li2023efficient} & 33.10/\textbf{0.9094} & 29.37/\textbf{0.8058} &  \textbf{28.53}/\textbf{0.8504} & DIV2K+Flickr2K & 20.2M   \\
    \hline
    HAT-L~\cite{chen2205activating} & \textbf{33.30}/0.9083 & \textbf{29.38}/0.8001 & 28.37/0.8447 & DIV2K+Flickr2K & 40.2M   \\
    \hline
    \end{tabular}
    }
    \label{Results}
\end{table*}

\section{Reconstruction Results}
To help readers intuitively know the performance of the aforementioned SISR models, we provide a detailed comparison of the reconstruction results of these models. 
Specifically, we collect 53 representative SISR models, including the most classic, latest, and SOTA SISR models.

In Table~\ref{Results}  we provide the reconstruction results, training datasets, and model parameters of these models. According to the results, we can find that: (1) Using a large dataset (e.g., DIV2K+Flickr2K) can make the model achieve better results; (2) It is not entirely correct that the more model parameters, the better the model performance. This means that unreasonably increasing the model size is not the best solution; (3) Transformer-based models show strong advantages, whether in lightweight models or large models; (4) Research on the tiny model (parameters less than 1000K) is still lacking. In the future, it is still important to explore more discriminative evaluation indicators and develop more effective SISR models.

\section{Remaining Issues and Future Directions}

It is true that the above models have achieved promising results and have greatly promoted the development of SISR. However, we cannot ignore that there are still many challenging issues in SISR. In this section, we point out some challenges and summarize some promising future directions.

\subsection{Lightweight SISR for Edge Devices}

With the huge development of the smart terminal market, research on lightweight SISR models has gained increasing attention. Although existing lightweight SISR models have achieved a good balance between model size and performance, we find that they still cannot be used in edge devices (e.g., smartphones, and smart cameras). This is because the model size and computational costs of these models still exceed the limits of edge devices.
Therefore, exploring lightweight SISR models that can be practical in use for edge devices has great research significance and commercial value. To achieve this, more efficient network structures and mechanisms are worthy of further exploration. Moreover, it is also necessary to use technologies like network binarization~\cite{ma2019efficient} and network quantization~\cite{li2020pams} to further reduce the model size. Therefore, combining lightweight SISR models with model compression schemes has great application value.

\subsection{Flexible and Adjustable SISR}

Although DL-based SISR models have achieved gratifying results, we notice a phenomenon that the structure of all these models must be consistent during training and testing. This greatly limits the flexibility of the model, making the same model difficult to be applied to different applications scenarios. In other words, training specially designed models to meet the requirements of different platforms in necessary for previous methods. However, it will require a great amount of manpower and material resources. Therefore, it is crucial for us to design a flexible and adjustable SISR model that can be deployed on different platforms without retraining while keeping good results.

\subsection{New Loss Functions and Assessment Methods}
In the past, most SISR models relied on L1 loss or MSE loss. Although some other new loss functions like content loss, texture loss, and adversarial loss have been proposed, they still cannot achieve a good balance between reconstruction accuracy and perceptual quality. Therefore, it remains an important research topic to explore new loss functions that can ease the perception-distortion trade-off. Meanwhile, some new assessment methods are subjective and unfair. Therefore, new assessment methods that can efficiently reflect image perception and distortion at the same time are also essential. 

\subsection{Mutual Promotion with High-Level Tasks}

As we all know, high-level computer vision tasks (e.g., image classification, image segmentation, and image analysis) are highly dependent on the quality of the input image, so SISR technology is usually used for pre-processing. Meanwhile, the quality of the SR images will greatly affect the accuracy of these tasks. Therefore, integrate image super-resolution with high-level tasks has become a hot topic in recent years. To achieve this, Zangeneh \emph{et al.}~\cite{zangeneh2020low} proposed a novel nonlinear coupled mapping architecture using two deep convolutional neural networks to project the low and high resolution face images into a common space to achieve low-resolution face recognition; Wang \emph{et al.}~\cite{wang2020dual} proposed a dual super-resolution learning method for semantic segmentation, which integrate image super-resolution with semantic segmentation into a end-to-end model; Xiang \emph{et al.}~\cite{xiang2021boosting} boosted high-level vision with joint compression artifacts reduction and super-resolution. Although these methods combine SR with high-level tasks and achieve good results, they focus more on the results of high-level vision tasks and ignore the use of feedback from other tasks to further improve the quality of SR images. Therefore, we recommend using the accuracy of high-level CV tasks as an evaluation indicator to measure the quality of the SR image. Meanwhile, we can design some loss functions related to high-level tasks, thus SISR and other tasks can promote and learn from each other.

\subsection{Efficient and Accurate Real SISR}

Real SISR is destined to become the future mainstream in this field. Therefore, it will inevitably become the focus of researchers in the next few years. On the one hand, a sufficiently large and accurate real image dataset is critical to Real SISR. To achieve this, in addition to the manual collection, we recommend using generative technology to simulate the images, as well as using the generative adversarial network to simulate enough degradation modes to build the large real dataset. On the other hand, considering the difficulty of constructing real image datasets, it is important to develop unsupervised learning-based SISR, meta-learning-based SISR, and blind SISR. Among them, unsupervised learning can make the models get rid of the dependence on the dataset, meta-learning can help models migrate from simulated datasets to real data with simple fine-tuning, and blind SISR can display or implicitly learn the degradation mode of the image, and then reconstruct high-quality SR images based on the learned degradation mode. Although plenty of blind SISR methods have been proposed, they always have unstable performance or have strict prerequisites. Therefore, combining them may bring new solutions for real SISR. 

\subsection{Efficient and Accurate Scale Arbitrary SISR}

SISR has seen its applications in diverse real-life scenarios and users. Currently, most DL-based SISR models can only be applied to one or a limited number of multiple upsampling factors. Therefore, it is necessary to develop a flexible and universal scale arbitrary SISR model that can be adapted to any scale, including asymmetric and non-integer scale factors. Although a few scale-arbitrary SISR methods have also been proposed, they tend to lack the flexibility to use and the simplicity to be implemented, which greatly limits their application scenarios. Therefore, it is of great significance to explore a simple and flexible CNN-based accurate scale-arbitrary SISR model like Bicubic.

\subsection{Consider the Characteristics of Different Images}

Although a series of models have been proposed for domain-specific applications, most of them directly transfer the SISR methods to these specific fields. This is the simplest and most feasible method, but it will also inhibit the model performance since they ignore the data structure characteristics of the domain-specific images. Therefore, fully mining and using the potential prior and data characteristics of the domain-specific images is beneficial for efficient and accurate domain-specific SISR model construction. In the future, it will be a trend to further optimize the existing SISR models based on prior knowledge and the characteristics of the domain-specific images. 

\section{Conclusion}

In this survey, we provided a comprehensive overview of DL-based SISR methods according to their targets, including reconstruction efficiency, reconstruction accuracy, perceptual quality, and other technologies that can further improve model performance. Meanwhile, we provided a detailed introduction to the related works of SISR and introduced a series of new tasks and domain-specific applications extended by SISR. In order to view the performance of each model more intuitively, we also provided a detailed comparison of reconstruction results. Moreover, we provided some underlying problems in SISR and introduced several new trends and future directions worthy of further exploration. We believe that the survey can help researchers better understand this field and further promote the development of this field.

\section{Acknowledgments}
This work is supported in part by the National Key R\&D Program of China under Grant 2021YFA1003004, in part by the National Natural Science Foundation of China under Grants 62301306, 62301601, in part by the Science and Technology Commission of Shanghai Municipality under Grant 23ZR1422200, 23YF1412800.


\bibliographystyle{ACM-Reference-Format}
\bibliography{sample-acmsmall}


\begin{thebibliography}{241}


\ifx \showCODEN    \undefined \def \showCODEN     #1{\unskip}     \fi
\ifx \showDOI      \undefined \def \showDOI       #1{#1}\fi
\ifx \showISBNx    \undefined \def \showISBNx     #1{\unskip}     \fi
\ifx \showISBNxiii \undefined \def \showISBNxiii  #1{\unskip}     \fi
\ifx \showISSN     \undefined \def \showISSN      #1{\unskip}     \fi
\ifx \showLCCN     \undefined \def \showLCCN      #1{\unskip}     \fi
\ifx \shownote     \undefined \def \shownote      #1{#1}          \fi
\ifx \showarticletitle \undefined \def \showarticletitle #1{#1}   \fi
\ifx \showURL      \undefined \def \showURL       {\relax}        \fi
\providecommand\bibfield[2]{#2}
\providecommand\bibinfo[2]{#2}
\providecommand\natexlab[1]{#1}
\providecommand\showeprint[2][]{arXiv:#2}

\bibitem[\protect\citeauthoryear{Adriana, Nicolas, Ebrahimi, Antoine, Carlo,
  and Yoshua}{Adriana et~al\mbox{.}}{2015}]%
        {romero2014fitnets}
\bibfield{author}{\bibinfo{person}{Romero Adriana}, \bibinfo{person}{Ballas
  Nicolas}, \bibinfo{person}{K~Samira Ebrahimi}, \bibinfo{person}{Chassang
  Antoine}, \bibinfo{person}{Gatta Carlo}, {and} \bibinfo{person}{Bengio
  Yoshua}.} \bibinfo{year}{2015}\natexlab{}.
\newblock \showarticletitle{Fitnets: Hints for thin deep nets}.
\newblock  (\bibinfo{year}{2015}).
\newblock


\bibitem[\protect\citeauthoryear{Agustsson and Timofte}{Agustsson and
  Timofte}{2017}]%
        {agustsson2017ntire}
\bibfield{author}{\bibinfo{person}{Eirikur Agustsson} {and}
  \bibinfo{person}{Radu Timofte}.} \bibinfo{year}{2017}\natexlab{}.
\newblock \showarticletitle{Ntire 2017 challenge on single image
  super-resolution: Dataset and study}. In \bibinfo{booktitle}{\emph{CVPRW}}.
\newblock


\bibitem[\protect\citeauthoryear{Ahn, Kang, and Sohn}{Ahn
  et~al\mbox{.}}{2018a}]%
        {ahn2018fast}
\bibfield{author}{\bibinfo{person}{Namhyuk Ahn}, \bibinfo{person}{Byungkon
  Kang}, {and} \bibinfo{person}{Kyung-Ah Sohn}.}
  \bibinfo{year}{2018}\natexlab{a}.
\newblock \showarticletitle{Fast, accurate, and lightweight super-resolution
  with cascading residual network}. In \bibinfo{booktitle}{\emph{ECCV}}.
\newblock


\bibitem[\protect\citeauthoryear{Ahn, Kang, and Sohn}{Ahn
  et~al\mbox{.}}{2018b}]%
        {ahn2018image}
\bibfield{author}{\bibinfo{person}{Namhyuk Ahn}, \bibinfo{person}{Byungkon
  Kang}, {and} \bibinfo{person}{Kyung-Ah Sohn}.}
  \bibinfo{year}{2018}\natexlab{b}.
\newblock \showarticletitle{Image super-resolution via progressive cascading
  residual network}. In \bibinfo{booktitle}{\emph{CVPRW}}.
\newblock


\bibitem[\protect\citeauthoryear{Ahn, Hu, Damianou, Lawrence, and Dai}{Ahn
  et~al\mbox{.}}{2019}]%
        {ahn2019variational}
\bibfield{author}{\bibinfo{person}{Sungsoo Ahn}, \bibinfo{person}{Shell~Xu Hu},
  \bibinfo{person}{Andreas Damianou}, \bibinfo{person}{Neil~D Lawrence}, {and}
  \bibinfo{person}{Zhenwen Dai}.} \bibinfo{year}{2019}\natexlab{}.
\newblock \showarticletitle{Variational information distillation for knowledge
  transfer}. In \bibinfo{booktitle}{\emph{CVPR}}.
\newblock


\bibitem[\protect\citeauthoryear{Anwar, Khan, and Barnes}{Anwar
  et~al\mbox{.}}{2020}]%
        {anwar2019deep}
\bibfield{author}{\bibinfo{person}{Saeed Anwar}, \bibinfo{person}{Salman Khan},
  {and} \bibinfo{person}{Nick Barnes}.} \bibinfo{year}{2020}\natexlab{}.
\newblock \showarticletitle{A deep journey into super-resolution: A survey}.
\newblock \bibinfo{journal}{\emph{Comput. Surveys}} \bibinfo{volume}{53},
  \bibinfo{number}{3} (\bibinfo{year}{2020}), \bibinfo{pages}{1--34}.
\newblock


\bibitem[\protect\citeauthoryear{Arbelaez, Maire, Fowlkes, and Malik}{Arbelaez
  et~al\mbox{.}}{2010}]%
        {arbelaez2011}
\bibfield{author}{\bibinfo{person}{Pablo Arbelaez}, \bibinfo{person}{Michael
  Maire}, \bibinfo{person}{Charless Fowlkes}, {and} \bibinfo{person}{Jitendra
  Malik}.} \bibinfo{year}{2010}\natexlab{}.
\newblock \showarticletitle{Contour detection and hierarchical image
  segmentation}.
\newblock \bibinfo{journal}{\emph{IEEE Transactions on Pattern Analysis and
  Machine Intelligence}} \bibinfo{volume}{33}, \bibinfo{number}{5}
  (\bibinfo{year}{2010}), \bibinfo{pages}{898--916}.
\newblock


\bibitem[\protect\citeauthoryear{Bevilacqua, Roumy, Guillemot, and
  Alberi-Morel}{Bevilacqua et~al\mbox{.}}{2012}]%
        {bevilacqua2012low}
\bibfield{author}{\bibinfo{person}{Marco Bevilacqua}, \bibinfo{person}{Aline
  Roumy}, \bibinfo{person}{Christine Guillemot}, {and}
  \bibinfo{person}{Marie~Line Alberi-Morel}.} \bibinfo{year}{2012}\natexlab{}.
\newblock \showarticletitle{Low-complexity single-image super-resolution based
  on nonnegative neighbor embedding}. In \bibinfo{booktitle}{\emph{BMVC}}.
\newblock


\bibitem[\protect\citeauthoryear{Blau, Mechrez, Timofte, Michaeli, and
  Zelnik-Manor}{Blau et~al\mbox{.}}{2018}]%
        {blau20182018}
\bibfield{author}{\bibinfo{person}{Yochai Blau}, \bibinfo{person}{Roey
  Mechrez}, \bibinfo{person}{Radu Timofte}, \bibinfo{person}{Tomer Michaeli},
  {and} \bibinfo{person}{Lihi Zelnik-Manor}.} \bibinfo{year}{2018}\natexlab{}.
\newblock \showarticletitle{The 2018 pirm challenge on perceptual image
  super-resolution}. In \bibinfo{booktitle}{\emph{ECCVW}}.
\newblock


\bibitem[\protect\citeauthoryear{Blau and Michaeli}{Blau and Michaeli}{2018}]%
        {blau2018perception}
\bibfield{author}{\bibinfo{person}{Yochai Blau} {and} \bibinfo{person}{Tomer
  Michaeli}.} \bibinfo{year}{2018}\natexlab{}.
\newblock \showarticletitle{The perception-distortion tradeoff}. In
  \bibinfo{booktitle}{\emph{CVPR}}.
\newblock


\bibitem[\protect\citeauthoryear{Bulat, Yang, and Tzimiropoulos}{Bulat
  et~al\mbox{.}}{2018}]%
        {bulat2018learn}
\bibfield{author}{\bibinfo{person}{Adrian Bulat}, \bibinfo{person}{Jing Yang},
  {and} \bibinfo{person}{Georgios Tzimiropoulos}.}
  \bibinfo{year}{2018}\natexlab{}.
\newblock \showarticletitle{To learn image super-resolution, use a GAN to learn
  how to do image degradation first}. In \bibinfo{booktitle}{\emph{ECCV}}.
\newblock


\bibitem[\protect\citeauthoryear{Cai, Zeng, Yong, Cao, and Zhang}{Cai
  et~al\mbox{.}}{2019}]%
        {cai2019toward}
\bibfield{author}{\bibinfo{person}{Jianrui Cai}, \bibinfo{person}{Hui Zeng},
  \bibinfo{person}{Hongwei Yong}, \bibinfo{person}{Zisheng Cao}, {and}
  \bibinfo{person}{Lei Zhang}.} \bibinfo{year}{2019}\natexlab{}.
\newblock \showarticletitle{Toward real-world single image super-resolution: A
  new benchmark and a new model}. In \bibinfo{booktitle}{\emph{ICCV}}.
\newblock


\bibitem[\protect\citeauthoryear{Cai, Lin, Lin, Wang, Zhang, Pfister, Timofte,
  and Van~Gool}{Cai et~al\mbox{.}}{2022}]%
        {cai2022mst++}
\bibfield{author}{\bibinfo{person}{Yuanhao Cai}, \bibinfo{person}{Jing Lin},
  \bibinfo{person}{Zudi Lin}, \bibinfo{person}{Haoqian Wang},
  \bibinfo{person}{Yulun Zhang}, \bibinfo{person}{Hanspeter Pfister},
  \bibinfo{person}{Radu Timofte}, {and} \bibinfo{person}{Luc Van~Gool}.}
  \bibinfo{year}{2022}\natexlab{}.
\newblock \showarticletitle{Mst++: Multi-stage spectral-wise transformer for
  efficient spectral reconstruction}. In \bibinfo{booktitle}{\emph{CVPR}}.
\newblock


\bibitem[\protect\citeauthoryear{Cao, Liu, Yang, Yu, Wang, Wang, Huang, Wang,
  Huang, Xu, et~al\mbox{.}}{Cao et~al\mbox{.}}{2015}]%
        {cao2015look}
\bibfield{author}{\bibinfo{person}{Chunshui Cao}, \bibinfo{person}{Xianming
  Liu}, \bibinfo{person}{Yi Yang}, \bibinfo{person}{Yinan Yu},
  \bibinfo{person}{Jiang Wang}, \bibinfo{person}{Zilei Wang},
  \bibinfo{person}{Yongzhen Huang}, \bibinfo{person}{Liang Wang},
  \bibinfo{person}{Chang Huang}, \bibinfo{person}{Wei Xu}, {et~al\mbox{.}}}
  \bibinfo{year}{2015}\natexlab{}.
\newblock \showarticletitle{Look and think twice: Capturing top-down visual
  attention with feedback convolutional neural networks}. In
  \bibinfo{booktitle}{\emph{ICCV}}.
\newblock


\bibitem[\protect\citeauthoryear{Cao and Liu}{Cao and Liu}{2019}]%
        {cao2019single}
\bibfield{author}{\bibinfo{person}{Feilong Cao} {and} \bibinfo{person}{Huan
  Liu}.} \bibinfo{year}{2019}\natexlab{}.
\newblock \showarticletitle{Single image super-resolution via multi-scale
  residual channel attention network}.
\newblock \bibinfo{journal}{\emph{Neurocomputing}}  \bibinfo{volume}{358}
  (\bibinfo{year}{2019}), \bibinfo{pages}{424--436}.
\newblock


\bibitem[\protect\citeauthoryear{Carreira, Agrawal, Fragkiadaki, and
  Malik}{Carreira et~al\mbox{.}}{2016}]%
        {carreira2016human}
\bibfield{author}{\bibinfo{person}{Joao Carreira}, \bibinfo{person}{Pulkit
  Agrawal}, \bibinfo{person}{Katerina Fragkiadaki}, {and}
  \bibinfo{person}{Jitendra Malik}.} \bibinfo{year}{2016}\natexlab{}.
\newblock \showarticletitle{Human pose estimation with iterative error
  feedback}. In \bibinfo{booktitle}{\emph{CVPR}}.
\newblock


\bibitem[\protect\citeauthoryear{Chang and Chien}{Chang and Chien}{2019}]%
        {chang2019multi}
\bibfield{author}{\bibinfo{person}{Chia-Yang Chang} {and}
  \bibinfo{person}{Shao-Yi Chien}.} \bibinfo{year}{2019}\natexlab{}.
\newblock \showarticletitle{Multi-scale dense network for single-image
  super-resolution}. In \bibinfo{booktitle}{\emph{ICASSP}}.
\newblock


\bibitem[\protect\citeauthoryear{Chang, Yeung, and Xiong}{Chang
  et~al\mbox{.}}{2004}]%
        {chang2004super}
\bibfield{author}{\bibinfo{person}{Hong Chang}, \bibinfo{person}{Dit-Yan
  Yeung}, {and} \bibinfo{person}{Yimin Xiong}.}
  \bibinfo{year}{2004}\natexlab{}.
\newblock \showarticletitle{Super-resolution through neighbor embedding}. In
  \bibinfo{booktitle}{\emph{CVPR}}.
\newblock


\bibitem[\protect\citeauthoryear{Chen, Shi, Qin, Li, Han, Yang, and Guo}{Chen
  et~al\mbox{.}}{2022c}]%
        {chen2022fe}
\bibfield{author}{\bibinfo{person}{Chaofeng Chen}, \bibinfo{person}{Xinyu Shi},
  \bibinfo{person}{Yipeng Qin}, \bibinfo{person}{Xiaoming Li},
  \bibinfo{person}{Xiaoguang Han}, \bibinfo{person}{Tao Yang}, {and}
  \bibinfo{person}{Shihui Guo}.} \bibinfo{year}{2022}\natexlab{c}.
\newblock \showarticletitle{Real-world blind super-resolution via feature
  matching with implicit high-resolution priors}. In
  \bibinfo{booktitle}{\emph{ACMMM}}.
\newblock


\bibitem[\protect\citeauthoryear{Chen, Xiong, Tian, Zha, and Wu}{Chen
  et~al\mbox{.}}{2019}]%
        {chen2019camera}
\bibfield{author}{\bibinfo{person}{Chang Chen}, \bibinfo{person}{Zhiwei Xiong},
  \bibinfo{person}{Xinmei Tian}, \bibinfo{person}{Zheng-Jun Zha}, {and}
  \bibinfo{person}{Feng Wu}.} \bibinfo{year}{2019}\natexlab{}.
\newblock \showarticletitle{Camera lens super-resolution}. In
  \bibinfo{booktitle}{\emph{CVPR}}.
\newblock


\bibitem[\protect\citeauthoryear{Chen, He, Qing, Wu, Ren, Sheriff, and
  Zhu}{Chen et~al\mbox{.}}{2022b}]%
        {chen2022real}
\bibfield{author}{\bibinfo{person}{Honggang Chen}, \bibinfo{person}{Xiaohai
  He}, \bibinfo{person}{Linbo Qing}, \bibinfo{person}{Yuanyuan Wu},
  \bibinfo{person}{Chao Ren}, \bibinfo{person}{Ray~E Sheriff}, {and}
  \bibinfo{person}{Ce Zhu}.} \bibinfo{year}{2022}\natexlab{b}.
\newblock \showarticletitle{Real-world single image super-resolution: A brief
  review}.
\newblock \bibinfo{journal}{\emph{Information Fusion}}  \bibinfo{volume}{79}
  (\bibinfo{year}{2022}), \bibinfo{pages}{124--145}.
\newblock


\bibitem[\protect\citeauthoryear{Chen, Wang, Guo, Xu, Deng, Liu, Ma, Xu, Xu,
  and Gao}{Chen et~al\mbox{.}}{2021}]%
        {chen2021pre}
\bibfield{author}{\bibinfo{person}{Hanting Chen}, \bibinfo{person}{Yunhe Wang},
  \bibinfo{person}{Tianyu Guo}, \bibinfo{person}{Chang Xu},
  \bibinfo{person}{Yiping Deng}, \bibinfo{person}{Zhenhua Liu},
  \bibinfo{person}{Siwei Ma}, \bibinfo{person}{Chunjing Xu},
  \bibinfo{person}{Chao Xu}, {and} \bibinfo{person}{Wen Gao}.}
  \bibinfo{year}{2021}\natexlab{}.
\newblock \showarticletitle{Pre-trained image processing transformer}. In
  \bibinfo{booktitle}{\emph{CVPR}}.
\newblock


\bibitem[\protect\citeauthoryear{Chen, Chu, Zhang, and Sun}{Chen
  et~al\mbox{.}}{2022a}]%
        {chen2022simple}
\bibfield{author}{\bibinfo{person}{Liangyu Chen}, \bibinfo{person}{Xiaojie
  Chu}, \bibinfo{person}{Xiangyu Zhang}, {and} \bibinfo{person}{Jian Sun}.}
  \bibinfo{year}{2022}\natexlab{a}.
\newblock \showarticletitle{Simple baselines for image restoration}. In
  \bibinfo{booktitle}{\emph{CECCV}}.
\newblock


\bibitem[\protect\citeauthoryear{Chen, Wang, Zhou, and Dong}{Chen
  et~al\mbox{.}}{2023a}]%
        {chen2205activating}
\bibfield{author}{\bibinfo{person}{X Chen}, \bibinfo{person}{X Wang},
  \bibinfo{person}{J Zhou}, {and} \bibinfo{person}{C Dong}.}
  \bibinfo{year}{2023}\natexlab{a}.
\newblock \showarticletitle{Activating More Pixels in Image Super-Resolution
  Transformer}.
\newblock  (\bibinfo{year}{2023}).
\newblock


\bibitem[\protect\citeauthoryear{Chen, Shi, Christodoulou, Xie, Zhou, and
  Li}{Chen et~al\mbox{.}}{2018}]%
        {chen2018efficient}
\bibfield{author}{\bibinfo{person}{Yuhua Chen}, \bibinfo{person}{Feng Shi},
  \bibinfo{person}{Anthony~G Christodoulou}, \bibinfo{person}{Yibin Xie},
  \bibinfo{person}{Zhengwei Zhou}, {and} \bibinfo{person}{Debiao Li}.}
  \bibinfo{year}{2018}\natexlab{}.
\newblock \showarticletitle{Efficient and accurate MRI super-resolution using a
  generative adversarial network and 3D multi-level densely connected network}.
  In \bibinfo{booktitle}{\emph{MICCAI}}.
\newblock


\bibitem[\protect\citeauthoryear{Chen, Zhang, Gu, Kong, Yang, and Yu}{Chen
  et~al\mbox{.}}{2023b}]%
        {chen2023dual}
\bibfield{author}{\bibinfo{person}{Zheng Chen}, \bibinfo{person}{Yulun Zhang},
  \bibinfo{person}{Jinjin Gu}, \bibinfo{person}{Linghe Kong},
  \bibinfo{person}{Xiaokang Yang}, {and} \bibinfo{person}{Fisher Yu}.}
  \bibinfo{year}{2023}\natexlab{b}.
\newblock \showarticletitle{Dual aggregation transformer for image
  super-resolution}. In \bibinfo{booktitle}{\emph{ICCV}}.
\newblock


\bibitem[\protect\citeauthoryear{Chen, Zhang, Gu, Kong, Yuan,
  et~al\mbox{.}}{Chen et~al\mbox{.}}{2022d}]%
        {chen2022cross}
\bibfield{author}{\bibinfo{person}{Zheng Chen}, \bibinfo{person}{Yulun Zhang},
  \bibinfo{person}{Jinjin Gu}, \bibinfo{person}{Linghe Kong},
  \bibinfo{person}{Xin Yuan}, {et~al\mbox{.}}}
  \bibinfo{year}{2022}\natexlab{d}.
\newblock \showarticletitle{Cross Aggregation Transformer for Image
  Restoration}.
\newblock  (\bibinfo{year}{2022}).
\newblock


\bibitem[\protect\citeauthoryear{Choi, Lee, and Yang}{Choi
  et~al\mbox{.}}{2023}]%
        {choi2023n}
\bibfield{author}{\bibinfo{person}{Haram Choi}, \bibinfo{person}{Jeongmin Lee},
  {and} \bibinfo{person}{Jihoon Yang}.} \bibinfo{year}{2023}\natexlab{}.
\newblock \showarticletitle{N-gram in swin transformers for efficient
  lightweight image super-resolution}. In \bibinfo{booktitle}{\emph{CVPR}}.
\newblock


\bibitem[\protect\citeauthoryear{Chollet}{Chollet}{2017}]%
        {chollet2017xception}
\bibfield{author}{\bibinfo{person}{Fran{\c{c}}ois Chollet}.}
  \bibinfo{year}{2017}\natexlab{}.
\newblock \showarticletitle{Xception: Deep learning with depthwise separable
  convolutions}. In \bibinfo{booktitle}{\emph{CVPR}}.
\newblock


\bibitem[\protect\citeauthoryear{Chu, Chen, and Yu}{Chu et~al\mbox{.}}{2022}]%
        {chu2022nafssr}
\bibfield{author}{\bibinfo{person}{Xiaojie Chu}, \bibinfo{person}{Liangyu
  Chen}, {and} \bibinfo{person}{Wenqing Yu}.} \bibinfo{year}{2022}\natexlab{}.
\newblock \showarticletitle{NAFSSR: stereo image super-resolution using
  NAFNet}. In \bibinfo{booktitle}{\emph{CVPR}}.
\newblock


\bibitem[\protect\citeauthoryear{Collobert and Weston}{Collobert and
  Weston}{2008}]%
        {collobert2008unified}
\bibfield{author}{\bibinfo{person}{Ronan Collobert} {and}
  \bibinfo{person}{Jason Weston}.} \bibinfo{year}{2008}\natexlab{}.
\newblock \showarticletitle{A unified architecture for natural language
  processing: Deep neural networks with multitask learning}. In
  \bibinfo{booktitle}{\emph{ICML}}.
\newblock


\bibitem[\protect\citeauthoryear{Dai, Li, Yi, Fang, and Zhang}{Dai
  et~al\mbox{.}}{2021}]%
        {dai2021feedback}
\bibfield{author}{\bibinfo{person}{Qinyan Dai}, \bibinfo{person}{Juncheng Li},
  \bibinfo{person}{Qiaosi Yi}, \bibinfo{person}{Faming Fang}, {and}
  \bibinfo{person}{Guixu Zhang}.} \bibinfo{year}{2021}\natexlab{}.
\newblock \showarticletitle{Feedback Network for Mutually Boosted Stereo Image
  Super-Resolution and Disparity Estimation}.
\newblock \bibinfo{journal}{\emph{ACMMM}} (\bibinfo{year}{2021}).
\newblock


\bibitem[\protect\citeauthoryear{Dai, Cai, Zhang, Xia, and Zhang}{Dai
  et~al\mbox{.}}{2019}]%
        {dai2019second}
\bibfield{author}{\bibinfo{person}{Tao Dai}, \bibinfo{person}{Jianrui Cai},
  \bibinfo{person}{Yongbing Zhang}, \bibinfo{person}{Shu-Tao Xia}, {and}
  \bibinfo{person}{Lei Zhang}.} \bibinfo{year}{2019}\natexlab{}.
\newblock \showarticletitle{Second-order attention network for single image
  super-resolution}. In \bibinfo{booktitle}{\emph{CVPR}}.
\newblock


\bibitem[\protect\citeauthoryear{Deng, Dong, Socher, Li, Li, and Fei-Fei}{Deng
  et~al\mbox{.}}{2009}]%
        {deng2009imagenet}
\bibfield{author}{\bibinfo{person}{Jia Deng}, \bibinfo{person}{Wei Dong},
  \bibinfo{person}{Richard Socher}, \bibinfo{person}{Li-Jia Li},
  \bibinfo{person}{Kai Li}, {and} \bibinfo{person}{Li Fei-Fei}.}
  \bibinfo{year}{2009}\natexlab{}.
\newblock \showarticletitle{Imagenet: A large-scale hierarchical image
  database}. In \bibinfo{booktitle}{\emph{CVPR}}.
\newblock


\bibitem[\protect\citeauthoryear{Devlin, Chang, Lee, and Toutanova}{Devlin
  et~al\mbox{.}}{2018}]%
        {devlin2018bert}
\bibfield{author}{\bibinfo{person}{Jacob Devlin}, \bibinfo{person}{Ming-Wei
  Chang}, \bibinfo{person}{Kenton Lee}, {and} \bibinfo{person}{Kristina
  Toutanova}.} \bibinfo{year}{2018}\natexlab{}.
\newblock \showarticletitle{Bert: Pre-training of deep bidirectional
  transformers for language understanding}.
\newblock \bibinfo{journal}{\emph{arXiv preprint arXiv:1810.04805}}
  (\bibinfo{year}{2018}).
\newblock


\bibitem[\protect\citeauthoryear{Ding, Ma, Wang, and Simoncelli}{Ding
  et~al\mbox{.}}{2020}]%
        {ding2020image}
\bibfield{author}{\bibinfo{person}{Keyan Ding}, \bibinfo{person}{Kede Ma},
  \bibinfo{person}{Shiqi Wang}, {and} \bibinfo{person}{Eero~P Simoncelli}.}
  \bibinfo{year}{2020}\natexlab{}.
\newblock \showarticletitle{Image quality assessment: Unifying structure and
  texture similarity}.
\newblock \bibinfo{journal}{\emph{IEEE Transactions on Pattern Analysis and
  Machine Intelligence}} \bibinfo{volume}{44}, \bibinfo{number}{5}
  (\bibinfo{year}{2020}), \bibinfo{pages}{2567--2581}.
\newblock


\bibitem[\protect\citeauthoryear{Dogan, Gu, and Timofte}{Dogan
  et~al\mbox{.}}{2019}]%
        {dogan2019exemplar}
\bibfield{author}{\bibinfo{person}{Berk Dogan}, \bibinfo{person}{Shuhang Gu},
  {and} \bibinfo{person}{Radu Timofte}.} \bibinfo{year}{2019}\natexlab{}.
\newblock \showarticletitle{Exemplar guided face image super-resolution without
  facial landmarks}. In \bibinfo{booktitle}{\emph{CVPRW}}.
\newblock


\bibitem[\protect\citeauthoryear{Dong, Loy, He, and Tang}{Dong
  et~al\mbox{.}}{2014}]%
        {dong2015image}
\bibfield{author}{\bibinfo{person}{Chao Dong}, \bibinfo{person}{Chen~Change
  Loy}, \bibinfo{person}{Kaiming He}, {and} \bibinfo{person}{Xiaoou Tang}.}
  \bibinfo{year}{2014}\natexlab{}.
\newblock \showarticletitle{Learning a deep convolutional network for image
  super-resolution}. In \bibinfo{booktitle}{\emph{ECCV}}.
\newblock


\bibitem[\protect\citeauthoryear{Dong, Loy, and Tang}{Dong
  et~al\mbox{.}}{2016}]%
        {dong2016accelerating}
\bibfield{author}{\bibinfo{person}{Chao Dong}, \bibinfo{person}{Chen~Change
  Loy}, {and} \bibinfo{person}{Xiaoou Tang}.} \bibinfo{year}{2016}\natexlab{}.
\newblock \showarticletitle{Accelerating the super-resolution convolutional
  neural network}. In \bibinfo{booktitle}{\emph{ECCV}}.
\newblock


\bibitem[\protect\citeauthoryear{Dong, Zhang, Shi, and Wu}{Dong
  et~al\mbox{.}}{2011}]%
        {dong2011image}
\bibfield{author}{\bibinfo{person}{Weisheng Dong}, \bibinfo{person}{Lei Zhang},
  \bibinfo{person}{Guangming Shi}, {and} \bibinfo{person}{Xiaolin Wu}.}
  \bibinfo{year}{2011}\natexlab{}.
\newblock \showarticletitle{Image deblurring and super-resolution by adaptive
  sparse domain selection and adaptive regularization}.
\newblock \bibinfo{journal}{\emph{IEEE Transactions on Image Processing}}
  \bibinfo{volume}{20}, \bibinfo{number}{7} (\bibinfo{year}{2011}),
  \bibinfo{pages}{1838--1857}.
\newblock


\bibitem[\protect\citeauthoryear{Dong, Wang, Sun, Jia, Gao, and Zhang}{Dong
  et~al\mbox{.}}{2020}]%
        {dong2020remote}
\bibfield{author}{\bibinfo{person}{Xiaoyu Dong}, \bibinfo{person}{Longguang
  Wang}, \bibinfo{person}{Xu Sun}, \bibinfo{person}{Xiuping Jia},
  \bibinfo{person}{Lianru Gao}, {and} \bibinfo{person}{Bing Zhang}.}
  \bibinfo{year}{2020}\natexlab{}.
\newblock \showarticletitle{Remote Sensing Image Super-Resolution Using
  Second-Order Multi-Scale Networks}.
\newblock \bibinfo{journal}{\emph{IEEE Transactions on Geoscience and Remote
  Sensing}} \bibinfo{volume}{59}, \bibinfo{number}{4} (\bibinfo{year}{2020}),
  \bibinfo{pages}{3473--3485}.
\newblock


\bibitem[\protect\citeauthoryear{Duchon}{Duchon}{1979}]%
        {duchon1979lan}
\bibfield{author}{\bibinfo{person}{Claude~E. Duchon}.}
  \bibinfo{year}{1979}\natexlab{}.
\newblock \showarticletitle{Lanczos Filtering in One and Two Dimensions}.
\newblock \bibinfo{journal}{\emph{Journal of Applied Meteorology and
  Climatology}} \bibinfo{volume}{18}, \bibinfo{number}{8}
  (\bibinfo{year}{1979}), \bibinfo{pages}{1016--1022}.
\newblock


\bibitem[\protect\citeauthoryear{Fang, Li, and Zeng}{Fang
  et~al\mbox{.}}{2020}]%
        {fang2020soft}
\bibfield{author}{\bibinfo{person}{Faming Fang}, \bibinfo{person}{Juncheng Li},
  {and} \bibinfo{person}{Tieyong Zeng}.} \bibinfo{year}{2020}\natexlab{}.
\newblock \showarticletitle{Soft-edge assisted network for single image
  super-resolution}.
\newblock \bibinfo{journal}{\emph{IEEE Transactions on Image Processing}}
  \bibinfo{volume}{29} (\bibinfo{year}{2020}), \bibinfo{pages}{4656--4668}.
\newblock


\bibitem[\protect\citeauthoryear{Feng, Yan, Fu, Chen, and Xu}{Feng
  et~al\mbox{.}}{2021}]%
        {feng2021task}
\bibfield{author}{\bibinfo{person}{Chun-Mei Feng}, \bibinfo{person}{Yunlu Yan},
  \bibinfo{person}{Huazhu Fu}, \bibinfo{person}{Li Chen}, {and}
  \bibinfo{person}{Yong Xu}.} \bibinfo{year}{2021}\natexlab{}.
\newblock \showarticletitle{Task transformer network for joint MRI
  reconstruction and super-resolution}. In \bibinfo{booktitle}{\emph{MICCAI}}.
\newblock


\bibitem[\protect\citeauthoryear{Fu, Zhang, Zheng, Zhang, and Huang}{Fu
  et~al\mbox{.}}{2019}]%
        {fu2019hyperspectral}
\bibfield{author}{\bibinfo{person}{Ying Fu}, \bibinfo{person}{Tao Zhang},
  \bibinfo{person}{Yinqiang Zheng}, \bibinfo{person}{Debing Zhang}, {and}
  \bibinfo{person}{Hua Huang}.} \bibinfo{year}{2019}\natexlab{}.
\newblock \showarticletitle{Hyperspectral image super-resolution with optimized
  RGB guidance}. In \bibinfo{booktitle}{\emph{CVPR}}.
\newblock


\bibitem[\protect\citeauthoryear{Fujimoto, Ogawa, Yamamoto, Matsui, Yamasaki,
  and Aizawa}{Fujimoto et~al\mbox{.}}{2016}]%
        {fujimoto2016manga109}
\bibfield{author}{\bibinfo{person}{Azuma Fujimoto}, \bibinfo{person}{Toru
  Ogawa}, \bibinfo{person}{Kazuyoshi Yamamoto}, \bibinfo{person}{Yusuke
  Matsui}, \bibinfo{person}{Toshihiko Yamasaki}, {and}
  \bibinfo{person}{Kiyoharu Aizawa}.} \bibinfo{year}{2016}\natexlab{}.
\newblock \showarticletitle{Manga109 dataset and creation of metadata}. In
  \bibinfo{booktitle}{\emph{MANPU}}.
\newblock


\bibitem[\protect\citeauthoryear{Fuoli, Van~Gool, and Timofte}{Fuoli
  et~al\mbox{.}}{2021}]%
        {fuoli2021fourier}
\bibfield{author}{\bibinfo{person}{Dario Fuoli}, \bibinfo{person}{Luc
  Van~Gool}, {and} \bibinfo{person}{Radu Timofte}.}
  \bibinfo{year}{2021}\natexlab{}.
\newblock \showarticletitle{Fourier space losses for efficient perceptual image
  super-resolution}. In \bibinfo{booktitle}{\emph{ICCV}}.
\newblock


\bibitem[\protect\citeauthoryear{Gao, Li, Li, Wu, Lu, and Yu}{Gao
  et~al\mbox{.}}{2022a}]%
        {gao2022feature}
\bibfield{author}{\bibinfo{person}{Guangwei Gao}, \bibinfo{person}{Wenjie Li},
  \bibinfo{person}{Juncheng Li}, \bibinfo{person}{Fei Wu},
  \bibinfo{person}{Huimin Lu}, {and} \bibinfo{person}{Yi Yu}.}
  \bibinfo{year}{2022}\natexlab{a}.
\newblock \showarticletitle{Feature distillation interaction weighting network
  for lightweight image super-resolution}. In \bibinfo{booktitle}{\emph{AAAI}}.
\newblock


\bibitem[\protect\citeauthoryear{Gao, Tang, Wu, Lu, and Yang}{Gao
  et~al\mbox{.}}{2023b}]%
        {gao2023jdsr}
\bibfield{author}{\bibinfo{person}{Guangwei Gao}, \bibinfo{person}{Lei Tang},
  \bibinfo{person}{Fei Wu}, \bibinfo{person}{Huimin Lu}, {and}
  \bibinfo{person}{Jian Yang}.} \bibinfo{year}{2023}\natexlab{b}.
\newblock \showarticletitle{JDSR-GAN: Constructing An Efficient Joint Learning
  Network for Masked Face Super-Resolution}.
\newblock \bibinfo{journal}{\emph{IEEE Transactions on Multimedia}}
  \bibinfo{volume}{25} (\bibinfo{year}{2023}), \bibinfo{pages}{1505--1512}.
\newblock


\bibitem[\protect\citeauthoryear{Gao, Wang, Li, Li, Yu, and Zeng}{Gao
  et~al\mbox{.}}{2022b}]%
        {gao2022lightweight}
\bibfield{author}{\bibinfo{person}{Guangwei Gao}, \bibinfo{person}{Zhengxue
  Wang}, \bibinfo{person}{Juncheng Li}, \bibinfo{person}{Wenjie Li},
  \bibinfo{person}{Yi Yu}, {and} \bibinfo{person}{Tieyong Zeng}.}
  \bibinfo{year}{2022}\natexlab{b}.
\newblock \showarticletitle{Lightweight bimodal network for single-image
  super-resolution via symmetric cnn and recursive transformer}.
\newblock \bibinfo{journal}{\emph{IJCAI}} (\bibinfo{year}{2022}).
\newblock


\bibitem[\protect\citeauthoryear{Gao, Zhao, Li, and Tong}{Gao
  et~al\mbox{.}}{2018}]%
        {gao2018image}
\bibfield{author}{\bibinfo{person}{Qinquan Gao}, \bibinfo{person}{Yan Zhao},
  \bibinfo{person}{Gen Li}, {and} \bibinfo{person}{Tong Tong}.}
  \bibinfo{year}{2018}\natexlab{}.
\newblock \showarticletitle{Image super-resolution using knowledge
  distillation}. In \bibinfo{booktitle}{\emph{ACCV}}.
\newblock


\bibitem[\protect\citeauthoryear{Gao, Liu, Zeng, Xu, Li, Luo, Liu, Zhen, and
  Zhang}{Gao et~al\mbox{.}}{2023a}]%
        {gao2023implicit}
\bibfield{author}{\bibinfo{person}{Sicheng Gao}, \bibinfo{person}{Xuhui Liu},
  \bibinfo{person}{Bohan Zeng}, \bibinfo{person}{Sheng Xu},
  \bibinfo{person}{Yanjing Li}, \bibinfo{person}{Xiaoyan Luo},
  \bibinfo{person}{Jianzhuang Liu}, \bibinfo{person}{Xiantong Zhen}, {and}
  \bibinfo{person}{Baochang Zhang}.} \bibinfo{year}{2023}\natexlab{a}.
\newblock \showarticletitle{Implicit Diffusion Models for Continuous
  Super-Resolution}.
\newblock  (\bibinfo{year}{2023}).
\newblock


\bibitem[\protect\citeauthoryear{Gatys, Ecker, and Bethge}{Gatys
  et~al\mbox{.}}{2015a}]%
        {gatys2015texture}
\bibfield{author}{\bibinfo{person}{Leon Gatys}, \bibinfo{person}{Alexander~S
  Ecker}, {and} \bibinfo{person}{Matthias Bethge}.}
  \bibinfo{year}{2015}\natexlab{a}.
\newblock \showarticletitle{Texture synthesis using convolutional neural
  networks}.
\newblock \bibinfo{journal}{\emph{NeurIPS}}.
\newblock


\bibitem[\protect\citeauthoryear{Gatys, Ecker, and Bethge}{Gatys
  et~al\mbox{.}}{2015b}]%
        {gatys2015neural}
\bibfield{author}{\bibinfo{person}{Leon~A Gatys}, \bibinfo{person}{Alexander~S
  Ecker}, {and} \bibinfo{person}{Matthias Bethge}.}
  \bibinfo{year}{2015}\natexlab{b}.
\newblock \showarticletitle{A neural algorithm of artistic style}.
\newblock \bibinfo{journal}{\emph{arXiv preprint arXiv:1508.06576}}
  (\bibinfo{year}{2015}).
\newblock


\bibitem[\protect\citeauthoryear{Georgescu, Ionescu, Miron, Savencu, Ristea,
  Verga, and Khan}{Georgescu et~al\mbox{.}}{2023}]%
        {georgescu2023multimodal}
\bibfield{author}{\bibinfo{person}{Mariana-Iuliana Georgescu},
  \bibinfo{person}{Radu~Tudor Ionescu}, \bibinfo{person}{Andreea-Iuliana
  Miron}, \bibinfo{person}{Olivian Savencu},
  \bibinfo{person}{Nicolae-C{\u{a}}t{\u{a}}lin Ristea},
  \bibinfo{person}{Nicolae Verga}, {and} \bibinfo{person}{Fahad~Shahbaz Khan}.}
  \bibinfo{year}{2023}\natexlab{}.
\newblock \showarticletitle{Multimodal multi-head convolutional attention with
  various kernel sizes for medical image super-resolution}. In
  \bibinfo{booktitle}{\emph{CVPR}}.
\newblock


\bibitem[\protect\citeauthoryear{Goodfellow, Pouget-Abadie, Mirza, Xu,
  Warde-Farley, Ozair, Courville, and Bengio}{Goodfellow et~al\mbox{.}}{2014}]%
        {goodfellow2014generative}
\bibfield{author}{\bibinfo{person}{Ian Goodfellow}, \bibinfo{person}{Jean
  Pouget-Abadie}, \bibinfo{person}{Mehdi Mirza}, \bibinfo{person}{Bing Xu},
  \bibinfo{person}{David Warde-Farley}, \bibinfo{person}{Sherjil Ozair},
  \bibinfo{person}{Aaron Courville}, {and} \bibinfo{person}{Yoshua Bengio}.}
  \bibinfo{year}{2014}\natexlab{}.
\newblock \showarticletitle{Generative adversarial nets}.
\newblock \bibinfo{journal}{\emph{NeurIPS}} (\bibinfo{year}{2014}).
\newblock


\bibitem[\protect\citeauthoryear{Gu, Lu, Zuo, and Dong}{Gu
  et~al\mbox{.}}{2019a}]%
        {gu2019blind}
\bibfield{author}{\bibinfo{person}{Jinjin Gu}, \bibinfo{person}{Hannan Lu},
  \bibinfo{person}{Wangmeng Zuo}, {and} \bibinfo{person}{Chao Dong}.}
  \bibinfo{year}{2019}\natexlab{a}.
\newblock \showarticletitle{Blind super-resolution with iterative kernel
  correction}. In \bibinfo{booktitle}{\emph{CVPR}}.
\newblock


\bibitem[\protect\citeauthoryear{Gu, Sun, Zhang, Fu, and Wang}{Gu
  et~al\mbox{.}}{2019b}]%
        {gu2019deep}
\bibfield{author}{\bibinfo{person}{Jun Gu}, \bibinfo{person}{Xian Sun},
  \bibinfo{person}{Yue Zhang}, \bibinfo{person}{Kun Fu}, {and}
  \bibinfo{person}{Lei Wang}.} \bibinfo{year}{2019}\natexlab{b}.
\newblock \showarticletitle{Deep residual squeeze and excitation network for
  remote sensing image super-resolution}.
\newblock \bibinfo{journal}{\emph{Remote Sensing}} \bibinfo{volume}{11},
  \bibinfo{number}{15} (\bibinfo{year}{2019}), \bibinfo{pages}{1817}.
\newblock


\bibitem[\protect\citeauthoryear{Gu, Wang, Xie, Dong, Li, Shan, and Cheng}{Gu
  et~al\mbox{.}}{2022}]%
        {gu2022vqfr}
\bibfield{author}{\bibinfo{person}{Yuchao Gu}, \bibinfo{person}{Xintao Wang},
  \bibinfo{person}{Liangbin Xie}, \bibinfo{person}{Chao Dong},
  \bibinfo{person}{Gen Li}, \bibinfo{person}{Ying Shan}, {and}
  \bibinfo{person}{Ming-Ming Cheng}.} \bibinfo{year}{2022}\natexlab{}.
\newblock \showarticletitle{Vqfr: Blind face restoration with vector-quantized
  dictionary and parallel decoder}. In \bibinfo{booktitle}{\emph{ECCV}}.
\newblock


\bibitem[\protect\citeauthoryear{Guo, Chen, Wang, Chen, Cao, Deng, Xu, and
  Tan}{Guo et~al\mbox{.}}{2020}]%
        {guo2020closed}
\bibfield{author}{\bibinfo{person}{Yong Guo}, \bibinfo{person}{Jian Chen},
  \bibinfo{person}{Jingdong Wang}, \bibinfo{person}{Qi Chen},
  \bibinfo{person}{Jiezhang Cao}, \bibinfo{person}{Zeshuai Deng},
  \bibinfo{person}{Yanwu Xu}, {and} \bibinfo{person}{Mingkui Tan}.}
  \bibinfo{year}{2020}\natexlab{}.
\newblock \showarticletitle{Closed-loop matters: Dual regression networks for
  single image super-resolution}. In \bibinfo{booktitle}{\emph{CVPR}}.
\newblock


\bibitem[\protect\citeauthoryear{Han, Chang, Liu, Yu, Witbrock, and Huang}{Han
  et~al\mbox{.}}{2018}]%
        {han2018image}
\bibfield{author}{\bibinfo{person}{Wei Han}, \bibinfo{person}{Shiyu Chang},
  \bibinfo{person}{Ding Liu}, \bibinfo{person}{Mo Yu}, \bibinfo{person}{Michael
  Witbrock}, {and} \bibinfo{person}{Thomas~S Huang}.}
  \bibinfo{year}{2018}\natexlab{}.
\newblock \showarticletitle{Image super-resolution via dual-state recurrent
  networks}. In \bibinfo{booktitle}{\emph{CVPR}}.
\newblock


\bibitem[\protect\citeauthoryear{Haris, Shakhnarovich, and Ukita}{Haris
  et~al\mbox{.}}{2020}]%
        {haris2019deep}
\bibfield{author}{\bibinfo{person}{Muhammad Haris}, \bibinfo{person}{Greg
  Shakhnarovich}, {and} \bibinfo{person}{Norimichi Ukita}.}
  \bibinfo{year}{2020}\natexlab{}.
\newblock \showarticletitle{Deep back-projectinetworks for single image
  super-resolution}.
\newblock \bibinfo{journal}{\emph{IEEE Transactions on Pattern Analysis and
  Machine Intelligence}} \bibinfo{volume}{43}, \bibinfo{number}{12}
  (\bibinfo{year}{2020}), \bibinfo{pages}{4323--4337}.
\newblock


\bibitem[\protect\citeauthoryear{Haut, Fernandez-Beltran, Paoletti, Plaza,
  Plaza, and Pla}{Haut et~al\mbox{.}}{2018}]%
        {haut2018new}
\bibfield{author}{\bibinfo{person}{Juan~Mario Haut}, \bibinfo{person}{Ruben
  Fernandez-Beltran}, \bibinfo{person}{Mercedes~E Paoletti},
  \bibinfo{person}{Javier Plaza}, \bibinfo{person}{Antonio Plaza}, {and}
  \bibinfo{person}{Filiberto Pla}.} \bibinfo{year}{2018}\natexlab{}.
\newblock \showarticletitle{A new deep generative network for unsupervised
  remote sensing single-image super-resolution}.
\newblock \bibinfo{journal}{\emph{IEEE Transactions on Geoscience and Remote
  sensing}} \bibinfo{volume}{56}, \bibinfo{number}{11} (\bibinfo{year}{2018}),
  \bibinfo{pages}{6792--6810}.
\newblock


\bibitem[\protect\citeauthoryear{He and Sun}{He and Sun}{2015}]%
        {he2015convolutional}
\bibfield{author}{\bibinfo{person}{Kaiming He} {and} \bibinfo{person}{Jian
  Sun}.} \bibinfo{year}{2015}\natexlab{}.
\newblock \showarticletitle{Convolutional neural networks at constrained time
  cost}. In \bibinfo{booktitle}{\emph{CVPR}}.
\newblock


\bibitem[\protect\citeauthoryear{He, Zhang, Ren, and Sun}{He
  et~al\mbox{.}}{2016}]%
        {he2016deep}
\bibfield{author}{\bibinfo{person}{Kaiming He}, \bibinfo{person}{Xiangyu
  Zhang}, \bibinfo{person}{Shaoqing Ren}, {and} \bibinfo{person}{Jian Sun}.}
  \bibinfo{year}{2016}\natexlab{}.
\newblock \showarticletitle{Deep residual learning for image recognition}. In
  \bibinfo{booktitle}{\emph{CVPR}}.
\newblock


\bibitem[\protect\citeauthoryear{Heusel, Ramsauer, Unterthiner, Nessler, and
  Hochreiter}{Heusel et~al\mbox{.}}{2017}]%
        {heusel2017gans}
\bibfield{author}{\bibinfo{person}{Martin Heusel}, \bibinfo{person}{Hubert
  Ramsauer}, \bibinfo{person}{Thomas Unterthiner}, \bibinfo{person}{Bernhard
  Nessler}, {and} \bibinfo{person}{Sepp Hochreiter}.}
  \bibinfo{year}{2017}\natexlab{}.
\newblock \showarticletitle{Gans trained by a two time-scale update rule
  converge to a local nash equilibrium}.
\newblock \bibinfo{journal}{\emph{NeurIPS}} (\bibinfo{year}{2017}).
\newblock


\bibitem[\protect\citeauthoryear{Hinton, Vinyals, and Dean}{Hinton
  et~al\mbox{.}}{2015}]%
        {hinton2015distilling}
\bibfield{author}{\bibinfo{person}{Geoffrey Hinton}, \bibinfo{person}{Oriol
  Vinyals}, {and} \bibinfo{person}{Jeff Dean}.}
  \bibinfo{year}{2015}\natexlab{}.
\newblock \showarticletitle{Distilling the knowledge in a neural network}.
\newblock \bibinfo{journal}{\emph{arXiv preprint arXiv:1503.02531}}
  (\bibinfo{year}{2015}).
\newblock


\bibitem[\protect\citeauthoryear{Ho, Jain, and Abbeel}{Ho
  et~al\mbox{.}}{2020}]%
        {ho2020denoising}
\bibfield{author}{\bibinfo{person}{Jonathan Ho}, \bibinfo{person}{Ajay Jain},
  {and} \bibinfo{person}{Pieter Abbeel}.} \bibinfo{year}{2020}\natexlab{}.
\newblock \showarticletitle{Denoising diffusion probabilistic models}.
\newblock \bibinfo{journal}{\emph{NeurIPS}} (\bibinfo{year}{2020}).
\newblock


\bibitem[\protect\citeauthoryear{Hu, Shen, and Sun}{Hu et~al\mbox{.}}{2018}]%
        {hu2018squeeze}
\bibfield{author}{\bibinfo{person}{Jie Hu}, \bibinfo{person}{Li Shen}, {and}
  \bibinfo{person}{Gang Sun}.} \bibinfo{year}{2018}\natexlab{}.
\newblock \showarticletitle{Squeeze-and-excitation networks}. In
  \bibinfo{booktitle}{\emph{CVPR}}.
\newblock


\bibitem[\protect\citeauthoryear{Hu, Mu, Zhang, Wang, Tan, and Sun}{Hu
  et~al\mbox{.}}{2019}]%
        {hu2019meta}
\bibfield{author}{\bibinfo{person}{Xuecai Hu}, \bibinfo{person}{Haoyuan Mu},
  \bibinfo{person}{Xiangyu Zhang}, \bibinfo{person}{Zilei Wang},
  \bibinfo{person}{Tieniu Tan}, {and} \bibinfo{person}{Jian Sun}.}
  \bibinfo{year}{2019}\natexlab{}.
\newblock \showarticletitle{Meta-SR: A magnification-arbitrary network for
  super-resolution}. In \bibinfo{booktitle}{\emph{CVPR}}.
\newblock


\bibitem[\protect\citeauthoryear{Hu, Zhang, Shan, Wang, Wang, and Tan}{Hu
  et~al\mbox{.}}{2020}]%
        {hu2020meta}
\bibfield{author}{\bibinfo{person}{Xuecai Hu}, \bibinfo{person}{Zhang Zhang},
  \bibinfo{person}{Caifeng Shan}, \bibinfo{person}{Zilei Wang},
  \bibinfo{person}{Liang Wang}, {and} \bibinfo{person}{Tieniu Tan}.}
  \bibinfo{year}{2020}\natexlab{}.
\newblock \showarticletitle{Meta-USR: A unified super-resolution network for
  multiple degradation parameters}.
\newblock \bibinfo{journal}{\emph{IEEE Transactions on Neural Networks and
  Learning Systems}} \bibinfo{volume}{32}, \bibinfo{number}{9}
  (\bibinfo{year}{2020}), \bibinfo{pages}{4151--4165}.
\newblock


\bibitem[\protect\citeauthoryear{Huang, Liu, Van Der~Maaten, and
  Weinberger}{Huang et~al\mbox{.}}{2017}]%
        {huang2017densely}
\bibfield{author}{\bibinfo{person}{Gao Huang}, \bibinfo{person}{Zhuang Liu},
  \bibinfo{person}{Laurens Van Der~Maaten}, {and} \bibinfo{person}{Kilian~Q
  Weinberger}.} \bibinfo{year}{2017}\natexlab{}.
\newblock \showarticletitle{Densely connected convolutional networks}. In
  \bibinfo{booktitle}{\emph{CVPR}}.
\newblock


\bibitem[\protect\citeauthoryear{Huang, Singh, and Ahuja}{Huang
  et~al\mbox{.}}{2015}]%
        {huang2015single}
\bibfield{author}{\bibinfo{person}{Jia-Bin Huang}, \bibinfo{person}{Abhishek
  Singh}, {and} \bibinfo{person}{Narendra Ahuja}.}
  \bibinfo{year}{2015}\natexlab{}.
\newblock \showarticletitle{Single image super-resolution from transformed
  self-exemplars}. In \bibinfo{booktitle}{\emph{CVPR}}.
\newblock


\bibitem[\protect\citeauthoryear{Hui, Gao, Yang, and Wang}{Hui
  et~al\mbox{.}}{2019}]%
        {hui2019lightweight}
\bibfield{author}{\bibinfo{person}{Zheng Hui}, \bibinfo{person}{Xinbo Gao},
  \bibinfo{person}{Yunchu Yang}, {and} \bibinfo{person}{Xiumei Wang}.}
  \bibinfo{year}{2019}\natexlab{}.
\newblock \showarticletitle{Lightweight image super-resolution with information
  multi-distillation network}. In \bibinfo{booktitle}{\emph{ACMMM}}.
\newblock


\bibitem[\protect\citeauthoryear{Hui, Wang, and Gao}{Hui et~al\mbox{.}}{2018}]%
        {hui2018fast}
\bibfield{author}{\bibinfo{person}{Zheng Hui}, \bibinfo{person}{Xiumei Wang},
  {and} \bibinfo{person}{Xinbo Gao}.} \bibinfo{year}{2018}\natexlab{}.
\newblock \showarticletitle{Fast and accurate single image super-resolution via
  information distillation network}. In \bibinfo{booktitle}{\emph{CVPR}}.
\newblock


\bibitem[\protect\citeauthoryear{Jeon, Baek, Choi, and Kim}{Jeon
  et~al\mbox{.}}{2018}]%
        {jeon2018enhancing}
\bibfield{author}{\bibinfo{person}{Daniel~S Jeon}, \bibinfo{person}{Seung-Hwan
  Baek}, \bibinfo{person}{Inchang Choi}, {and} \bibinfo{person}{Min~H Kim}.}
  \bibinfo{year}{2018}\natexlab{}.
\newblock \showarticletitle{Enhancing the spatial resolution of stereo images
  using a parallax prior}. In \bibinfo{booktitle}{\emph{CVPR}}.
\newblock


\bibitem[\protect\citeauthoryear{{Jian Sun}, {Zongben Xu}, and {Heung-Yeung
  Shum}}{{Jian Sun} et~al\mbox{.}}{2008}]%
        {sun2008grad}
\bibfield{author}{\bibinfo{person}{{Jian Sun}}, \bibinfo{person}{{Zongben Xu}},
  {and} \bibinfo{person}{{Heung-Yeung Shum}}.} \bibinfo{year}{2008}\natexlab{}.
\newblock \showarticletitle{Image super-resolution using gradient profile
  prior}. In \bibinfo{booktitle}{\emph{CVPR}}.
\newblock


\bibitem[\protect\citeauthoryear{Jiang, Sun, Liu, and Ma}{Jiang
  et~al\mbox{.}}{2020}]%
        {jiang2020learning}
\bibfield{author}{\bibinfo{person}{Junjun Jiang}, \bibinfo{person}{He Sun},
  \bibinfo{person}{Xianming Liu}, {and} \bibinfo{person}{Jiayi Ma}.}
  \bibinfo{year}{2020}\natexlab{}.
\newblock \showarticletitle{Learning spatial-spectral prior for
  super-resolution of hyperspectral imagery}.
\newblock \bibinfo{journal}{\emph{IEEE Transactions on Computational Imaging}}
  \bibinfo{volume}{6} (\bibinfo{year}{2020}), \bibinfo{pages}{1082--1096}.
\newblock


\bibitem[\protect\citeauthoryear{Jinjin, Haoming, Haoyu, Xiaoxing, Ren, and
  Chao}{Jinjin et~al\mbox{.}}{2020}]%
        {jinjin2020pipal}
\bibfield{author}{\bibinfo{person}{Gu Jinjin}, \bibinfo{person}{Cai Haoming},
  \bibinfo{person}{Chen Haoyu}, \bibinfo{person}{Ye Xiaoxing},
  \bibinfo{person}{Jimmy~S Ren}, {and} \bibinfo{person}{Dong Chao}.}
  \bibinfo{year}{2020}\natexlab{}.
\newblock \showarticletitle{Pipal: a large-scale image quality assessment
  dataset for perceptual image restoration}. In
  \bibinfo{booktitle}{\emph{ECCV}}.
\newblock


\bibitem[\protect\citeauthoryear{Johnson, Alahi, and Fei-Fei}{Johnson
  et~al\mbox{.}}{2016}]%
        {johnson2016perceptual}
\bibfield{author}{\bibinfo{person}{Justin Johnson}, \bibinfo{person}{Alexandre
  Alahi}, {and} \bibinfo{person}{Li Fei-Fei}.} \bibinfo{year}{2016}\natexlab{}.
\newblock \showarticletitle{Perceptual losses for real-time style transfer and
  super-resolution}. In \bibinfo{booktitle}{\emph{ECCV}}.
\newblock


\bibitem[\protect\citeauthoryear{Jolicoeur-Martineau}{Jolicoeur-Martineau}{2018}]%
        {jolicoeur2018relativistic}
\bibfield{author}{\bibinfo{person}{Alexia Jolicoeur-Martineau}.}
  \bibinfo{year}{2018}\natexlab{}.
\newblock \showarticletitle{The relativistic discriminator: a key element
  missing from standard GAN}.
\newblock \bibinfo{journal}{\emph{arXiv preprint arXiv:1807.00734}}
  (\bibinfo{year}{2018}).
\newblock


\bibitem[\protect\citeauthoryear{Karras, Laine, and Aila}{Karras
  et~al\mbox{.}}{2019}]%
        {karras2019style}
\bibfield{author}{\bibinfo{person}{Tero Karras}, \bibinfo{person}{Samuli
  Laine}, {and} \bibinfo{person}{Timo Aila}.} \bibinfo{year}{2019}\natexlab{}.
\newblock \showarticletitle{A style-based generator architecture for generative
  adversarial networks}. In \bibinfo{booktitle}{\emph{CVPR}}.
\newblock


\bibitem[\protect\citeauthoryear{Kim, Kwon~Lee, and Mu~Lee}{Kim
  et~al\mbox{.}}{2016a}]%
        {kim2016accurate}
\bibfield{author}{\bibinfo{person}{Jiwon Kim}, \bibinfo{person}{Jung Kwon~Lee},
  {and} \bibinfo{person}{Kyoung Mu~Lee}.} \bibinfo{year}{2016}\natexlab{a}.
\newblock \showarticletitle{Accurate image super-resolution using very deep
  convolutional networks}. In \bibinfo{booktitle}{\emph{CVPR}}.
\newblock


\bibitem[\protect\citeauthoryear{Kim, Kwon~Lee, and Mu~Lee}{Kim
  et~al\mbox{.}}{2016b}]%
        {kim2016deeply}
\bibfield{author}{\bibinfo{person}{Jiwon Kim}, \bibinfo{person}{Jung Kwon~Lee},
  {and} \bibinfo{person}{Kyoung Mu~Lee}.} \bibinfo{year}{2016}\natexlab{b}.
\newblock \showarticletitle{Deeply-recursive convolutional network for image
  super-resolution}. In \bibinfo{booktitle}{\emph{CVPR}}.
\newblock


\bibitem[\protect\citeauthoryear{{Kim} and {Kwon}}{{Kim} and {Kwon}}{2010}]%
        {kim2010sparse}
\bibfield{author}{\bibinfo{person}{K.~I. {Kim}} {and} \bibinfo{person}{Y.
  {Kwon}}.} \bibinfo{year}{2010}\natexlab{}.
\newblock \showarticletitle{Single-Image Super-Resolution Using Sparse
  Regression and Natural Image Prior}.
\newblock \bibinfo{journal}{\emph{IEEE Transactions on Pattern Analysis and
  Machine Intelligence}} \bibinfo{volume}{32}, \bibinfo{number}{6}
  (\bibinfo{year}{2010}), \bibinfo{pages}{1127--1133}.
\newblock


\bibitem[\protect\citeauthoryear{Krizhevsky, Sutskever, and Hinton}{Krizhevsky
  et~al\mbox{.}}{2012}]%
        {krizhevsky2012imagenet}
\bibfield{author}{\bibinfo{person}{Alex Krizhevsky}, \bibinfo{person}{Ilya
  Sutskever}, {and} \bibinfo{person}{Geoffrey~E Hinton}.}
  \bibinfo{year}{2012}\natexlab{}.
\newblock \showarticletitle{Imagenet classification with deep convolutional
  neural networks}.
\newblock \bibinfo{journal}{\emph{NeurIPS}}.
\newblock


\bibitem[\protect\citeauthoryear{Lai, Huang, Ahuja, and Yang}{Lai
  et~al\mbox{.}}{2017a}]%
        {lai2017deep}
\bibfield{author}{\bibinfo{person}{Wei-Sheng Lai}, \bibinfo{person}{Jia-Bin
  Huang}, \bibinfo{person}{Narendra Ahuja}, {and} \bibinfo{person}{Ming-Hsuan
  Yang}.} \bibinfo{year}{2017}\natexlab{a}.
\newblock \showarticletitle{Deep laplacian pyramid networks for fast and
  accurate super-resolution}. In \bibinfo{booktitle}{\emph{CVPR}}.
\newblock


\bibitem[\protect\citeauthoryear{Lai, Huang, Ahuja, and Yang}{Lai
  et~al\mbox{.}}{2017b}]%
        {lai2017fast}
\bibfield{author}{\bibinfo{person}{Wei-Sheng Lai}, \bibinfo{person}{Jia-Bin
  Huang}, \bibinfo{person}{Narendra Ahuja}, {and} \bibinfo{person}{Ming-Hsuan
  Yang}.} \bibinfo{year}{2017}\natexlab{b}.
\newblock \showarticletitle{Deep laplacian pyramid networks for fast and
  accurate super-resolution}. In \bibinfo{booktitle}{\emph{CVPR}}.
\newblock


\bibitem[\protect\citeauthoryear{Lan, Sun, Liu, Lu, Pang, and Luo}{Lan
  et~al\mbox{.}}{2020}]%
        {lan2020madnet}
\bibfield{author}{\bibinfo{person}{Rushi Lan}, \bibinfo{person}{Long Sun},
  \bibinfo{person}{Zhenbing Liu}, \bibinfo{person}{Huimin Lu},
  \bibinfo{person}{Cheng Pang}, {and} \bibinfo{person}{Xiaonan Luo}.}
  \bibinfo{year}{2020}\natexlab{}.
\newblock \showarticletitle{Madnet: A fast and lightweight network for
  single-image super resolution}.
\newblock \bibinfo{journal}{\emph{IEEE Transactions on Cybernetics}}
  \bibinfo{volume}{51}, \bibinfo{number}{3} (\bibinfo{year}{2020}),
  \bibinfo{pages}{1443--1453}.
\newblock


\bibitem[\protect\citeauthoryear{LeCun, Bengio, and Hinton}{LeCun
  et~al\mbox{.}}{2015}]%
        {lecun2015deep}
\bibfield{author}{\bibinfo{person}{Y. LeCun}, \bibinfo{person}{Y. Bengio},
  {and} \bibinfo{person}{G. Hinton}.} \bibinfo{year}{2015}\natexlab{}.
\newblock \showarticletitle{Deep learning}.
\newblock \bibinfo{journal}{\emph{Nature}} \bibinfo{volume}{521},
  \bibinfo{number}{7553} (\bibinfo{year}{2015}), \bibinfo{pages}{436--444}.
\newblock


\bibitem[\protect\citeauthoryear{Ledig, Theis, Husz{\'a}r, Caballero,
  Cunningham, Acosta, Aitken, Tejani, Totz, Wang, et~al\mbox{.}}{Ledig
  et~al\mbox{.}}{2017}]%
        {ledig2017photo}
\bibfield{author}{\bibinfo{person}{Christian Ledig}, \bibinfo{person}{Lucas
  Theis}, \bibinfo{person}{Ferenc Husz{\'a}r}, \bibinfo{person}{Jose
  Caballero}, \bibinfo{person}{Andrew Cunningham}, \bibinfo{person}{Alejandro
  Acosta}, \bibinfo{person}{Andrew~P Aitken}, \bibinfo{person}{Alykhan Tejani},
  \bibinfo{person}{Johannes Totz}, \bibinfo{person}{Zehan Wang},
  {et~al\mbox{.}}} \bibinfo{year}{2017}\natexlab{}.
\newblock \showarticletitle{Photo-realistic single image super-resolution using
  a generative adversarial network}. In \bibinfo{booktitle}{\emph{CVPR}}.
\newblock


\bibitem[\protect\citeauthoryear{Lee, Liu, Wu, and Luo}{Lee
  et~al\mbox{.}}{2020b}]%
        {CelebAMask-HQ}
\bibfield{author}{\bibinfo{person}{Cheng-Han Lee}, \bibinfo{person}{Ziwei Liu},
  \bibinfo{person}{Lingyun Wu}, {and} \bibinfo{person}{Ping Luo}.}
  \bibinfo{year}{2020}\natexlab{b}.
\newblock \showarticletitle{MaskGAN: Towards Diverse and Interactive Facial
  Image Manipulation}. In \bibinfo{booktitle}{\emph{CVPR}}.
\newblock


\bibitem[\protect\citeauthoryear{Lee, Lee, Kim, and Ham}{Lee
  et~al\mbox{.}}{2020a}]%
        {lee2020learning}
\bibfield{author}{\bibinfo{person}{Wonkyung Lee}, \bibinfo{person}{Junghyup
  Lee}, \bibinfo{person}{Dohyung Kim}, {and} \bibinfo{person}{Bumsub Ham}.}
  \bibinfo{year}{2020}\natexlab{a}.
\newblock \showarticletitle{Learning with privileged information for efficient
  image super-resolution}. In \bibinfo{booktitle}{\emph{ECCV}}.
\newblock


\bibitem[\protect\citeauthoryear{Lei and Shi}{Lei and Shi}{2021}]%
        {lei2021hybrid}
\bibfield{author}{\bibinfo{person}{Sen Lei} {and} \bibinfo{person}{Zhenwei
  Shi}.} \bibinfo{year}{2021}\natexlab{}.
\newblock \showarticletitle{Hybrid-scale self-similarity exploitation for
  remote sensing image super-resolution}.
\newblock \bibinfo{journal}{\emph{IEEE Transactions on Geoscience and Remote
  Sensing}}  \bibinfo{volume}{60} (\bibinfo{year}{2021}),
  \bibinfo{pages}{1--10}.
\newblock


\bibitem[\protect\citeauthoryear{Li, Yan, Lin, Zheng, Zhang, Yang, and Ji}{Li
  et~al\mbox{.}}{2020b}]%
        {li2020pams}
\bibfield{author}{\bibinfo{person}{Huixia Li}, \bibinfo{person}{Chenqian Yan},
  \bibinfo{person}{Shaohui Lin}, \bibinfo{person}{Xiawu Zheng},
  \bibinfo{person}{Baochang Zhang}, \bibinfo{person}{Fan Yang}, {and}
  \bibinfo{person}{Rongrong Ji}.} \bibinfo{year}{2020}\natexlab{b}.
\newblock \showarticletitle{Pams: Quantized super-resolution via parameterized
  max scale}. In \bibinfo{booktitle}{\emph{ECCV}}.
\newblock


\bibitem[\protect\citeauthoryear{Li, Yang, Chang, Chen, Feng, Xu, Li, and
  Chen}{Li et~al\mbox{.}}{2022b}]%
        {li2022srdiff}
\bibfield{author}{\bibinfo{person}{Haoying Li}, \bibinfo{person}{Yifan Yang},
  \bibinfo{person}{Meng Chang}, \bibinfo{person}{Shiqi Chen},
  \bibinfo{person}{Huajun Feng}, \bibinfo{person}{Zhihai Xu},
  \bibinfo{person}{Qi Li}, {and} \bibinfo{person}{Yueting Chen}.}
  \bibinfo{year}{2022}\natexlab{b}.
\newblock \showarticletitle{Srdiff: Single image super-resolution with
  diffusion probabilistic models}.
\newblock \bibinfo{journal}{\emph{Neurocomputing}}  \bibinfo{volume}{479}
  (\bibinfo{year}{2022}), \bibinfo{pages}{47--59}.
\newblock


\bibitem[\protect\citeauthoryear{Li, Fang, Li, Mei, and Zhang}{Li
  et~al\mbox{.}}{2020a}]%
        {li2020mdcn}
\bibfield{author}{\bibinfo{person}{Juncheng Li}, \bibinfo{person}{Faming Fang},
  \bibinfo{person}{Jiaqian Li}, \bibinfo{person}{Kangfu Mei}, {and}
  \bibinfo{person}{Guixu Zhang}.} \bibinfo{year}{2020}\natexlab{a}.
\newblock \showarticletitle{MDCN: Multi-scale dense cross network for image
  super-resolution}.
\newblock \bibinfo{journal}{\emph{IEEE Transactions on Circuits and Systems for
  Video Technology}} \bibinfo{volume}{31}, \bibinfo{number}{7}
  (\bibinfo{year}{2020}), \bibinfo{pages}{2547--2561}.
\newblock


\bibitem[\protect\citeauthoryear{Li, Fang, Mei, and Zhang}{Li
  et~al\mbox{.}}{2018a}]%
        {Li_2018_ECCV}
\bibfield{author}{\bibinfo{person}{Juncheng Li}, \bibinfo{person}{Faming Fang},
  \bibinfo{person}{Kangfu Mei}, {and} \bibinfo{person}{Guixu Zhang}.}
  \bibinfo{year}{2018}\natexlab{a}.
\newblock \showarticletitle{Multi-scale Residual Network for Image
  Super-Resolution}. In \bibinfo{booktitle}{\emph{ECCV}}.
\newblock


\bibitem[\protect\citeauthoryear{Li, Yuan, Mei, and Fang}{Li
  et~al\mbox{.}}{2019b}]%
        {Li_2019_ICCV}
\bibfield{author}{\bibinfo{person}{Juncheng Li}, \bibinfo{person}{Yiting Yuan},
  \bibinfo{person}{Kangfu Mei}, {and} \bibinfo{person}{Faming Fang}.}
  \bibinfo{year}{2019}\natexlab{b}.
\newblock \showarticletitle{Lightweight and Accurate Recursive Fractal Network
  for Image Super-Resolution}. In \bibinfo{booktitle}{\emph{ICCVW}}.
\newblock


\bibitem[\protect\citeauthoryear{Li, Ma, and Zhang}{Li et~al\mbox{.}}{2023d}]%
        {li2023lightweight}
\bibfield{author}{\bibinfo{person}{Meng Li}, \bibinfo{person}{Bo Ma}, {and}
  \bibinfo{person}{Yulin Zhang}.} \bibinfo{year}{2023}\natexlab{d}.
\newblock \showarticletitle{Lightweight Image Super-Resolution with Pyramid
  Clustering Transformer}.
\newblock \bibinfo{journal}{\emph{IEEE Transactions on Circuits and Systems for
  Video Technology}} (\bibinfo{year}{2023}).
\newblock


\bibitem[\protect\citeauthoryear{Li, Li, Gao, Deng, Zhou, Yang, and Qi}{Li
  et~al\mbox{.}}{2023c}]%
        {li2023cross}
\bibfield{author}{\bibinfo{person}{Wenjie Li}, \bibinfo{person}{Juncheng Li},
  \bibinfo{person}{Guangwei Gao}, \bibinfo{person}{Weihong Deng},
  \bibinfo{person}{Jiantao Zhou}, \bibinfo{person}{Jian Yang}, {and}
  \bibinfo{person}{Guo-Jun Qi}.} \bibinfo{year}{2023}\natexlab{c}.
\newblock \showarticletitle{Cross-receptive focused inference network for
  lightweight image super-resolution}.
\newblock \bibinfo{journal}{\emph{IEEE Transactions on Multimedia}}
  \bibinfo{volume}{26} (\bibinfo{year}{2023}), \bibinfo{pages}{864--877}.
\newblock


\bibitem[\protect\citeauthoryear{Li, Dong, Tang, and Pan}{Li
  et~al\mbox{.}}{2023a}]%
        {li2023dlgsanet}
\bibfield{author}{\bibinfo{person}{Xiang Li}, \bibinfo{person}{Jiangxin Dong},
  \bibinfo{person}{Jinhui Tang}, {and} \bibinfo{person}{Jinshan Pan}.}
  \bibinfo{year}{2023}\natexlab{a}.
\newblock \showarticletitle{DLGSANet: lightweight dynamic local and global
  self-attention networks for image super-resolution}. In
  \bibinfo{booktitle}{\emph{ICCV}}.
\newblock


\bibitem[\protect\citeauthoryear{Li, Fan, Xiang, Demandolx, Ranjan, Timofte,
  and Van~Gool}{Li et~al\mbox{.}}{2023b}]%
        {li2023efficient}
\bibfield{author}{\bibinfo{person}{Yawei Li}, \bibinfo{person}{Yuchen Fan},
  \bibinfo{person}{Xiaoyu Xiang}, \bibinfo{person}{Denis Demandolx},
  \bibinfo{person}{Rakesh Ranjan}, \bibinfo{person}{Radu Timofte}, {and}
  \bibinfo{person}{Luc Van~Gool}.} \bibinfo{year}{2023}\natexlab{b}.
\newblock \showarticletitle{Efficient and explicit modelling of image
  hierarchies for image restoration}. In \bibinfo{booktitle}{\emph{CVPR}}.
\newblock


\bibitem[\protect\citeauthoryear{Li, Zhang, Dingl, Wei, and Zhang}{Li
  et~al\mbox{.}}{2018b}]%
        {li2018single}
\bibfield{author}{\bibinfo{person}{Yong Li}, \bibinfo{person}{Lei Zhang},
  \bibinfo{person}{Chen Dingl}, \bibinfo{person}{Wei Wei}, {and}
  \bibinfo{person}{Yanning Zhang}.} \bibinfo{year}{2018}\natexlab{b}.
\newblock \showarticletitle{Single hyperspectral image super-resolution with
  grouped deep recursive residual network}. In
  \bibinfo{booktitle}{\emph{BigMM}}.
\newblock


\bibitem[\protect\citeauthoryear{Li, Liu, Chen, Cai, Gu, Qiao, and Dong}{Li
  et~al\mbox{.}}{2022a}]%
        {li2022blueprint}
\bibfield{author}{\bibinfo{person}{Zheyuan Li}, \bibinfo{person}{Yingqi Liu},
  \bibinfo{person}{Xiangyu Chen}, \bibinfo{person}{Haoming Cai},
  \bibinfo{person}{Jinjin Gu}, \bibinfo{person}{Yu Qiao}, {and}
  \bibinfo{person}{Chao Dong}.} \bibinfo{year}{2022}\natexlab{a}.
\newblock \showarticletitle{Blueprint separable residual network for efficient
  image super-resolution}. In \bibinfo{booktitle}{\emph{CVPR}}.
\newblock


\bibitem[\protect\citeauthoryear{Li, Yang, Liu, Yang, Jeon, and Wu}{Li
  et~al\mbox{.}}{2019a}]%
        {li2019feedback}
\bibfield{author}{\bibinfo{person}{Zhen Li}, \bibinfo{person}{Jinglei Yang},
  \bibinfo{person}{Zheng Liu}, \bibinfo{person}{Xiaomin Yang},
  \bibinfo{person}{Gwanggil Jeon}, {and} \bibinfo{person}{Wei Wu}.}
  \bibinfo{year}{2019}\natexlab{a}.
\newblock \showarticletitle{Feedback network for image super-resolution}. In
  \bibinfo{booktitle}{\emph{CVPR}}.
\newblock


\bibitem[\protect\citeauthoryear{Liang, Cao, Sun, Zhang, Van~Gool, and
  Timofte}{Liang et~al\mbox{.}}{2021}]%
        {liang2021swinir}
\bibfield{author}{\bibinfo{person}{Jingyun Liang}, \bibinfo{person}{Jiezhang
  Cao}, \bibinfo{person}{Guolei Sun}, \bibinfo{person}{Kai Zhang},
  \bibinfo{person}{Luc Van~Gool}, {and} \bibinfo{person}{Radu Timofte}.}
  \bibinfo{year}{2021}\natexlab{}.
\newblock \showarticletitle{SwinIR: Image restoration using swin transformer}.
  In \bibinfo{booktitle}{\emph{ICCVW}}.
\newblock


\bibitem[\protect\citeauthoryear{Lim, Son, Kim, Nah, and Lee}{Lim
  et~al\mbox{.}}{2017}]%
        {lim2017enhanced}
\bibfield{author}{\bibinfo{person}{Bee Lim}, \bibinfo{person}{Sanghyun Son},
  \bibinfo{person}{Heewon Kim}, \bibinfo{person}{Seungjun Nah}, {and}
  \bibinfo{person}{Kyoung~Mu Lee}.} \bibinfo{year}{2017}\natexlab{}.
\newblock \showarticletitle{Enhanced deep residual networks for single image
  super-resolution}. In \bibinfo{booktitle}{\emph{CVPRW}}.
\newblock


\bibitem[\protect\citeauthoryear{Lin, Luo, Hong, Qu, Xie, and Wu}{Lin
  et~al\mbox{.}}{2023b}]%
        {lin2023memory}
\bibfield{author}{\bibinfo{person}{Jin Lin}, \bibinfo{person}{Xiaotong Luo},
  \bibinfo{person}{Ming Hong}, \bibinfo{person}{Yanyun Qu},
  \bibinfo{person}{Yuan Xie}, {and} \bibinfo{person}{Zongze Wu}.}
  \bibinfo{year}{2023}\natexlab{b}.
\newblock \showarticletitle{Memory-Friendly Scalable Super-Resolution via
  Rewinding Lottery Ticket Hypothesis}. In \bibinfo{booktitle}{\emph{CVPR}}.
\newblock


\bibitem[\protect\citeauthoryear{Lin, He, Chen, Lyu, Fei, Dai, Ouyang, Qiao,
  and Dong}{Lin et~al\mbox{.}}{2023a}]%
        {lin2023diffbir}
\bibfield{author}{\bibinfo{person}{Xinqi Lin}, \bibinfo{person}{Jingwen He},
  \bibinfo{person}{Ziyan Chen}, \bibinfo{person}{Zhaoyang Lyu},
  \bibinfo{person}{Ben Fei}, \bibinfo{person}{Bo Dai}, \bibinfo{person}{Wanli
  Ouyang}, \bibinfo{person}{Yu Qiao}, {and} \bibinfo{person}{Chao Dong}.}
  \bibinfo{year}{2023}\natexlab{a}.
\newblock \showarticletitle{Diffbir: Towards blind image restoration with
  generative diffusion prior}.
\newblock \bibinfo{journal}{\emph{arXiv preprint arXiv:2308.15070}}
  (\bibinfo{year}{2023}).
\newblock


\bibitem[\protect\citeauthoryear{Liu, Liu, Gu, Qiao, and Dong}{Liu
  et~al\mbox{.}}{2022b}]%
        {liu2022blind}
\bibfield{author}{\bibinfo{person}{Anran Liu}, \bibinfo{person}{Yihao Liu},
  \bibinfo{person}{Jinjin Gu}, \bibinfo{person}{Yu Qiao}, {and}
  \bibinfo{person}{Chao Dong}.} \bibinfo{year}{2022}\natexlab{b}.
\newblock \showarticletitle{Blind image super-resolution: A survey and beyond}.
\newblock \bibinfo{journal}{\emph{IEEE transactions on pattern analysis and
  machine intelligence}} \bibinfo{volume}{45}, \bibinfo{number}{5}
  (\bibinfo{year}{2022}), \bibinfo{pages}{5461--5480}.
\newblock


\bibitem[\protect\citeauthoryear{Liu, Li, and Yuan}{Liu et~al\mbox{.}}{2021a}]%
        {liu2021spectral}
\bibfield{author}{\bibinfo{person}{Denghong Liu}, \bibinfo{person}{Jie Li},
  {and} \bibinfo{person}{Qiangqiang Yuan}.} \bibinfo{year}{2021}\natexlab{a}.
\newblock \showarticletitle{A Spectral Grouping and Attention-Driven Residual
  Dense Network for Hyperspectral Image Super-Resolution}.
\newblock \bibinfo{journal}{\emph{IEEE Transactions on Geoscience and Remote
  Sensing}} \bibinfo{volume}{59}, \bibinfo{number}{9} (\bibinfo{year}{2021}),
  \bibinfo{pages}{7711--7725}.
\newblock


\bibitem[\protect\citeauthoryear{Liu, Wen, Fan, Loy, and Huang}{Liu
  et~al\mbox{.}}{2018}]%
        {liu2018non}
\bibfield{author}{\bibinfo{person}{Ding Liu}, \bibinfo{person}{Bihan Wen},
  \bibinfo{person}{Yuchen Fan}, \bibinfo{person}{Chen~Change Loy}, {and}
  \bibinfo{person}{Thomas~S Huang}.} \bibinfo{year}{2018}\natexlab{}.
\newblock \showarticletitle{Non-local recurrent network for image restoration}.
\newblock \bibinfo{journal}{\emph{NeurIPS}} (\bibinfo{year}{2018}).
\newblock


\bibitem[\protect\citeauthoryear{Liu, Tang, and Wu}{Liu et~al\mbox{.}}{2020a}]%
        {liu2020residual}
\bibfield{author}{\bibinfo{person}{Jie Liu}, \bibinfo{person}{Jie Tang}, {and}
  \bibinfo{person}{Gangshan Wu}.} \bibinfo{year}{2020}\natexlab{a}.
\newblock \showarticletitle{Residual feature distillation network for
  lightweight image super-resolution}. In \bibinfo{booktitle}{\emph{ECCVW}}.
\newblock


\bibitem[\protect\citeauthoryear{Liu, Zhang, Tang, Tang, and Wu}{Liu
  et~al\mbox{.}}{2020b}]%
        {liu2020residuala}
\bibfield{author}{\bibinfo{person}{Jie Liu}, \bibinfo{person}{Wenjie Zhang},
  \bibinfo{person}{Yuting Tang}, \bibinfo{person}{Jie Tang}, {and}
  \bibinfo{person}{Gangshan Wu}.} \bibinfo{year}{2020}\natexlab{b}.
\newblock \showarticletitle{Residual feature aggregation network for image
  super-resolution}. In \bibinfo{booktitle}{\emph{CVPR}}.
\newblock


\bibitem[\protect\citeauthoryear{Liu, Feng, Wang, Han, and Zeng}{Liu
  et~al\mbox{.}}{2022a}]%
        {liu2022dual}
\bibfield{author}{\bibinfo{person}{Ziyu Liu}, \bibinfo{person}{Ruyi Feng},
  \bibinfo{person}{Lizhe Wang}, \bibinfo{person}{Wei Han}, {and}
  \bibinfo{person}{Tieyong Zeng}.} \bibinfo{year}{2022}\natexlab{a}.
\newblock \showarticletitle{Dual learning-based graph neural network for remote
  sensing image super-resolution}.
\newblock \bibinfo{journal}{\emph{IEEE Transactions on Geoscience and Remote
  Sensing}}  \bibinfo{volume}{60} (\bibinfo{year}{2022}),
  \bibinfo{pages}{1--14}.
\newblock


\bibitem[\protect\citeauthoryear{Liu, Lin, Cao, Hu, Wei, Zhang, Lin, and
  Guo}{Liu et~al\mbox{.}}{2021b}]%
        {liu2021swin}
\bibfield{author}{\bibinfo{person}{Ze Liu}, \bibinfo{person}{Yutong Lin},
  \bibinfo{person}{Yue Cao}, \bibinfo{person}{Han Hu}, \bibinfo{person}{Yixuan
  Wei}, \bibinfo{person}{Zheng Zhang}, \bibinfo{person}{Stephen Lin}, {and}
  \bibinfo{person}{Baining Guo}.} \bibinfo{year}{2021}\natexlab{b}.
\newblock \showarticletitle{Swin transformer: Hierarchical vision transformer
  using shifted windows}.
\newblock \bibinfo{journal}{\emph{CVPR}} (\bibinfo{year}{2021}).
\newblock


\bibitem[\protect\citeauthoryear{Liu, Luo, Wang, and Tang}{Liu
  et~al\mbox{.}}{2015}]%
        {liu2015deep}
\bibfield{author}{\bibinfo{person}{Ziwei Liu}, \bibinfo{person}{Ping Luo},
  \bibinfo{person}{Xiaogang Wang}, {and} \bibinfo{person}{Xiaoou Tang}.}
  \bibinfo{year}{2015}\natexlab{}.
\newblock \showarticletitle{Deep learning face attributes in the wild}. In
  \bibinfo{booktitle}{\emph{ICCV}}.
\newblock


\bibitem[\protect\citeauthoryear{Lu, Liu, Li, and Zhang}{Lu
  et~al\mbox{.}}{2021}]%
        {lu2021efficient}
\bibfield{author}{\bibinfo{person}{Zhisheng Lu}, \bibinfo{person}{Hong Liu},
  \bibinfo{person}{Juncheng Li}, {and} \bibinfo{person}{Linlin Zhang}.}
  \bibinfo{year}{2021}\natexlab{}.
\newblock \showarticletitle{Transformer for Single Image Super-Resolution}.
\newblock \bibinfo{journal}{\emph{CVPRW}} (\bibinfo{year}{2021}).
\newblock


\bibitem[\protect\citeauthoryear{Luo, Ai, Liang, Liu, Xie, Qu, and Fu}{Luo
  et~al\mbox{.}}{2024}]%
        {luo2024adaformer}
\bibfield{author}{\bibinfo{person}{Xiaotong Luo}, \bibinfo{person}{Zekun Ai},
  \bibinfo{person}{Qiuyuan Liang}, \bibinfo{person}{Ding Liu},
  \bibinfo{person}{Yuan Xie}, \bibinfo{person}{Yanyun Qu}, {and}
  \bibinfo{person}{Yun Fu}.} \bibinfo{year}{2024}\natexlab{}.
\newblock \showarticletitle{AdaFormer: Efficient Transformer with Adaptive
  Token Sparsification for Image Super-resolution}. In
  \bibinfo{booktitle}{\emph{AAAI}}.
\newblock


\bibitem[\protect\citeauthoryear{Luo, Dai, Zhang, Xie, Liu, Qu, Fu, and
  Zhang}{Luo et~al\mbox{.}}{2022a}]%
        {luo2022adjustable}
\bibfield{author}{\bibinfo{person}{Xiaotong Luo}, \bibinfo{person}{Mingliang
  Dai}, \bibinfo{person}{Yulun Zhang}, \bibinfo{person}{Yuan Xie},
  \bibinfo{person}{Ding Liu}, \bibinfo{person}{Yanyun Qu}, \bibinfo{person}{Yun
  Fu}, {and} \bibinfo{person}{Junping Zhang}.}
  \bibinfo{year}{2022}\natexlab{a}.
\newblock \showarticletitle{Adjustable Memory-efficient Image Super-resolution
  via Individual Kernel Sparsity}. In \bibinfo{booktitle}{\emph{ACMMM}}.
\newblock


\bibitem[\protect\citeauthoryear{Luo, Liang, Liu, and Qu}{Luo
  et~al\mbox{.}}{2021}]%
        {luo2021boosting}
\bibfield{author}{\bibinfo{person}{Xiaotong Luo}, \bibinfo{person}{Qiuyuan
  Liang}, \bibinfo{person}{Ding Liu}, {and} \bibinfo{person}{Yanyun Qu}.}
  \bibinfo{year}{2021}\natexlab{}.
\newblock \showarticletitle{Boosting lightweight single image super-resolution
  via joint-distillation}. In \bibinfo{booktitle}{\emph{ACMMM}}.
\newblock


\bibitem[\protect\citeauthoryear{Luo, Qu, Xie, Zhang, Li, and Fu}{Luo
  et~al\mbox{.}}{2022b}]%
        {luo2022lattice}
\bibfield{author}{\bibinfo{person}{Xiaotong Luo}, \bibinfo{person}{Yanyun Qu},
  \bibinfo{person}{Yuan Xie}, \bibinfo{person}{Yulun Zhang},
  \bibinfo{person}{Cuihua Li}, {and} \bibinfo{person}{Yun Fu}.}
  \bibinfo{year}{2022}\natexlab{b}.
\newblock \showarticletitle{Lattice network for lightweight image restoration}.
\newblock \bibinfo{journal}{\emph{IEEE Transactions on Pattern Analysis and
  Machine Intelligence}} \bibinfo{volume}{45}, \bibinfo{number}{4}
  (\bibinfo{year}{2022}), \bibinfo{pages}{4826--4842}.
\newblock


\bibitem[\protect\citeauthoryear{Luo, Huang, Li, Wang, and Tan}{Luo
  et~al\mbox{.}}{2020}]%
        {luo2020unfolding}
\bibfield{author}{\bibinfo{person}{Zhengxiong Luo}, \bibinfo{person}{Yan
  Huang}, \bibinfo{person}{Shang Li}, \bibinfo{person}{Liang Wang}, {and}
  \bibinfo{person}{Tieniu Tan}.} \bibinfo{year}{2020}\natexlab{}.
\newblock \showarticletitle{Unfolding the alternating optimization for blind
  super resolution}.
\newblock \bibinfo{journal}{\emph{NeurIPS}} (\bibinfo{year}{2020}).
\newblock


\bibitem[\protect\citeauthoryear{Ma, Rao, Cheng, Chen, Lu, and Zhou}{Ma
  et~al\mbox{.}}{2020}]%
        {ma2020structure}
\bibfield{author}{\bibinfo{person}{Cheng Ma}, \bibinfo{person}{Yongming Rao},
  \bibinfo{person}{Yean Cheng}, \bibinfo{person}{Ce Chen},
  \bibinfo{person}{Jiwen Lu}, {and} \bibinfo{person}{Jie Zhou}.}
  \bibinfo{year}{2020}\natexlab{}.
\newblock \showarticletitle{Structure-preserving super resolution with gradient
  guidance}. In \bibinfo{booktitle}{\emph{CVPR}}.
\newblock


\bibitem[\protect\citeauthoryear{Ma, Yang, Yang, and Yang}{Ma
  et~al\mbox{.}}{2017}]%
        {ma2017learning}
\bibfield{author}{\bibinfo{person}{Chao Ma}, \bibinfo{person}{Chih-Yuan Yang},
  \bibinfo{person}{Xiaokang Yang}, {and} \bibinfo{person}{Ming-Hsuan Yang}.}
  \bibinfo{year}{2017}\natexlab{}.
\newblock \showarticletitle{Learning a no-reference quality metric for
  single-image super-resolution}.
\newblock \bibinfo{journal}{\emph{Computer Vision and Image Understanding}}
  \bibinfo{volume}{158} (\bibinfo{year}{2017}), \bibinfo{pages}{1--16}.
\newblock


\bibitem[\protect\citeauthoryear{Ma, Duanmu, Wu, Wang, Yong, Li, and Zhang}{Ma
  et~al\mbox{.}}{2016}]%
        {ma2016waterloo}
\bibfield{author}{\bibinfo{person}{Kede Ma}, \bibinfo{person}{Zhengfang
  Duanmu}, \bibinfo{person}{Qingbo Wu}, \bibinfo{person}{Zhou Wang},
  \bibinfo{person}{Hongwei Yong}, \bibinfo{person}{Hongliang Li}, {and}
  \bibinfo{person}{Lei Zhang}.} \bibinfo{year}{2016}\natexlab{}.
\newblock \showarticletitle{Waterloo exploration database: New challenges for
  image quality assessment models}.
\newblock \bibinfo{journal}{\emph{IEEE Transactions on Image Processing}}
  \bibinfo{volume}{26}, \bibinfo{number}{2} (\bibinfo{year}{2016}),
  \bibinfo{pages}{1004--1016}.
\newblock


\bibitem[\protect\citeauthoryear{Ma, Xiong, Hu, and Ma}{Ma
  et~al\mbox{.}}{2019}]%
        {ma2019efficient}
\bibfield{author}{\bibinfo{person}{Yinglan Ma}, \bibinfo{person}{Hongyu Xiong},
  \bibinfo{person}{Zhe Hu}, {and} \bibinfo{person}{Lizhuang Ma}.}
  \bibinfo{year}{2019}\natexlab{}.
\newblock \showarticletitle{Efficient super resolution using binarized neural
  network}. In \bibinfo{booktitle}{\emph{CVPRW}}.
\newblock


\bibitem[\protect\citeauthoryear{Maeda}{Maeda}{2020}]%
        {maeda2020unpaired}
\bibfield{author}{\bibinfo{person}{Shunta Maeda}.}
  \bibinfo{year}{2020}\natexlab{}.
\newblock \showarticletitle{Unpaired image super-resolution using
  pseudo-supervision}. In \bibinfo{booktitle}{\emph{CVPR}}.
\newblock


\bibitem[\protect\citeauthoryear{Martin, Fowlkes, Tal, and Malik}{Martin
  et~al\mbox{.}}{2001}]%
        {martin2001database}
\bibfield{author}{\bibinfo{person}{David Martin}, \bibinfo{person}{Charless
  Fowlkes}, \bibinfo{person}{Doron Tal}, {and} \bibinfo{person}{Jitendra
  Malik}.} \bibinfo{year}{2001}\natexlab{}.
\newblock \showarticletitle{A database of human segmented natural images and
  its application to evaluating segmentation algorithms and measuring
  ecological statistics}. In \bibinfo{booktitle}{\emph{ICCV}}.
\newblock


\bibitem[\protect\citeauthoryear{Mei, Jiang, Li, Liu, Ye, and Wang}{Mei
  et~al\mbox{.}}{2019}]%
        {mei2019deep}
\bibfield{author}{\bibinfo{person}{Kangfu Mei}, \bibinfo{person}{Aiwen Jiang},
  \bibinfo{person}{Juncheng Li}, \bibinfo{person}{Bo Liu},
  \bibinfo{person}{Jihua Ye}, {and} \bibinfo{person}{Mingwen Wang}.}
  \bibinfo{year}{2019}\natexlab{}.
\newblock \showarticletitle{Deep residual refining based pseudo-multi-frame
  network for effective single image super-resolution}.
\newblock \bibinfo{journal}{\emph{IET Image Processing}} \bibinfo{volume}{13},
  \bibinfo{number}{4} (\bibinfo{year}{2019}), \bibinfo{pages}{591--599}.
\newblock


\bibitem[\protect\citeauthoryear{Mei, Jiang, Li, Ye, and Wang}{Mei
  et~al\mbox{.}}{2018}]%
        {mei2018effective}
\bibfield{author}{\bibinfo{person}{Kangfu Mei}, \bibinfo{person}{Aiwen Jiang},
  \bibinfo{person}{Juncheng Li}, \bibinfo{person}{Jihua Ye}, {and}
  \bibinfo{person}{Mingwen Wang}.} \bibinfo{year}{2018}\natexlab{}.
\newblock \showarticletitle{An Effective Single-Image Super-Resolution Model
  Using Squeeze-and-Excitation Networks}. In
  \bibinfo{booktitle}{\emph{NeurIPS}}.
\newblock


\bibitem[\protect\citeauthoryear{Mei, Yuan, Ji, Zhang, Wan, and Du}{Mei
  et~al\mbox{.}}{2017}]%
        {mei2017hyperspectral}
\bibfield{author}{\bibinfo{person}{Shaohui Mei}, \bibinfo{person}{Xin Yuan},
  \bibinfo{person}{Jingyu Ji}, \bibinfo{person}{Yifan Zhang},
  \bibinfo{person}{Shuai Wan}, {and} \bibinfo{person}{Qian Du}.}
  \bibinfo{year}{2017}\natexlab{}.
\newblock \showarticletitle{Hyperspectral image spatial super-resolution via 3D
  full convolutional neural network}.
\newblock \bibinfo{journal}{\emph{Remote Sensing}} \bibinfo{volume}{9},
  \bibinfo{number}{11} (\bibinfo{year}{2017}), \bibinfo{pages}{1139}.
\newblock


\bibitem[\protect\citeauthoryear{Mei, Fan, Zhou, Huang, Huang, and Shi}{Mei
  et~al\mbox{.}}{2020}]%
        {mei2020image}
\bibfield{author}{\bibinfo{person}{Yiqun Mei}, \bibinfo{person}{Yuchen Fan},
  \bibinfo{person}{Yuqian Zhou}, \bibinfo{person}{Lichao Huang},
  \bibinfo{person}{Thomas~S Huang}, {and} \bibinfo{person}{Honghui Shi}.}
  \bibinfo{year}{2020}\natexlab{}.
\newblock \showarticletitle{Image super-resolution with cross-scale non-local
  attention and exhaustive self-exemplars mining}. In
  \bibinfo{booktitle}{\emph{CVPR}}.
\newblock


\bibitem[\protect\citeauthoryear{Mittal, Moorthy, and Bovik}{Mittal
  et~al\mbox{.}}{2012a}]%
        {mittal2012no}
\bibfield{author}{\bibinfo{person}{Anish Mittal},
  \bibinfo{person}{Anush~Krishna Moorthy}, {and} \bibinfo{person}{Alan~Conrad
  Bovik}.} \bibinfo{year}{2012}\natexlab{a}.
\newblock \showarticletitle{No-reference image quality assessment in the
  spatial domain}.
\newblock \bibinfo{journal}{\emph{IEEE Transactions on Image Processing}}
  \bibinfo{volume}{21}, \bibinfo{number}{12} (\bibinfo{year}{2012}),
  \bibinfo{pages}{4695--4708}.
\newblock


\bibitem[\protect\citeauthoryear{Mittal, Soundararajan, and Bovik}{Mittal
  et~al\mbox{.}}{2012b}]%
        {mittal2012making}
\bibfield{author}{\bibinfo{person}{Anish Mittal}, \bibinfo{person}{Rajiv
  Soundararajan}, {and} \bibinfo{person}{Alan~C Bovik}.}
  \bibinfo{year}{2012}\natexlab{b}.
\newblock \showarticletitle{Making a ?completely blind? image quality
  analyzer}.
\newblock \bibinfo{journal}{\emph{IEEE Signal Processing Letters}}
  \bibinfo{volume}{20}, \bibinfo{number}{3} (\bibinfo{year}{2012}),
  \bibinfo{pages}{209--212}.
\newblock


\bibitem[\protect\citeauthoryear{Muqeet, Hwang, Yang, Kang, Kim, and
  Bae}{Muqeet et~al\mbox{.}}{2020}]%
        {muqeet2020multi}
\bibfield{author}{\bibinfo{person}{Abdul Muqeet}, \bibinfo{person}{Jiwon
  Hwang}, \bibinfo{person}{Subin Yang}, \bibinfo{person}{JungHeum Kang},
  \bibinfo{person}{Yongwoo Kim}, {and} \bibinfo{person}{Sung-Ho Bae}.}
  \bibinfo{year}{2020}\natexlab{}.
\newblock \showarticletitle{Multi-attention based ultra lightweight image
  super-resolution}. In \bibinfo{booktitle}{\emph{ECCV}}.
\newblock


\bibitem[\protect\citeauthoryear{Niu, Wen, Ren, Zhang, Yang, Wang, Zhang, Cao,
  and Shen}{Niu et~al\mbox{.}}{2020}]%
        {niu2020single}
\bibfield{author}{\bibinfo{person}{Ben Niu}, \bibinfo{person}{Weilei Wen},
  \bibinfo{person}{Wenqi Ren}, \bibinfo{person}{Xiangde Zhang},
  \bibinfo{person}{Lianping Yang}, \bibinfo{person}{Shuzhen Wang},
  \bibinfo{person}{Kaihao Zhang}, \bibinfo{person}{Xiaochun Cao}, {and}
  \bibinfo{person}{Haifeng Shen}.} \bibinfo{year}{2020}\natexlab{}.
\newblock \showarticletitle{Single image super-resolution via a holistic
  attention network}. In \bibinfo{booktitle}{\emph{ECCV}}.
\newblock


\bibitem[\protect\citeauthoryear{Park, Son, Cho, Hong, and Lee}{Park
  et~al\mbox{.}}{2018}]%
        {park2018srfeat}
\bibfield{author}{\bibinfo{person}{Seong-Jin Park}, \bibinfo{person}{Hyeongseok
  Son}, \bibinfo{person}{Sunghyun Cho}, \bibinfo{person}{Ki-Sang Hong}, {and}
  \bibinfo{person}{Seungyong Lee}.} \bibinfo{year}{2018}\natexlab{}.
\newblock \showarticletitle{Srfeat: Single image super-resolution with feature
  discrimination}. In \bibinfo{booktitle}{\emph{ECCV}}.
\newblock


\bibitem[\protect\citeauthoryear{Peng, Lin, Liao, Chellappa, and Zhou}{Peng
  et~al\mbox{.}}{2020}]%
        {peng2020saint}
\bibfield{author}{\bibinfo{person}{Cheng Peng}, \bibinfo{person}{Wei-An Lin},
  \bibinfo{person}{Haofu Liao}, \bibinfo{person}{Rama Chellappa}, {and}
  \bibinfo{person}{S~Kevin Zhou}.} \bibinfo{year}{2020}\natexlab{}.
\newblock \showarticletitle{SAINT: spatially aware interpolation network for
  medical slice synthesis}. In \bibinfo{booktitle}{\emph{CVPR}}.
\newblock


\bibitem[\protect\citeauthoryear{Qin, Huang, and Wen}{Qin
  et~al\mbox{.}}{2020}]%
        {qin2020multi}
\bibfield{author}{\bibinfo{person}{Jinghui Qin}, \bibinfo{person}{Yongjie
  Huang}, {and} \bibinfo{person}{Wushao Wen}.} \bibinfo{year}{2020}\natexlab{}.
\newblock \showarticletitle{Multi-scale feature fusion residual network for
  Single Image Super-Resolution}.
\newblock \bibinfo{journal}{\emph{Neurocomputing}}  \bibinfo{volume}{379}
  (\bibinfo{year}{2020}), \bibinfo{pages}{334--342}.
\newblock


\bibitem[\protect\citeauthoryear{Rad, Bozorgtabar, Marti, Basler, Ekenel, and
  Thiran}{Rad et~al\mbox{.}}{2019}]%
        {rad2019srobb}
\bibfield{author}{\bibinfo{person}{Mohammad~Saeed Rad}, \bibinfo{person}{Behzad
  Bozorgtabar}, \bibinfo{person}{Urs-Viktor Marti}, \bibinfo{person}{Max
  Basler}, \bibinfo{person}{Hazim~Kemal Ekenel}, {and}
  \bibinfo{person}{Jean-Philippe Thiran}.} \bibinfo{year}{2019}\natexlab{}.
\newblock \showarticletitle{Srobb: Targeted perceptual loss for single image
  super-resolution}. In \bibinfo{booktitle}{\emph{ICCV}}.
\newblock


\bibitem[\protect\citeauthoryear{Radford, Metz, and Chintala}{Radford
  et~al\mbox{.}}{2015}]%
        {radford2015unsupervised}
\bibfield{author}{\bibinfo{person}{Alec Radford}, \bibinfo{person}{Luke Metz},
  {and} \bibinfo{person}{Soumith Chintala}.} \bibinfo{year}{2015}\natexlab{}.
\newblock \showarticletitle{Unsupervised representation learning with deep
  convolutional generative adversarial networks}.
\newblock \bibinfo{journal}{\emph{arXiv preprint arXiv:1511.06434}}
  (\bibinfo{year}{2015}).
\newblock


\bibitem[\protect\citeauthoryear{Radford, Wu, Child, Luan, Amodei, Sutskever,
  et~al\mbox{.}}{Radford et~al\mbox{.}}{2019}]%
        {radford2019language}
\bibfield{author}{\bibinfo{person}{Alec Radford}, \bibinfo{person}{Jeffrey Wu},
  \bibinfo{person}{Rewon Child}, \bibinfo{person}{David Luan},
  \bibinfo{person}{Dario Amodei}, \bibinfo{person}{Ilya Sutskever},
  {et~al\mbox{.}}} \bibinfo{year}{2019}\natexlab{}.
\newblock \showarticletitle{Language models are unsupervised multitask
  learners}.
\newblock \bibinfo{journal}{\emph{OpenAI Blog}} (\bibinfo{year}{2019}).
\newblock


\bibitem[\protect\citeauthoryear{Rickard, Basedow, Zalewski, Silverglate, and
  Landers}{Rickard et~al\mbox{.}}{1993}]%
        {rickard1993hydice}
\bibfield{author}{\bibinfo{person}{Lee~J Rickard}, \bibinfo{person}{Robert~W
  Basedow}, \bibinfo{person}{Edward~F Zalewski}, \bibinfo{person}{Peter~R
  Silverglate}, {and} \bibinfo{person}{Mark Landers}.}
  \bibinfo{year}{1993}\natexlab{}.
\newblock \showarticletitle{HYDICE: An airborne system for hyperspectral
  imaging}. In \bibinfo{booktitle}{\emph{Imaging Spectrometry of the
  Terrestrial Environment}}, Vol.~\bibinfo{volume}{1937}.
  \bibinfo{pages}{173--179}.
\newblock


\bibitem[\protect\citeauthoryear{Rombach, Blattmann, Lorenz, Esser, and
  Ommer}{Rombach et~al\mbox{.}}{2022}]%
        {rombach2022high}
\bibfield{author}{\bibinfo{person}{Robin Rombach}, \bibinfo{person}{Andreas
  Blattmann}, \bibinfo{person}{Dominik Lorenz}, \bibinfo{person}{Patrick
  Esser}, {and} \bibinfo{person}{Bj{\"o}rn Ommer}.}
  \bibinfo{year}{2022}\natexlab{}.
\newblock \showarticletitle{High-resolution image synthesis with latent
  diffusion models}. In \bibinfo{booktitle}{\emph{CVPR}}.
\newblock


\bibitem[\protect\citeauthoryear{Saharia, Ho, Chan, Salimans, Fleet, and
  Norouzi}{Saharia et~al\mbox{.}}{2022}]%
        {saharia2022image}
\bibfield{author}{\bibinfo{person}{Chitwan Saharia}, \bibinfo{person}{Jonathan
  Ho}, \bibinfo{person}{William Chan}, \bibinfo{person}{Tim Salimans},
  \bibinfo{person}{David~J Fleet}, {and} \bibinfo{person}{Mohammad Norouzi}.}
  \bibinfo{year}{2022}\natexlab{}.
\newblock \showarticletitle{Image super-resolution via iterative refinement}.
\newblock \bibinfo{journal}{\emph{IEEE Transactions on Pattern Analysis and
  Machine Intelligence}} \bibinfo{volume}{45}, \bibinfo{number}{4}
  (\bibinfo{year}{2022}), \bibinfo{pages}{4713--4726}.
\newblock


\bibitem[\protect\citeauthoryear{Shaham, Dekel, and Michaeli}{Shaham
  et~al\mbox{.}}{2019}]%
        {shaham2019singan}
\bibfield{author}{\bibinfo{person}{Tamar~Rott Shaham}, \bibinfo{person}{Tali
  Dekel}, {and} \bibinfo{person}{Tomer Michaeli}.}
  \bibinfo{year}{2019}\natexlab{}.
\newblock \showarticletitle{Singan: Learning a generative model from a single
  natural image}. In \bibinfo{booktitle}{\emph{ICCV}}.
\newblock


\bibitem[\protect\citeauthoryear{Shang, Shan, Liu, and Zhang}{Shang
  et~al\mbox{.}}{2023}]%
        {shang2023resdiff}
\bibfield{author}{\bibinfo{person}{Shuyao Shang}, \bibinfo{person}{Zhengyang
  Shan}, \bibinfo{person}{Guangxing Liu}, {and} \bibinfo{person}{Jinglin
  Zhang}.} \bibinfo{year}{2023}\natexlab{}.
\newblock \showarticletitle{ResDiff: Combining CNN and Diffusion Model for
  Image Super-Resolution}.
\newblock  (\bibinfo{year}{2023}).
\newblock


\bibitem[\protect\citeauthoryear{Shen, Yu, Wang, Yang, Xue, and Hu}{Shen
  et~al\mbox{.}}{2019}]%
        {shen2019multipath}
\bibfield{author}{\bibinfo{person}{Mingyu Shen}, \bibinfo{person}{Pengfei Yu},
  \bibinfo{person}{Ronggui Wang}, \bibinfo{person}{Juan Yang},
  \bibinfo{person}{Lixia Xue}, {and} \bibinfo{person}{Min Hu}.}
  \bibinfo{year}{2019}\natexlab{}.
\newblock \showarticletitle{Multipath feedforward network for single image
  super-resolution}.
\newblock \bibinfo{journal}{\emph{Multimedia Tools and Applications}}
  \bibinfo{volume}{78} (\bibinfo{year}{2019}), \bibinfo{pages}{19621--19640}.
\newblock


\bibitem[\protect\citeauthoryear{Shi, Caballero, Husz{\'a}r, Totz, Aitken,
  Bishop, Rueckert, and Wang}{Shi et~al\mbox{.}}{2016}]%
        {shi2016real}
\bibfield{author}{\bibinfo{person}{Wenzhe Shi}, \bibinfo{person}{Jose
  Caballero}, \bibinfo{person}{Ferenc Husz{\'a}r}, \bibinfo{person}{Johannes
  Totz}, \bibinfo{person}{Andrew~P Aitken}, \bibinfo{person}{Rob Bishop},
  \bibinfo{person}{Daniel Rueckert}, {and} \bibinfo{person}{Zehan Wang}.}
  \bibinfo{year}{2016}\natexlab{}.
\newblock \showarticletitle{Real-time single image and video super-resolution
  using an efficient sub-pixel convolutional neural network}. In
  \bibinfo{booktitle}{\emph{CVPR}}.
\newblock


\bibitem[\protect\citeauthoryear{Shocher, Cohen, and Irani}{Shocher
  et~al\mbox{.}}{2018}]%
        {shocher2018zero}
\bibfield{author}{\bibinfo{person}{Assaf Shocher}, \bibinfo{person}{Nadav
  Cohen}, {and} \bibinfo{person}{Michal Irani}.}
  \bibinfo{year}{2018}\natexlab{}.
\newblock \showarticletitle{?zero-shot? super-resolution using deep internal
  learning}. In \bibinfo{booktitle}{\emph{CVPR}}.
\newblock


\bibitem[\protect\citeauthoryear{Simonyan and Zisserman}{Simonyan and
  Zisserman}{2014}]%
        {simonyan2014very}
\bibfield{author}{\bibinfo{person}{Karen Simonyan} {and}
  \bibinfo{person}{Andrew Zisserman}.} \bibinfo{year}{2014}\natexlab{}.
\newblock \showarticletitle{Very deep convolutional networks for large-scale
  image recognition}.
\newblock \bibinfo{journal}{\emph{arXiv preprint arXiv:1409.1556}}
  (\bibinfo{year}{2014}).
\newblock


\bibitem[\protect\citeauthoryear{Sun, Zhang, Jiang, and Fu}{Sun
  et~al\mbox{.}}{2023}]%
        {sun2022hybrid}
\bibfield{author}{\bibinfo{person}{Bin Sun}, \bibinfo{person}{Yulun Zhang},
  \bibinfo{person}{Songyao Jiang}, {and} \bibinfo{person}{Yun Fu}.}
  \bibinfo{year}{2023}\natexlab{}.
\newblock \showarticletitle{Hybrid pixel-unshuffled network for lightweight
  image super-resolution}.
\newblock  (\bibinfo{year}{2023}).
\newblock


\bibitem[\protect\citeauthoryear{Sun, Pan, and Tang}{Sun et~al\mbox{.}}{2022}]%
        {sun2022shufflemixer}
\bibfield{author}{\bibinfo{person}{Long Sun}, \bibinfo{person}{Jinshan Pan},
  {and} \bibinfo{person}{Jinhui Tang}.} \bibinfo{year}{2022}\natexlab{}.
\newblock \showarticletitle{Shufflemixer: An efficient convnet for image
  super-resolution}.
\newblock  (\bibinfo{year}{2022}).
\newblock


\bibitem[\protect\citeauthoryear{Szegedy, Ioffe, Vanhoucke, and Alemi}{Szegedy
  et~al\mbox{.}}{2017}]%
        {szegedy2017inception}
\bibfield{author}{\bibinfo{person}{Christian Szegedy}, \bibinfo{person}{Sergey
  Ioffe}, \bibinfo{person}{Vincent Vanhoucke}, {and}
  \bibinfo{person}{Alexander~A Alemi}.} \bibinfo{year}{2017}\natexlab{}.
\newblock \showarticletitle{Inception-v4, inception-resnet and the impact of
  residual connections on learning}. In \bibinfo{booktitle}{\emph{AAAI}}.
\newblock


\bibitem[\protect\citeauthoryear{Szegedy, Vanhoucke, Ioffe, Shlens, and
  Wojna}{Szegedy et~al\mbox{.}}{2016}]%
        {szegedy2016rethinking}
\bibfield{author}{\bibinfo{person}{Christian Szegedy}, \bibinfo{person}{Vincent
  Vanhoucke}, \bibinfo{person}{Sergey Ioffe}, \bibinfo{person}{Jon Shlens},
  {and} \bibinfo{person}{Zbigniew Wojna}.} \bibinfo{year}{2016}\natexlab{}.
\newblock \showarticletitle{Rethinking the inception architecture for computer
  vision}. In \bibinfo{booktitle}{\emph{CVPR}}.
\newblock


\bibitem[\protect\citeauthoryear{Tai, Yang, and Liu}{Tai
  et~al\mbox{.}}{2017a}]%
        {tai2017image}
\bibfield{author}{\bibinfo{person}{Ying Tai}, \bibinfo{person}{Jian Yang},
  {and} \bibinfo{person}{Xiaoming Liu}.} \bibinfo{year}{2017}\natexlab{a}.
\newblock \showarticletitle{Image super-resolution via deep recursive residual
  network}. In \bibinfo{booktitle}{\emph{CVPR}}.
\newblock


\bibitem[\protect\citeauthoryear{Tai, Yang, Liu, and Xu}{Tai
  et~al\mbox{.}}{2017b}]%
        {tai2017memnet}
\bibfield{author}{\bibinfo{person}{Ying Tai}, \bibinfo{person}{Jian Yang},
  \bibinfo{person}{Xiaoming Liu}, {and} \bibinfo{person}{Chunyan Xu}.}
  \bibinfo{year}{2017}\natexlab{b}.
\newblock \showarticletitle{Memnet: A persistent memory network for image
  restoration}. In \bibinfo{booktitle}{\emph{CVPR}}.
\newblock


\bibitem[\protect\citeauthoryear{Timofte, Agustsson, Van~Gool, Yang, and
  Zhang}{Timofte et~al\mbox{.}}{2017}]%
        {timofte2017ntire}
\bibfield{author}{\bibinfo{person}{Radu Timofte}, \bibinfo{person}{Eirikur
  Agustsson}, \bibinfo{person}{Luc Van~Gool}, \bibinfo{person}{Ming-Hsuan
  Yang}, {and} \bibinfo{person}{Lei Zhang}.} \bibinfo{year}{2017}\natexlab{}.
\newblock \showarticletitle{Ntire 2017 challenge on single image
  super-resolution: Methods and results}. In \bibinfo{booktitle}{\emph{CVPRW}}.
\newblock


\bibitem[\protect\citeauthoryear{Timofte, Rothe, and Van~Gool}{Timofte
  et~al\mbox{.}}{2016}]%
        {timofte2016seven}
\bibfield{author}{\bibinfo{person}{Radu Timofte}, \bibinfo{person}{Rasmus
  Rothe}, {and} \bibinfo{person}{Luc Van~Gool}.}
  \bibinfo{year}{2016}\natexlab{}.
\newblock \showarticletitle{Seven ways to improve example-based single image
  super resolution}. In \bibinfo{booktitle}{\emph{CVPR}}.
\newblock


\bibitem[\protect\citeauthoryear{Tong, Li, Liu, and Gao}{Tong
  et~al\mbox{.}}{2017}]%
        {tong2017image}
\bibfield{author}{\bibinfo{person}{Tong Tong}, \bibinfo{person}{Gen Li},
  \bibinfo{person}{Xiejie Liu}, {and} \bibinfo{person}{Qinquan Gao}.}
  \bibinfo{year}{2017}\natexlab{}.
\newblock \showarticletitle{Image super-resolution using dense skip
  connections}. In \bibinfo{booktitle}{\emph{ICCV}}.
\newblock


\bibitem[\protect\citeauthoryear{Van~Duong, Huu, Yim, and Jeon}{Van~Duong
  et~al\mbox{.}}{2023}]%
        {van2023light}
\bibfield{author}{\bibinfo{person}{Vinh Van~Duong},
  \bibinfo{person}{Thuc~Nguyen Huu}, \bibinfo{person}{Jonghoon Yim}, {and}
  \bibinfo{person}{Byeungwoo Jeon}.} \bibinfo{year}{2023}\natexlab{}.
\newblock \showarticletitle{Light field image super-resolution network via
  joint spatial-angular and epipolar information}.
\newblock \bibinfo{journal}{\emph{IEEE Transactions on Computational Imaging}}
  \bibinfo{volume}{9} (\bibinfo{year}{2023}), \bibinfo{pages}{350--366}.
\newblock


\bibitem[\protect\citeauthoryear{Vaswani, Shazeer, Parmar, Uszkoreit, Jones,
  Gomez, Kaiser, and Polosukhin}{Vaswani et~al\mbox{.}}{2017}]%
        {vaswani2017attention}
\bibfield{author}{\bibinfo{person}{Ashish Vaswani}, \bibinfo{person}{Noam
  Shazeer}, \bibinfo{person}{Niki Parmar}, \bibinfo{person}{Jakob Uszkoreit},
  \bibinfo{person}{Llion Jones}, \bibinfo{person}{Aidan~N Gomez},
  \bibinfo{person}{{\L}ukasz Kaiser}, {and} \bibinfo{person}{Illia
  Polosukhin}.} \bibinfo{year}{2017}\natexlab{}.
\newblock \showarticletitle{Attention is all you need}. In
  \bibinfo{booktitle}{\emph{NeurIPS}}.
\newblock


\bibitem[\protect\citeauthoryear{Wang, Li, and Shi}{Wang
  et~al\mbox{.}}{2019a}]%
        {wang1904lightweight}
\bibfield{author}{\bibinfo{person}{C Wang}, \bibinfo{person}{Z Li}, {and}
  \bibinfo{person}{J Shi}.} \bibinfo{year}{2019}\natexlab{a}.
\newblock \showarticletitle{Lightweight image super-resolution with adaptive
  weighted learning network}.
\newblock \bibinfo{journal}{\emph{arXiv preprint arXiv:1904.02358}}
  (\bibinfo{year}{2019}).
\newblock


\bibitem[\protect\citeauthoryear{Wang, Chen, Ni, Liu, and Liu}{Wang
  et~al\mbox{.}}{2023a}]%
        {wang2023omni}
\bibfield{author}{\bibinfo{person}{Hang Wang}, \bibinfo{person}{Xuanhong Chen},
  \bibinfo{person}{Bingbing Ni}, \bibinfo{person}{Yutian Liu}, {and}
  \bibinfo{person}{Jinfan Liu}.} \bibinfo{year}{2023}\natexlab{a}.
\newblock \showarticletitle{Omni aggregation networks for lightweight image
  super-resolution}. In \bibinfo{booktitle}{\emph{CVPR}}.
\newblock


\bibitem[\protect\citeauthoryear{Wang, Yue, Zhou, Chan, and Loy}{Wang
  et~al\mbox{.}}{2023d}]%
        {wang2023exploiting}
\bibfield{author}{\bibinfo{person}{Jianyi Wang}, \bibinfo{person}{Zongsheng
  Yue}, \bibinfo{person}{Shangchen Zhou}, \bibinfo{person}{Kelvin~CK Chan},
  {and} \bibinfo{person}{Chen~Change Loy}.} \bibinfo{year}{2023}\natexlab{d}.
\newblock \showarticletitle{Exploiting Diffusion Prior for Real-World Image
  Super-Resolution}.
\newblock \bibinfo{journal}{\emph{arXiv preprint arXiv:2305.07015}}
  (\bibinfo{year}{2023}).
\newblock


\bibitem[\protect\citeauthoryear{Wang, Dong, Wang, Ying, Lin, An, and Guo}{Wang
  et~al\mbox{.}}{2021a}]%
        {wang2021exploring}
\bibfield{author}{\bibinfo{person}{Longguang Wang}, \bibinfo{person}{Xiaoyu
  Dong}, \bibinfo{person}{Yingqian Wang}, \bibinfo{person}{Xinyi Ying},
  \bibinfo{person}{Zaiping Lin}, \bibinfo{person}{Wei An}, {and}
  \bibinfo{person}{Yulan Guo}.} \bibinfo{year}{2021}\natexlab{a}.
\newblock \showarticletitle{Exploring sparsity in image super-resolution for
  efficient inference}. In \bibinfo{booktitle}{\emph{CVPR}}.
\newblock


\bibitem[\protect\citeauthoryear{Wang, Guo, Wang, Liang, Lin, Yang, and
  An}{Wang et~al\mbox{.}}{2020b}]%
        {wang2020parallax}
\bibfield{author}{\bibinfo{person}{Longguang Wang}, \bibinfo{person}{Yulan
  Guo}, \bibinfo{person}{Yingqian Wang}, \bibinfo{person}{Zhengfa Liang},
  \bibinfo{person}{Zaiping Lin}, \bibinfo{person}{Jungang Yang}, {and}
  \bibinfo{person}{Wei An}.} \bibinfo{year}{2020}\natexlab{b}.
\newblock \showarticletitle{Parallax attention for unsupervised stereo
  correspondence learning}.
\newblock \bibinfo{journal}{\emph{IEEE Transactions on Pattern Analysis and
  Machine Intelligence}} (\bibinfo{year}{2020}).
\newblock


\bibitem[\protect\citeauthoryear{Wang, Li, Zhu, Tian, and Shan}{Wang
  et~al\mbox{.}}{2020c}]%
        {wang2020dual}
\bibfield{author}{\bibinfo{person}{Li Wang}, \bibinfo{person}{Dong Li},
  \bibinfo{person}{Yousong Zhu}, \bibinfo{person}{Lu Tian}, {and}
  \bibinfo{person}{Yi Shan}.} \bibinfo{year}{2020}\natexlab{c}.
\newblock \showarticletitle{Dual super-resolution learning for semantic
  segmentation}. In \bibinfo{booktitle}{\emph{CVPR}}.
\newblock


\bibitem[\protect\citeauthoryear{Wang, Wang, Dong, Xu, Yang, An, and Guo}{Wang
  et~al\mbox{.}}{2021d}]%
        {wang2021unsupervised}
\bibfield{author}{\bibinfo{person}{Longguang Wang}, \bibinfo{person}{Yingqian
  Wang}, \bibinfo{person}{Xiaoyu Dong}, \bibinfo{person}{Qingyu Xu},
  \bibinfo{person}{Jungang Yang}, \bibinfo{person}{Wei An}, {and}
  \bibinfo{person}{Yulan Guo}.} \bibinfo{year}{2021}\natexlab{d}.
\newblock \showarticletitle{Unsupervised Degradation Representation Learning
  for Blind Super-Resolution}. In \bibinfo{booktitle}{\emph{CVPR}}.
\newblock


\bibitem[\protect\citeauthoryear{Wang, Wang, Liang, Lin, Yang, An, and
  Guo}{Wang et~al\mbox{.}}{2019c}]%
        {wang2019learning}
\bibfield{author}{\bibinfo{person}{Longguang Wang}, \bibinfo{person}{Yingqian
  Wang}, \bibinfo{person}{Zhengfa Liang}, \bibinfo{person}{Zaiping Lin},
  \bibinfo{person}{Jungang Yang}, \bibinfo{person}{Wei An}, {and}
  \bibinfo{person}{Yulan Guo}.} \bibinfo{year}{2019}\natexlab{c}.
\newblock \showarticletitle{Learning parallax attention for stereo image
  super-resolution}. In \bibinfo{booktitle}{\emph{CVPR}}.
\newblock


\bibitem[\protect\citeauthoryear{Wang, Wang, Lin, Yang, An, and Guo}{Wang
  et~al\mbox{.}}{2021e}]%
        {wanglearning}
\bibfield{author}{\bibinfo{person}{Longguang Wang}, \bibinfo{person}{Yingqian
  Wang}, \bibinfo{person}{Zaiping Lin}, \bibinfo{person}{Jungang Yang},
  \bibinfo{person}{Wei An}, {and} \bibinfo{person}{Yulan Guo}.}
  \bibinfo{year}{2021}\natexlab{e}.
\newblock \showarticletitle{Learning a single network for scale-arbitrary
  super-resolution}.
\newblock  (\bibinfo{year}{2021}).
\newblock


\bibitem[\protect\citeauthoryear{Wang, Girshick, Gupta, and He}{Wang
  et~al\mbox{.}}{2018a}]%
        {wang2018non}
\bibfield{author}{\bibinfo{person}{Xiaolong Wang}, \bibinfo{person}{Ross
  Girshick}, \bibinfo{person}{Abhinav Gupta}, {and} \bibinfo{person}{Kaiming
  He}.} \bibinfo{year}{2018}\natexlab{a}.
\newblock \showarticletitle{Non-local neural networks}. In
  \bibinfo{booktitle}{\emph{CVPR}}.
\newblock


\bibitem[\protect\citeauthoryear{Wang, Li, Zhang, and Shan}{Wang
  et~al\mbox{.}}{2021c}]%
        {wang2021towards}
\bibfield{author}{\bibinfo{person}{Xintao Wang}, \bibinfo{person}{Yu Li},
  \bibinfo{person}{Honglun Zhang}, {and} \bibinfo{person}{Ying Shan}.}
  \bibinfo{year}{2021}\natexlab{c}.
\newblock \showarticletitle{Towards real-world blind face restoration with
  generative facial prior}. In \bibinfo{booktitle}{\emph{CVPR}}.
\newblock


\bibitem[\protect\citeauthoryear{Wang, Yu, Dong, and Loy}{Wang
  et~al\mbox{.}}{2018d}]%
        {wang2018recovering}
\bibfield{author}{\bibinfo{person}{Xintao Wang}, \bibinfo{person}{Ke Yu},
  \bibinfo{person}{Chao Dong}, {and} \bibinfo{person}{Chen~Change Loy}.}
  \bibinfo{year}{2018}\natexlab{d}.
\newblock \showarticletitle{Recovering realistic texture in image
  super-resolution by deep spatial feature transform}. In
  \bibinfo{booktitle}{\emph{CVPR}}.
\newblock


\bibitem[\protect\citeauthoryear{Wang, Yu, Wu, Gu, Liu, Dong, Qiao, and
  Loy}{Wang et~al\mbox{.}}{2018e}]%
        {wang2018esrgan}
\bibfield{author}{\bibinfo{person}{Xintao Wang}, \bibinfo{person}{Ke Yu},
  \bibinfo{person}{Shixiang Wu}, \bibinfo{person}{Jinjin Gu},
  \bibinfo{person}{Yihao Liu}, \bibinfo{person}{Chao Dong}, \bibinfo{person}{Yu
  Qiao}, {and} \bibinfo{person}{Chen~Change Loy}.}
  \bibinfo{year}{2018}\natexlab{e}.
\newblock \showarticletitle{Esrgan: Enhanced super-resolution generative
  adversarial networks}. In \bibinfo{booktitle}{\emph{ECCV}}.
\newblock


\bibitem[\protect\citeauthoryear{Wang, Lin, Luo, Tai, Zhang, and Xie}{Wang
  et~al\mbox{.}}{2023b}]%
        {wang2023high}
\bibfield{author}{\bibinfo{person}{Yanbo Wang}, \bibinfo{person}{Chuming Lin},
  \bibinfo{person}{Donghao Luo}, \bibinfo{person}{Ying Tai},
  \bibinfo{person}{Zhizhong Zhang}, {and} \bibinfo{person}{Yuan Xie}.}
  \bibinfo{year}{2023}\natexlab{b}.
\newblock \showarticletitle{High-Resolution GAN Inversion for Degraded Images
  in Large Diverse Datasets}.
\newblock \bibinfo{journal}{\emph{AAAI}} (\bibinfo{year}{2023}).
\newblock


\bibitem[\protect\citeauthoryear{Wang, Liu, Zhang, Hou, Sun, and Tan}{Wang
  et~al\mbox{.}}{2018b}]%
        {wang2018lfnet}
\bibfield{author}{\bibinfo{person}{Yunlong Wang}, \bibinfo{person}{Fei Liu},
  \bibinfo{person}{Kunbo Zhang}, \bibinfo{person}{Guangqi Hou},
  \bibinfo{person}{Zhenan Sun}, {and} \bibinfo{person}{Tieniu Tan}.}
  \bibinfo{year}{2018}\natexlab{b}.
\newblock \showarticletitle{LFNet: A novel bidirectional recurrent
  convolutional neural network for light-field image super-resolution}.
\newblock \bibinfo{journal}{\emph{IEEE Transactions on Image Processing}}
  \bibinfo{volume}{27}, \bibinfo{number}{9} (\bibinfo{year}{2018}),
  \bibinfo{pages}{4274--4286}.
\newblock


\bibitem[\protect\citeauthoryear{Wang, Perazzi, McWilliams, Sorkine-Hornung,
  Sorkine-Hornung, and Schroers}{Wang et~al\mbox{.}}{2018c}]%
        {wang2018fully}
\bibfield{author}{\bibinfo{person}{Yifan Wang}, \bibinfo{person}{Federico
  Perazzi}, \bibinfo{person}{Brian McWilliams}, \bibinfo{person}{Alexander
  Sorkine-Hornung}, \bibinfo{person}{Olga Sorkine-Hornung}, {and}
  \bibinfo{person}{Christopher Schroers}.} \bibinfo{year}{2018}\natexlab{c}.
\newblock \showarticletitle{A fully progressive approach to single-image
  super-resolution}. In \bibinfo{booktitle}{\emph{CVPRW}}.
\newblock


\bibitem[\protect\citeauthoryear{Wang, Shao, Lu, Wu, and Wang}{Wang
  et~al\mbox{.}}{2023c}]%
        {wang2023remote}
\bibfield{author}{\bibinfo{person}{Yu Wang}, \bibinfo{person}{Zhenfeng Shao},
  \bibinfo{person}{Tao Lu}, \bibinfo{person}{Changzhi Wu}, {and}
  \bibinfo{person}{Jiaming Wang}.} \bibinfo{year}{2023}\natexlab{c}.
\newblock \showarticletitle{Remote sensing image super-resolution via
  multiscale enhancement network}.
\newblock \bibinfo{journal}{\emph{IEEE Geoscience and Remote Sensing Letters}}
  \bibinfo{volume}{20} (\bibinfo{year}{2023}), \bibinfo{pages}{1--5}.
\newblock


\bibitem[\protect\citeauthoryear{Wang, Teng, He, Feng, and Zhang}{Wang
  et~al\mbox{.}}{2019b}]%
        {wang2019ct}
\bibfield{author}{\bibinfo{person}{Yukai Wang}, \bibinfo{person}{Qizhi Teng},
  \bibinfo{person}{Xiaohai He}, \bibinfo{person}{Junxi Feng}, {and}
  \bibinfo{person}{Tingrong Zhang}.} \bibinfo{year}{2019}\natexlab{b}.
\newblock \showarticletitle{CT-image of rock samples super resolution using 3D
  convolutional neural network}.
\newblock \bibinfo{journal}{\emph{Computers \& Geosciences}}
  \bibinfo{volume}{133} (\bibinfo{year}{2019}), \bibinfo{pages}{104314}.
\newblock


\bibitem[\protect\citeauthoryear{Wang, Wang, Wu, Yang, An, Yu, and Guo}{Wang
  et~al\mbox{.}}{2022b}]%
        {wang2022disentangling}
\bibfield{author}{\bibinfo{person}{Yingqian Wang}, \bibinfo{person}{Longguang
  Wang}, \bibinfo{person}{Gaochang Wu}, \bibinfo{person}{Jungang Yang},
  \bibinfo{person}{Wei An}, \bibinfo{person}{Jingyi Yu}, {and}
  \bibinfo{person}{Yulan Guo}.} \bibinfo{year}{2022}\natexlab{b}.
\newblock \showarticletitle{Disentangling light fields for super-resolution and
  disparity estimation}.
\newblock \bibinfo{journal}{\emph{IEEE Transactions on Pattern Analysis and
  Machine Intelligence}} \bibinfo{volume}{45}, \bibinfo{number}{1}
  (\bibinfo{year}{2022}), \bibinfo{pages}{425--443}.
\newblock


\bibitem[\protect\citeauthoryear{Wang, Wang, Yang, An, and Guo}{Wang
  et~al\mbox{.}}{2019d}]%
        {wang2019flickr1024}
\bibfield{author}{\bibinfo{person}{Yingqian Wang}, \bibinfo{person}{Longguang
  Wang}, \bibinfo{person}{Jungang Yang}, \bibinfo{person}{Wei An}, {and}
  \bibinfo{person}{Yulan Guo}.} \bibinfo{year}{2019}\natexlab{d}.
\newblock \showarticletitle{Flickr1024: A large-scale dataset for stereo image
  super-resolution}. In \bibinfo{booktitle}{\emph{ICCVW}}.
\newblock


\bibitem[\protect\citeauthoryear{Wang, Wang, Yang, An, Yu, and Guo}{Wang
  et~al\mbox{.}}{2020d}]%
        {wang2020spatial}
\bibfield{author}{\bibinfo{person}{Yingqian Wang}, \bibinfo{person}{Longguang
  Wang}, \bibinfo{person}{Jungang Yang}, \bibinfo{person}{Wei An},
  \bibinfo{person}{Jingyi Yu}, {and} \bibinfo{person}{Yulan Guo}.}
  \bibinfo{year}{2020}\natexlab{d}.
\newblock \showarticletitle{Spatial-angular interaction for light field image
  super-resolution}. In \bibinfo{booktitle}{\emph{ECCV}}.
\newblock


\bibitem[\protect\citeauthoryear{Wang, Yang, Wang, Ying, Wu, An, and Guo}{Wang
  et~al\mbox{.}}{2020e}]%
        {wang2020light}
\bibfield{author}{\bibinfo{person}{Yingqian Wang}, \bibinfo{person}{Jungang
  Yang}, \bibinfo{person}{Longguang Wang}, \bibinfo{person}{Xinyi Ying},
  \bibinfo{person}{Tianhao Wu}, \bibinfo{person}{Wei An}, {and}
  \bibinfo{person}{Yulan Guo}.} \bibinfo{year}{2020}\natexlab{e}.
\newblock \showarticletitle{Light field image super-resolution using deformable
  convolution}.
\newblock \bibinfo{journal}{\emph{IEEE Transactions on Image Processing}}
  \bibinfo{volume}{30} (\bibinfo{year}{2020}), \bibinfo{pages}{1057--1071}.
\newblock


\bibitem[\protect\citeauthoryear{Wang, Ying, Wang, Yang, An, and Guo}{Wang
  et~al\mbox{.}}{2021f}]%
        {wang2021symmetric}
\bibfield{author}{\bibinfo{person}{Yingqian Wang}, \bibinfo{person}{Xinyi
  Ying}, \bibinfo{person}{Longguang Wang}, \bibinfo{person}{Jungang Yang},
  \bibinfo{person}{Wei An}, {and} \bibinfo{person}{Yulan Guo}.}
  \bibinfo{year}{2021}\natexlab{f}.
\newblock \showarticletitle{Symmetric parallax attention for stereo image
  super-resolution}. In \bibinfo{booktitle}{\emph{CVPR}}.
\newblock


\bibitem[\protect\citeauthoryear{Wang and Bovik}{Wang and Bovik}{2009}]%
        {wang2009mean}
\bibfield{author}{\bibinfo{person}{Zhou Wang} {and} \bibinfo{person}{Alan~C
  Bovik}.} \bibinfo{year}{2009}\natexlab{}.
\newblock \showarticletitle{Mean squared error: Love it or leave it? A new look
  at signal fidelity measures}.
\newblock \bibinfo{journal}{\emph{IEEE Signal Processing Magazine}}
  \bibinfo{volume}{26}, \bibinfo{number}{1} (\bibinfo{year}{2009}),
  \bibinfo{pages}{98--117}.
\newblock


\bibitem[\protect\citeauthoryear{Wang, Bovik, Sheikh, and Simoncelli}{Wang
  et~al\mbox{.}}{2004}]%
        {wang2004image}
\bibfield{author}{\bibinfo{person}{Zhou Wang}, \bibinfo{person}{Alan~C Bovik},
  \bibinfo{person}{Hamid~R Sheikh}, {and} \bibinfo{person}{Eero~P Simoncelli}.}
  \bibinfo{year}{2004}\natexlab{}.
\newblock \showarticletitle{Image quality assessment: from error visibility to
  structural similarity}.
\newblock \bibinfo{journal}{\emph{IEEE Transactions on Image Processing}}
  \bibinfo{volume}{13}, \bibinfo{number}{4} (\bibinfo{year}{2004}),
  \bibinfo{pages}{600--612}.
\newblock


\bibitem[\protect\citeauthoryear{Wang, Chen, and Hoi}{Wang
  et~al\mbox{.}}{2020a}]%
        {wang2020deep}
\bibfield{author}{\bibinfo{person}{Zhihao Wang}, \bibinfo{person}{Jian Chen},
  {and} \bibinfo{person}{Steven~CH Hoi}.} \bibinfo{year}{2020}\natexlab{a}.
\newblock \showarticletitle{Deep learning for image super-resolution: A
  survey}.
\newblock \bibinfo{journal}{\emph{IEEE Transactions on Pattern Analysis and
  Machine Intelligence}} \bibinfo{volume}{43}, \bibinfo{number}{10}
  (\bibinfo{year}{2020}), \bibinfo{pages}{3365--3387}.
\newblock


\bibitem[\protect\citeauthoryear{Wang, Cun, Bao, Zhou, Liu, and Li}{Wang
  et~al\mbox{.}}{2022a}]%
        {wang2022uformer}
\bibfield{author}{\bibinfo{person}{Zhendong Wang}, \bibinfo{person}{Xiaodong
  Cun}, \bibinfo{person}{Jianmin Bao}, \bibinfo{person}{Wengang Zhou},
  \bibinfo{person}{Jianzhuang Liu}, {and} \bibinfo{person}{Houqiang Li}.}
  \bibinfo{year}{2022}\natexlab{a}.
\newblock \showarticletitle{Uformer: A general u-shaped transformer for image
  restoration}. In \bibinfo{booktitle}{\emph{CVPR}}.
\newblock


\bibitem[\protect\citeauthoryear{Wang, Gao, Li, Yu, and Lu}{Wang
  et~al\mbox{.}}{2021b}]%
        {wang2021lightweight}
\bibfield{author}{\bibinfo{person}{Zhengxue Wang}, \bibinfo{person}{Guangwei
  Gao}, \bibinfo{person}{Juncheng Li}, \bibinfo{person}{Yi Yu}, {and}
  \bibinfo{person}{Huimin Lu}.} \bibinfo{year}{2021}\natexlab{b}.
\newblock \showarticletitle{Lightweight Image Super-Resolution with Multi-scale
  Feature Interaction Network}. In \bibinfo{booktitle}{\emph{ICME}}.
\newblock


\bibitem[\protect\citeauthoryear{Wang, Zhang, Zhang, Zheng, Zhou, Zhang, and
  Wang}{Wang et~al\mbox{.}}{2023e}]%
        {wang2023dr2}
\bibfield{author}{\bibinfo{person}{Zhixin Wang}, \bibinfo{person}{Ziying
  Zhang}, \bibinfo{person}{Xiaoyun Zhang}, \bibinfo{person}{Huangjie Zheng},
  \bibinfo{person}{Mingyuan Zhou}, \bibinfo{person}{Ya Zhang}, {and}
  \bibinfo{person}{Yanfeng Wang}.} \bibinfo{year}{2023}\natexlab{e}.
\newblock \showarticletitle{DR2: Diffusion-based Robust Degradation Remover for
  Blind Face Restoration}. In \bibinfo{booktitle}{\emph{CVPR}}.
\newblock


\bibitem[\protect\citeauthoryear{Wei, Xie, Lu, Zhan, Ye, Zuo, and Lin}{Wei
  et~al\mbox{.}}{2020}]%
        {wei2020component}
\bibfield{author}{\bibinfo{person}{Pengxu Wei}, \bibinfo{person}{Ziwei Xie},
  \bibinfo{person}{Hannan Lu}, \bibinfo{person}{Zongyuan Zhan},
  \bibinfo{person}{Qixiang Ye}, \bibinfo{person}{Wangmeng Zuo}, {and}
  \bibinfo{person}{Liang Lin}.} \bibinfo{year}{2020}\natexlab{}.
\newblock \showarticletitle{Component divide-and-conquer for real-world image
  super-resolution}. In \bibinfo{booktitle}{\emph{ECCV}}.
\newblock


\bibitem[\protect\citeauthoryear{Wei, Gu, Li, Timofte, Jin, and Song}{Wei
  et~al\mbox{.}}{2021}]%
        {wei2021unsupervised}
\bibfield{author}{\bibinfo{person}{Yunxuan Wei}, \bibinfo{person}{Shuhang Gu},
  \bibinfo{person}{Yawei Li}, \bibinfo{person}{Radu Timofte},
  \bibinfo{person}{Longcun Jin}, {and} \bibinfo{person}{Hengjie Song}.}
  \bibinfo{year}{2021}\natexlab{}.
\newblock \showarticletitle{Unsupervised real-world image super resolution via
  domain-distance aware training}. In \bibinfo{booktitle}{\emph{CVPR}}.
\newblock


\bibitem[\protect\citeauthoryear{Wolf, Lugmayr, Danelljan, Van~Gool, and
  Timofte}{Wolf et~al\mbox{.}}{2021}]%
        {wolf2021deflow}
\bibfield{author}{\bibinfo{person}{Valentin Wolf}, \bibinfo{person}{Andreas
  Lugmayr}, \bibinfo{person}{Martin Danelljan}, \bibinfo{person}{Luc Van~Gool},
  {and} \bibinfo{person}{Radu Timofte}.} \bibinfo{year}{2021}\natexlab{}.
\newblock \showarticletitle{Deflow: Learning complex image degradations from
  unpaired data with conditional flows}. In \bibinfo{booktitle}{\emph{CVPR}}.
\newblock


\bibitem[\protect\citeauthoryear{Xia, Hang, Tian, Yang, Liao, and Zhou}{Xia
  et~al\mbox{.}}{2022}]%
        {xia2022efficient}
\bibfield{author}{\bibinfo{person}{Bin Xia}, \bibinfo{person}{Yucheng Hang},
  \bibinfo{person}{Yapeng Tian}, \bibinfo{person}{Wenming Yang},
  \bibinfo{person}{Qingmin Liao}, {and} \bibinfo{person}{Jie Zhou}.}
  \bibinfo{year}{2022}\natexlab{}.
\newblock \showarticletitle{Efficient non-local contrastive attention for image
  super-resolution}. In \bibinfo{booktitle}{\emph{AAAI}}.
\newblock


\bibitem[\protect\citeauthoryear{Xia, Zhang, Wang, Wang, Wu, Tian, Yang, and
  Van~Gool}{Xia et~al\mbox{.}}{2023}]%
        {xia2023diffir}
\bibfield{author}{\bibinfo{person}{Bin Xia}, \bibinfo{person}{Yulun Zhang},
  \bibinfo{person}{Shiyin Wang}, \bibinfo{person}{Yitong Wang},
  \bibinfo{person}{Xinglong Wu}, \bibinfo{person}{Yapeng Tian},
  \bibinfo{person}{Wenming Yang}, {and} \bibinfo{person}{Luc Van~Gool}.}
  \bibinfo{year}{2023}\natexlab{}.
\newblock \showarticletitle{Diffir: Efficient diffusion model for image
  restoration}.
\newblock \bibinfo{journal}{\emph{ICCV}} (\bibinfo{year}{2023}).
\newblock


\bibitem[\protect\citeauthoryear{Xiang, Lin, and Allebach}{Xiang
  et~al\mbox{.}}{2021}]%
        {xiang2021boosting}
\bibfield{author}{\bibinfo{person}{Xiaoyu Xiang}, \bibinfo{person}{Qian Lin},
  {and} \bibinfo{person}{Jan~P Allebach}.} \bibinfo{year}{2021}\natexlab{}.
\newblock \showarticletitle{Boosting high-level vision with joint compression
  artifacts reduction and super-resolution}. In
  \bibinfo{booktitle}{\emph{ICPR}}.
\newblock


\bibitem[\protect\citeauthoryear{Xu, Tseng, Tseng, Kuo, and Tsai}{Xu
  et~al\mbox{.}}{2020}]%
        {xu2020unified}
\bibfield{author}{\bibinfo{person}{Yu-Syuan Xu}, \bibinfo{person}{Shou-Yao~Roy
  Tseng}, \bibinfo{person}{Yu Tseng}, \bibinfo{person}{Hsien-Kai Kuo}, {and}
  \bibinfo{person}{Yi-Min Tsai}.} \bibinfo{year}{2020}\natexlab{}.
\newblock \showarticletitle{Unified dynamic convolutional network for
  super-resolution with variational degradations}. In
  \bibinfo{booktitle}{\emph{CVPR}}.
\newblock


\bibitem[\protect\citeauthoryear{Yang, Yang, Fu, Lu, and Guo}{Yang
  et~al\mbox{.}}{2020}]%
        {yang2020learning}
\bibfield{author}{\bibinfo{person}{Fuzhi Yang}, \bibinfo{person}{Huan Yang},
  \bibinfo{person}{Jianlong Fu}, \bibinfo{person}{Hongtao Lu}, {and}
  \bibinfo{person}{Baining Guo}.} \bibinfo{year}{2020}\natexlab{}.
\newblock \showarticletitle{Learning texture transformer network for image
  super-resolution}. In \bibinfo{booktitle}{\emph{CVPR}}.
\newblock


\bibitem[\protect\citeauthoryear{Yang, Wright, Huang, and Ma}{Yang
  et~al\mbox{.}}{2010}]%
        {yang2010image}
\bibfield{author}{\bibinfo{person}{Jianchao Yang}, \bibinfo{person}{John
  Wright}, \bibinfo{person}{Thomas~S Huang}, {and} \bibinfo{person}{Yi Ma}.}
  \bibinfo{year}{2010}\natexlab{}.
\newblock \showarticletitle{Image super-resolution via sparse representation}.
\newblock \bibinfo{journal}{\emph{IEEE Transactions on Image Processing}}
  \bibinfo{volume}{19}, \bibinfo{number}{11} (\bibinfo{year}{2010}),
  \bibinfo{pages}{2861--2873}.
\newblock


\bibitem[\protect\citeauthoryear{Yang, Feng, Yang, Zhao, Liu, Guo, and
  Yan}{Yang et~al\mbox{.}}{2017}]%
        {yang2017deep}
\bibfield{author}{\bibinfo{person}{Wenhan Yang}, \bibinfo{person}{Jiashi Feng},
  \bibinfo{person}{Jianchao Yang}, \bibinfo{person}{Fang Zhao},
  \bibinfo{person}{Jiaying Liu}, \bibinfo{person}{Zongming Guo}, {and}
  \bibinfo{person}{Shuicheng Yan}.} \bibinfo{year}{2017}\natexlab{}.
\newblock \showarticletitle{Deep edge guided recurrent residual learning for
  image super-resolution}.
\newblock \bibinfo{journal}{\emph{IEEE Transactions on Image Processing}}
  \bibinfo{volume}{26}, \bibinfo{number}{12} (\bibinfo{year}{2017}),
  \bibinfo{pages}{5895--5907}.
\newblock


\bibitem[\protect\citeauthoryear{Ye, Lin, Huang, Fan, Shi, Xie, and Qu}{Ye
  et~al\mbox{.}}{2023}]%
        {ye2023hardware}
\bibfield{author}{\bibinfo{person}{Fangchen Ye}, \bibinfo{person}{Jin Lin},
  \bibinfo{person}{Hongzhan Huang}, \bibinfo{person}{Jianping Fan},
  \bibinfo{person}{Zhongchao Shi}, \bibinfo{person}{Yuan Xie}, {and}
  \bibinfo{person}{Yanyun Qu}.} \bibinfo{year}{2023}\natexlab{}.
\newblock \showarticletitle{Hardware-friendly Scalable Image Super Resolution
  with Progressive Structured Sparsity}. In \bibinfo{booktitle}{\emph{ACMMM}}.
\newblock


\bibitem[\protect\citeauthoryear{Ying, Wang, Wang, Sheng, An, and Guo}{Ying
  et~al\mbox{.}}{2020}]%
        {ying2020stereo}
\bibfield{author}{\bibinfo{person}{Xinyi Ying}, \bibinfo{person}{Yingqian
  Wang}, \bibinfo{person}{Longguang Wang}, \bibinfo{person}{Weidong Sheng},
  \bibinfo{person}{Wei An}, {and} \bibinfo{person}{Yulan Guo}.}
  \bibinfo{year}{2020}\natexlab{}.
\newblock \showarticletitle{A stereo attention module for stereo image
  super-resolution}.
\newblock \bibinfo{journal}{\emph{IEEE Signal Processing Letters}}
  \bibinfo{volume}{27} (\bibinfo{year}{2020}), \bibinfo{pages}{496--500}.
\newblock


\bibitem[\protect\citeauthoryear{Yoon, Jeon, Yoo, Lee, and Kweon}{Yoon
  et~al\mbox{.}}{2017}]%
        {yoon2017light}
\bibfield{author}{\bibinfo{person}{Youngjin Yoon}, \bibinfo{person}{Hae-Gon
  Jeon}, \bibinfo{person}{Donggeun Yoo}, \bibinfo{person}{Joon-Young Lee},
  {and} \bibinfo{person}{In~So Kweon}.} \bibinfo{year}{2017}\natexlab{}.
\newblock \showarticletitle{Light-field image super-resolution using
  convolutional neural network}.
\newblock \bibinfo{journal}{\emph{IEEE Signal Processing Letters}}
  \bibinfo{volume}{24}, \bibinfo{number}{6} (\bibinfo{year}{2017}),
  \bibinfo{pages}{848--852}.
\newblock


\bibitem[\protect\citeauthoryear{Yoon, Jeon, Yoo, Lee, and So~Kweon}{Yoon
  et~al\mbox{.}}{2015}]%
        {yoon2015learning}
\bibfield{author}{\bibinfo{person}{Youngjin Yoon}, \bibinfo{person}{Hae-Gon
  Jeon}, \bibinfo{person}{Donggeun Yoo}, \bibinfo{person}{Joon-Young Lee},
  {and} \bibinfo{person}{In So~Kweon}.} \bibinfo{year}{2015}\natexlab{}.
\newblock \showarticletitle{Learning a deep convolutional network for
  light-field image super-resolution}. In \bibinfo{booktitle}{\emph{ICCVW}}.
\newblock


\bibitem[\protect\citeauthoryear{Yu, Li, Koh, Zhang, Pang, Qin, Ku, Xu,
  Baldridge, and Wu}{Yu et~al\mbox{.}}{2021}]%
        {yu2021vector}
\bibfield{author}{\bibinfo{person}{Jiahui Yu}, \bibinfo{person}{Xin Li},
  \bibinfo{person}{Jing~Yu Koh}, \bibinfo{person}{Han Zhang},
  \bibinfo{person}{Ruoming Pang}, \bibinfo{person}{James Qin},
  \bibinfo{person}{Alexander Ku}, \bibinfo{person}{Yuanzhong Xu},
  \bibinfo{person}{Jason Baldridge}, {and} \bibinfo{person}{Yonghui Wu}.}
  \bibinfo{year}{2021}\natexlab{}.
\newblock \showarticletitle{Vector-quantized image modeling with improved
  VQGAN}.
\newblock \bibinfo{journal}{\emph{arXiv preprint arXiv:2110.04627}}
  (\bibinfo{year}{2021}).
\newblock


\bibitem[\protect\citeauthoryear{Yu and Porikli}{Yu and Porikli}{2017}]%
        {yu2017hallucinating}
\bibfield{author}{\bibinfo{person}{Xin Yu} {and} \bibinfo{person}{Fatih
  Porikli}.} \bibinfo{year}{2017}\natexlab{}.
\newblock \showarticletitle{Hallucinating very low-resolution unaligned and
  noisy face images by transformative discriminative autoencoders}. In
  \bibinfo{booktitle}{\emph{CVPR}}.
\newblock


\bibitem[\protect\citeauthoryear{Yuan, Liu, Zhang, Zhang, Dong, and Lin}{Yuan
  et~al\mbox{.}}{2018}]%
        {yuan2018unsupervised}
\bibfield{author}{\bibinfo{person}{Yuan Yuan}, \bibinfo{person}{Siyuan Liu},
  \bibinfo{person}{Jiawei Zhang}, \bibinfo{person}{Yongbing Zhang},
  \bibinfo{person}{Chao Dong}, {and} \bibinfo{person}{Liang Lin}.}
  \bibinfo{year}{2018}\natexlab{}.
\newblock \showarticletitle{Unsupervised image super-resolution using
  cycle-in-cycle generative adversarial networks}. In
  \bibinfo{booktitle}{\emph{CVPRW}}.
\newblock


\bibitem[\protect\citeauthoryear{Yue, Sun, Yang, and Wu}{Yue
  et~al\mbox{.}}{2013}]%
        {yue2013landmark}
\bibfield{author}{\bibinfo{person}{Huanjing Yue}, \bibinfo{person}{Xiaoyan
  Sun}, \bibinfo{person}{Jingyu Yang}, {and} \bibinfo{person}{Feng Wu}.}
  \bibinfo{year}{2013}\natexlab{}.
\newblock \showarticletitle{Landmark image super-resolution by retrieving web
  images}.
\newblock \bibinfo{journal}{\emph{IEEE Transactions on Image Processing}}
  \bibinfo{volume}{22}, \bibinfo{number}{12} (\bibinfo{year}{2013}),
  \bibinfo{pages}{4865--4878}.
\newblock


\bibitem[\protect\citeauthoryear{Zamir, Arora, Khan, Hayat, Khan, and
  Yang}{Zamir et~al\mbox{.}}{2022}]%
        {zamir2022restormer}
\bibfield{author}{\bibinfo{person}{Syed~Waqas Zamir}, \bibinfo{person}{Aditya
  Arora}, \bibinfo{person}{Salman Khan}, \bibinfo{person}{Munawar Hayat},
  \bibinfo{person}{Fahad~Shahbaz Khan}, {and} \bibinfo{person}{Ming-Hsuan
  Yang}.} \bibinfo{year}{2022}\natexlab{}.
\newblock \showarticletitle{Restormer: Efficient transformer for
  high-resolution image restoration}. In \bibinfo{booktitle}{\emph{CVPR}}.
\newblock


\bibitem[\protect\citeauthoryear{Zangeneh, Rahmati, and Mohsenzadeh}{Zangeneh
  et~al\mbox{.}}{2020}]%
        {zangeneh2020low}
\bibfield{author}{\bibinfo{person}{Erfan Zangeneh}, \bibinfo{person}{Mohammad
  Rahmati}, {and} \bibinfo{person}{Yalda Mohsenzadeh}.}
  \bibinfo{year}{2020}\natexlab{}.
\newblock \showarticletitle{Low resolution face recognition using a two-branch
  deep convolutional neural network architecture}.
\newblock \bibinfo{journal}{\emph{Expert Systems with Applications}}
  \bibinfo{volume}{139} (\bibinfo{year}{2020}), \bibinfo{pages}{112854}.
\newblock


\bibitem[\protect\citeauthoryear{Zeyde, Elad, and Protter}{Zeyde
  et~al\mbox{.}}{2010}]%
        {zeyde2010single}
\bibfield{author}{\bibinfo{person}{Roman Zeyde}, \bibinfo{person}{Michael
  Elad}, {and} \bibinfo{person}{Matan Protter}.}
  \bibinfo{year}{2010}\natexlab{}.
\newblock \showarticletitle{On single image scale-up using
  sparse-representations}. In \bibinfo{booktitle}{\emph{ICCS}}.
\newblock


\bibitem[\protect\citeauthoryear{Zhang, Shao, Li, and Shen}{Zhang
  et~al\mbox{.}}{2020}]%
        {zhang2020remote}
\bibfield{author}{\bibinfo{person}{Dongyang Zhang}, \bibinfo{person}{Jie Shao},
  \bibinfo{person}{Xinyao Li}, {and} \bibinfo{person}{Heng~Tao Shen}.}
  \bibinfo{year}{2020}\natexlab{}.
\newblock \showarticletitle{Remote sensing image super-resolution via mixed
  high-order attention network}.
\newblock \bibinfo{journal}{\emph{IEEE Transactions on Geoscience and Remote
  Sensing}} \bibinfo{volume}{59}, \bibinfo{number}{6} (\bibinfo{year}{2020}),
  \bibinfo{pages}{5183--5196}.
\newblock


\bibitem[\protect\citeauthoryear{Zhang, Zhang, Cheng, Hsu, Qiao, Liu, and
  Zhang}{Zhang et~al\mbox{.}}{2018e}]%
        {zhang2018super}
\bibfield{author}{\bibinfo{person}{Kaipeng Zhang}, \bibinfo{person}{Zhanpeng
  Zhang}, \bibinfo{person}{Chia-Wen Cheng}, \bibinfo{person}{Winston~H Hsu},
  \bibinfo{person}{Yu Qiao}, \bibinfo{person}{Wei Liu}, {and}
  \bibinfo{person}{Tong Zhang}.} \bibinfo{year}{2018}\natexlab{e}.
\newblock \showarticletitle{Super-identity convolutional neural network for
  face hallucination}. In \bibinfo{booktitle}{\emph{ECCV}}.
\newblock


\bibitem[\protect\citeauthoryear{Zhang, Zuo, and Zhang}{Zhang
  et~al\mbox{.}}{2018f}]%
        {zhang2018learning}
\bibfield{author}{\bibinfo{person}{Kai Zhang}, \bibinfo{person}{Wangmeng Zuo},
  {and} \bibinfo{person}{Lei Zhang}.} \bibinfo{year}{2018}\natexlab{f}.
\newblock \showarticletitle{Learning a single convolutional super-resolution
  network for multiple degradations}. In \bibinfo{booktitle}{\emph{CVPR}}.
\newblock


\bibitem[\protect\citeauthoryear{Zhang, Li, Zhou, Zhao, and Gu}{Zhang
  et~al\mbox{.}}{2024}]%
        {zhang2024transcending}
\bibfield{author}{\bibinfo{person}{Leheng Zhang}, \bibinfo{person}{Yawei Li},
  \bibinfo{person}{Xingyu Zhou}, \bibinfo{person}{Xiaorui Zhao}, {and}
  \bibinfo{person}{Shuhang Gu}.} \bibinfo{year}{2024}\natexlab{}.
\newblock \showarticletitle{Transcending the Limit of Local Window: Advanced
  Super-Resolution Transformer with Adaptive Token Dictionary}.
\newblock  (\bibinfo{year}{2024}).
\newblock


\bibitem[\protect\citeauthoryear{Zhang, Zhang, Zhang, Guo, Gao, and
  Zhang}{Zhang et~al\mbox{.}}{2023}]%
        {zhang2023essaformer}
\bibfield{author}{\bibinfo{person}{Mingjin Zhang}, \bibinfo{person}{Chi Zhang},
  \bibinfo{person}{Qiming Zhang}, \bibinfo{person}{Jie Guo},
  \bibinfo{person}{Xinbo Gao}, {and} \bibinfo{person}{Jing Zhang}.}
  \bibinfo{year}{2023}\natexlab{}.
\newblock \showarticletitle{ESSAformer: Efficient Transformer for Hyperspectral
  Image Super-resolution}. In \bibinfo{booktitle}{\emph{CVPR}}.
\newblock


\bibitem[\protect\citeauthoryear{Zhang, Isola, Efros, Shechtman, and
  Wang}{Zhang et~al\mbox{.}}{2018a}]%
        {zhang2018unreasonable}
\bibfield{author}{\bibinfo{person}{Richard Zhang}, \bibinfo{person}{Phillip
  Isola}, \bibinfo{person}{Alexei~A Efros}, \bibinfo{person}{Eli Shechtman},
  {and} \bibinfo{person}{Oliver Wang}.} \bibinfo{year}{2018}\natexlab{a}.
\newblock \showarticletitle{The unreasonable effectiveness of deep features as
  a perceptual metric}. In \bibinfo{booktitle}{\emph{CVPR}}.
\newblock


\bibitem[\protect\citeauthoryear{Zhang, Liu, Dong, and Qiao}{Zhang
  et~al\mbox{.}}{2019b}]%
        {zhang2019ranksrgan}
\bibfield{author}{\bibinfo{person}{Wenlong Zhang}, \bibinfo{person}{Yihao Liu},
  \bibinfo{person}{Chao Dong}, {and} \bibinfo{person}{Yu Qiao}.}
  \bibinfo{year}{2019}\natexlab{b}.
\newblock \showarticletitle{Ranksrgan: Generative adversarial networks with
  ranker for image super-resolution}. In \bibinfo{booktitle}{\emph{ICCV}}.
\newblock


\bibitem[\protect\citeauthoryear{Zhang, Zeng, Guo, and Zhang}{Zhang
  et~al\mbox{.}}{2022}]%
        {zhang2022efficient}
\bibfield{author}{\bibinfo{person}{Xindong Zhang}, \bibinfo{person}{Hui Zeng},
  \bibinfo{person}{Shi Guo}, {and} \bibinfo{person}{Lei Zhang}.}
  \bibinfo{year}{2022}\natexlab{}.
\newblock \showarticletitle{Efficient long-range attention network for image
  super-resolution}. In \bibinfo{booktitle}{\emph{ECCV}}.
\newblock


\bibitem[\protect\citeauthoryear{Zhang, Zeng, and Zhang}{Zhang
  et~al\mbox{.}}{2021c}]%
        {zhang2021edge}
\bibfield{author}{\bibinfo{person}{Xindong Zhang}, \bibinfo{person}{Hui Zeng},
  {and} \bibinfo{person}{Lei Zhang}.} \bibinfo{year}{2021}\natexlab{c}.
\newblock \showarticletitle{Edge-oriented convolution block for real-time super
  resolution on mobile devices}. In \bibinfo{booktitle}{\emph{ACMMM}}.
\newblock


\bibitem[\protect\citeauthoryear{Zhang, Chen, Chen, Deng, Xu, and Wang}{Zhang
  et~al\mbox{.}}{2021a}]%
        {zhang2021data}
\bibfield{author}{\bibinfo{person}{Yiman Zhang}, \bibinfo{person}{Hanting
  Chen}, \bibinfo{person}{Xinghao Chen}, \bibinfo{person}{Yiping Deng},
  \bibinfo{person}{Chunjing Xu}, {and} \bibinfo{person}{Yunhe Wang}.}
  \bibinfo{year}{2021}\natexlab{a}.
\newblock \showarticletitle{Data-free knowledge distillation for image
  super-resolution}. In \bibinfo{booktitle}{\emph{CVPR}}.
\newblock


\bibitem[\protect\citeauthoryear{Zhang, Li, Li, Wang, Zhong, and Fu}{Zhang
  et~al\mbox{.}}{2018b}]%
        {Zhang_2018_ECCV}
\bibfield{author}{\bibinfo{person}{Yulun Zhang}, \bibinfo{person}{Kunpeng Li},
  \bibinfo{person}{Kai Li}, \bibinfo{person}{Lichen Wang},
  \bibinfo{person}{Bineng Zhong}, {and} \bibinfo{person}{Yun Fu}.}
  \bibinfo{year}{2018}\natexlab{b}.
\newblock \showarticletitle{Image Super-Resolution Using Very Deep Residual
  Channel Attention Networks}. In \bibinfo{booktitle}{\emph{ECCV}}.
\newblock


\bibitem[\protect\citeauthoryear{Zhang, Li, Li, Wang, Zhong, and Fu}{Zhang
  et~al\mbox{.}}{2018c}]%
        {zhang2018image}
\bibfield{author}{\bibinfo{person}{Yulun Zhang}, \bibinfo{person}{Kunpeng Li},
  \bibinfo{person}{Kai Li}, \bibinfo{person}{Lichen Wang},
  \bibinfo{person}{Bineng Zhong}, {and} \bibinfo{person}{Yun Fu}.}
  \bibinfo{year}{2018}\natexlab{c}.
\newblock \showarticletitle{Image super-resolution using very deep residual
  channel attention networks}. In \bibinfo{booktitle}{\emph{ECCV}}.
\newblock


\bibitem[\protect\citeauthoryear{Zhang, Li, Li, Zhong, and Fu}{Zhang
  et~al\mbox{.}}{2019a}]%
        {zhang2019residual}
\bibfield{author}{\bibinfo{person}{Yulun Zhang}, \bibinfo{person}{Kunpeng Li},
  \bibinfo{person}{Kai Li}, \bibinfo{person}{Bineng Zhong}, {and}
  \bibinfo{person}{Yun Fu}.} \bibinfo{year}{2019}\natexlab{a}.
\newblock \showarticletitle{Residual non-local attention networks for image
  restoration}.
\newblock \bibinfo{journal}{\emph{arXiv preprint arXiv:1903.10082}}
  (\bibinfo{year}{2019}).
\newblock


\bibitem[\protect\citeauthoryear{Zhang, Tian, Kong, Zhong, and Fu}{Zhang
  et~al\mbox{.}}{2018d}]%
        {zhang2018residual}
\bibfield{author}{\bibinfo{person}{Yulun Zhang}, \bibinfo{person}{Yapeng Tian},
  \bibinfo{person}{Yu Kong}, \bibinfo{person}{Bineng Zhong}, {and}
  \bibinfo{person}{Yun Fu}.} \bibinfo{year}{2018}\natexlab{d}.
\newblock \showarticletitle{Residual dense network for image super-resolution}.
  In \bibinfo{booktitle}{\emph{CVPR}}.
\newblock


\bibitem[\protect\citeauthoryear{Zhang, Wei, Qin, Wang, Pfister, and Fu}{Zhang
  et~al\mbox{.}}{2021b}]%
        {zhang2021context}
\bibfield{author}{\bibinfo{person}{Yulun Zhang}, \bibinfo{person}{Donglai Wei},
  \bibinfo{person}{Can Qin}, \bibinfo{person}{Huan Wang},
  \bibinfo{person}{Hanspeter Pfister}, {and} \bibinfo{person}{Yun Fu}.}
  \bibinfo{year}{2021}\natexlab{b}.
\newblock \showarticletitle{Context reasoning attention network for image
  super-resolution}. In \bibinfo{booktitle}{\emph{ICCV}}.
\newblock


\bibitem[\protect\citeauthoryear{Zhang, Wang, Lin, and Qi}{Zhang
  et~al\mbox{.}}{2019c}]%
        {zhang2019image}
\bibfield{author}{\bibinfo{person}{Zhifei Zhang}, \bibinfo{person}{Zhaowen
  Wang}, \bibinfo{person}{Zhe Lin}, {and} \bibinfo{person}{Hairong Qi}.}
  \bibinfo{year}{2019}\natexlab{c}.
\newblock \showarticletitle{Image super-resolution by neural texture transfer}.
  In \bibinfo{booktitle}{\emph{CVPR}}.
\newblock


\bibitem[\protect\citeauthoryear{Zhao, Zhang, Zhang, and Zou}{Zhao
  et~al\mbox{.}}{2019}]%
        {zhao2019channel}
\bibfield{author}{\bibinfo{person}{Xiaole Zhao}, \bibinfo{person}{Yulun Zhang},
  \bibinfo{person}{Tao Zhang}, {and} \bibinfo{person}{Xueming Zou}.}
  \bibinfo{year}{2019}\natexlab{}.
\newblock \showarticletitle{Channel splitting network for single MR image
  super-resolution}.
\newblock \bibinfo{journal}{\emph{IEEE Transactions on Image Processing}}
  \bibinfo{volume}{28}, \bibinfo{number}{11} (\bibinfo{year}{2019}),
  \bibinfo{pages}{5649--5662}.
\newblock


\bibitem[\protect\citeauthoryear{Zheng, Ji, Wang, Liu, and Fang}{Zheng
  et~al\mbox{.}}{2018}]%
        {zheng2018crossnet}
\bibfield{author}{\bibinfo{person}{Haitian Zheng}, \bibinfo{person}{Mengqi Ji},
  \bibinfo{person}{Haoqian Wang}, \bibinfo{person}{Yebin Liu}, {and}
  \bibinfo{person}{Lu Fang}.} \bibinfo{year}{2018}\natexlab{}.
\newblock \showarticletitle{Crossnet: An end-to-end reference-based super
  resolution network using cross-scale warping}. In
  \bibinfo{booktitle}{\emph{ECCV}}.
\newblock


\bibitem[\protect\citeauthoryear{Zhou, Fan, Cao, Jiang, and Yin}{Zhou
  et~al\mbox{.}}{2015}]%
        {zhou2015learning}
\bibfield{author}{\bibinfo{person}{Erjin Zhou}, \bibinfo{person}{Haoqiang Fan},
  \bibinfo{person}{Zhimin Cao}, \bibinfo{person}{Yuning Jiang}, {and}
  \bibinfo{person}{Qi Yin}.} \bibinfo{year}{2015}\natexlab{}.
\newblock \showarticletitle{Learning face hallucination in the wild}. In
  \bibinfo{booktitle}{\emph{AAAI}}.
\newblock


\bibitem[\protect\citeauthoryear{Zhou, Cai, Gu, Li, Liu, Chen, Qiao, and
  Dong}{Zhou et~al\mbox{.}}{2022a}]%
        {zhou2022efficient}
\bibfield{author}{\bibinfo{person}{Lin Zhou}, \bibinfo{person}{Haoming Cai},
  \bibinfo{person}{Jinjin Gu}, \bibinfo{person}{Zheyuan Li},
  \bibinfo{person}{Yingqi Liu}, \bibinfo{person}{Xiangyu Chen},
  \bibinfo{person}{Yu Qiao}, {and} \bibinfo{person}{Chao Dong}.}
  \bibinfo{year}{2022}\natexlab{a}.
\newblock \showarticletitle{Efficient image super-resolution using
  vast-receptive-field attention}. In \bibinfo{booktitle}{\emph{ECCV}}.
\newblock


\bibitem[\protect\citeauthoryear{Zhou, Chan, Li, and Loy}{Zhou
  et~al\mbox{.}}{2022b}]%
        {zhou2022towards}
\bibfield{author}{\bibinfo{person}{Shangchen Zhou}, \bibinfo{person}{Kelvin
  Chan}, \bibinfo{person}{Chongyi Li}, {and} \bibinfo{person}{Chen~Change
  Loy}.} \bibinfo{year}{2022}\natexlab{b}.
\newblock \showarticletitle{Towards robust blind face restoration with codebook
  lookup transformer}.
\newblock \bibinfo{journal}{\emph{NeurIPS}} (\bibinfo{year}{2022}).
\newblock


\bibitem[\protect\citeauthoryear{Zhou, Li, Guo, Bai, Cheng, and Hou}{Zhou
  et~al\mbox{.}}{2023}]%
        {zhou2023srformer}
\bibfield{author}{\bibinfo{person}{Yupeng Zhou}, \bibinfo{person}{Zhen Li},
  \bibinfo{person}{Chun-Le Guo}, \bibinfo{person}{Song Bai},
  \bibinfo{person}{Ming-Ming Cheng}, {and} \bibinfo{person}{Qibin Hou}.}
  \bibinfo{year}{2023}\natexlab{}.
\newblock \showarticletitle{Srformer: Permuted self-attention for single image
  super-resolution}. In \bibinfo{booktitle}{\emph{ICCV}}.
\newblock


\bibitem[\protect\citeauthoryear{Zhu, Zhu, Chu, Zhang, Ji, Wang, and Tai}{Zhu
  et~al\mbox{.}}{2022}]%
        {zhu2022blind}
\bibfield{author}{\bibinfo{person}{Feida Zhu}, \bibinfo{person}{Junwei Zhu},
  \bibinfo{person}{Wenqing Chu}, \bibinfo{person}{Xinyi Zhang},
  \bibinfo{person}{Xiaozhong Ji}, \bibinfo{person}{Chengjie Wang}, {and}
  \bibinfo{person}{Ying Tai}.} \bibinfo{year}{2022}\natexlab{}.
\newblock \showarticletitle{Blind face restoration via integrating face shape
  and generative priors}. In \bibinfo{booktitle}{\emph{CVPR}}.
\newblock


\bibitem[\protect\citeauthoryear{Zhu, Park, Isola, and Efros}{Zhu
  et~al\mbox{.}}{2017}]%
        {zhu2017unpaired}
\bibfield{author}{\bibinfo{person}{Jun-Yan Zhu}, \bibinfo{person}{Taesung
  Park}, \bibinfo{person}{Phillip Isola}, {and} \bibinfo{person}{Alexei~A
  Efros}.} \bibinfo{year}{2017}\natexlab{}.
\newblock \showarticletitle{Unpaired image-to-image translation using
  cycle-consistent adversarial networks}. In \bibinfo{booktitle}{\emph{ICCV}}.
\newblock


\bibitem[\protect\citeauthoryear{Zhu, Li, and Li}{Zhu et~al\mbox{.}}{2023}]%
        {zhu2023attention}
\bibfield{author}{\bibinfo{person}{Qiang Zhu}, \bibinfo{person}{Pengfei Li},
  {and} \bibinfo{person}{Qianhui Li}.} \bibinfo{year}{2023}\natexlab{}.
\newblock \showarticletitle{Attention Retractable Frequency Fusion Transformer
  for Image Super Resolution}. In \bibinfo{booktitle}{\emph{CVPR}}.
\newblock


\bibitem[\protect\citeauthoryear{Zhu, Liu, Loy, and Tang}{Zhu
  et~al\mbox{.}}{2016}]%
        {zhu2016deep}
\bibfield{author}{\bibinfo{person}{Shizhan Zhu}, \bibinfo{person}{Sifei Liu},
  \bibinfo{person}{Chen~Change Loy}, {and} \bibinfo{person}{Xiaoou Tang}.}
  \bibinfo{year}{2016}\natexlab{}.
\newblock \showarticletitle{Deep cascaded bi-network for face hallucination}.
  In \bibinfo{booktitle}{\emph{ECCV}}.
\newblock


\bibitem[\protect\citeauthoryear{Zontak and Irani}{Zontak and Irani}{2011}]%
        {zontak2011internal}
\bibfield{author}{\bibinfo{person}{Maria Zontak} {and} \bibinfo{person}{Michal
  Irani}.} \bibinfo{year}{2011}\natexlab{}.
\newblock \showarticletitle{Internal statistics of a single natural image}. In
  \bibinfo{booktitle}{\emph{CVPR}}.
\newblock


\end{thebibliography}

\end{document}